\documentclass[24pt, usenatbib]{mn2e}
\usepackage{graphicx}
\usepackage{times}
\usepackage{bm}
\usepackage{epsf}
\usepackage{amsmath}
\usepackage{amssymb}
\usepackage{subfig}
\usepackage{float}
\usepackage[usenames,dvipsnames]{color}
\date{\today,~ $ $Revision: 0.9 $ $}
\def\la{\langle}
\def\ra{\rangle}
\def\n{\noindent}
\def\be{\begin{equation*}}
\def\ee{\end{equation*}}
\def\ben{\begin{eqnarray*}}
\def\een{\end{eqnarray*}}
\def\nn{\nonumber}
\def\oh{\hat\Omega}
\def\myC{{\cal C}}

\def\rk{{\rm K}}

\def\br{{\bf r}}
\def\inc{{\int_0^{r_s}}}
\def\rH{\rm H}

\def\myC{{\cal C}}

\def\ho{{\hat \Omega}}

\def\2p{{(2\pi)^2}}

\def\be{\begin{equation}}
\def\ee{\end{equation}}
\def\beq{\begin{equation}}
\def\eeq{\end{equation}}
\def\ben{\begin{eqnarray}}
\def\een{\end{eqnarray}}

\def\oh{{\hat\Omega}}
\def\nn{{\nonumber}}

\def\rT{{\rm T_0}}
\newcommand{\beqa}{\begin{eqnarray}}
\newcommand{\eeqa}{\end{eqnarray}}
\newcommand{\edth}{\,\eth\,}
\newcommand{\edthbar}{\,\overline{\eth}\,}

\def\ex1{{\int {d^2 {\bf l} \over (2
\pi)^2}~ {\rm P}_{\pi} { \big ( {l\over d_A(r)} \big )} b_l^2(\theta_b)}}

\def\kmin1{{\int_0^{r_s}\; \omega_{\rm SZ}(r)\; dr }}

\def\ex{{\cal I}_{\theta_b}}
\def\kmin{{\Theta_m}}

\begin{document}

\onecolumn
\title{Reconstructing the Thermal Sunyaev-Zel'dovich Effect in 3D}

\author[G. Pratten and D. Munshi]
{Geraint Pratten$^{1}$, Dipak Munshi$^{2}$\\
$^{1}$School of Physics and Astronomy, Cardiff University, Queen's
Buildings, 5 The Parade, Cardiff, CF24 3AA, UK\\
$^{2}$ Astronomy Centre, School of Mathematical and Physical Sciences, University of Sussex, Brighton BN1 9QH, UK}
\maketitle
\begin{abstract}
The thermal Sunyaev-Zel'dovich (tSZ) effect measures the line-of-sight projection of the thermal pressure of free electrons and lacks any redshift information. By cross-correlating the
tSZ effect with an external cosmological tracer we can recover a good fraction of this lost information.
Weak lensing (WL) is thought to provide an unbiased probe of the dark Universe, with many WL surveys having sky coverage that overlaps with tSZ surveys.
Generalising the tomographic approach, we advocate the use of the
spherical Fourier-Bessel (sFB) expansion to perform an analysis of the cross-correlation
between the projected (2D) tSZ Compton $y$-parameter maps and 3D weak lensing convergence maps. We use redshift dependent linear biasing and the halo model as a tool to investigate the tSZ-WL cross-correlations in 3D. We use the Press-Schechter (PS) and the Sheth-Tormen (ST) mass-functions in
our calculations, finding that the results are quite sensitive to detailed modelling. We provide detailed 
analysis of surveys with photometric and spectroscopic redshifts. The
signal-to-noise $(S/N)$ of the cross-spectra $\myC_{\ell} (k)$ for individual 3D modes, defined by the
radial and tangential wave numbers $( k ; \ell )$, remains comparable to, but below, unity though optimal
binning is expected to improve this. The results presented can be generalised to
analyse other CMB secondaries, such as the kinetic Sunyaev-Zel'dovich (kSZ) effect.
\end{abstract}
\begin{keywords}: Cosmology-- Thermal Sunyaev Zel'dovich Surveys -- Methods: analytical, statistical, numerical
\end{keywords}
\section{Introduction}
Only 50\% of baryons consistent with cosmic microwave background radiation (CMBR) and 
big bang nucleosynthesis (BBN) observations have been detected observationally \citep{FP04,FP06}. The validation of standard
cosmological model relies on our ability to detect the missing baryons observationally \citep{Breg07}.
The cosmological simulations suggest that majority of the IGM are in the form of a warm-hot intergalactic medium (WHIM)
with temperature $10^5{\rm K}<{\rm T}<10^7{\rm K}$ \citep{CO99,Dave01,CO06}. It is also believed that WHIMs reside in moderately overdense 
structures such as the filaments. Being collisionally ionized, these baryons do not leave any
footprints in the Lyman-$\alpha$ absorption systems. The emission from WHIMs in either UV or X-ray
are too weak to be detected given the sensitivity of current instruments and detection in X-ray given is also unfeasible given the low level of emission from WHIM. However, the baryons in the cosmic web do have sufficient velocity and column density 
to produce a detectable CMB secondary effect also known as the kinetic Sunyaev Zeldovich (kSZ) effect \citep{SZ80}. 

Secondary anisotropies arise at all angular scales; the largest secondary anisotropy
at the arcminute scale is the thermal Sunyaev-Zeldovich (tSZ) effect. 
The tSZ effect is caused by the thermal motion of electrons mainly
from hot ionized gas in galaxy clusters where as the {\em kinetic} Sunyaev Zeldovich (kSZ) effect is attributed to the
bulk motion of electrons in an ionized medium \citep{SZ72,SZ80}.
The tSZ can be separated from CMB maps using spectral information.
Along with weak lensing of CMB, the kSZ is the most dominant secondary
contribution at arcminute scales after the removal of tSZ effect. This is because the primary CMB
is sub-dominant on these scales as a result of Silk-damping. Although the tSZ is capable of overwhelming the CMB primaries on cluster scales, the blind detection of the tSZ effect on a random direction in the sky is difficult as the CMB primaries dominate on angular scales larger than that of the clusters. The tSZ and kSZ are both promising probes of the ionized fractions of the baryons with the majority of the tSZ effect being caused by electrons in virialized collapsed objects \cite{WSP09,HH09} with overdensities that can be considerably high $\delta > 100$. 

A detailed mapping and understanding of the SZ effect is of particular interest to cosmology and astrophysics as it is thought that the SZ effect will be a powerful method to detect galaxy clusters at high redshifts. One of the key and central features to the tSZ effect is that the efficiency of the free electron distribution in generating the tSZ effect seems to be independent of redshift. Cosmological expansion introduces an energy loss of $1+z$ to a photon emitted from a source at redshift $z$ but the scattering of these CMB photons off electrons increases the energy of the photons by a factor of $1+z$. These two effects cancel allowing us to use the tSZ effect as a probe for galaxy clusters at high redshifts. In addition, the tSZ effect will be a powerful probe of the thermal history of our Universe as it is a direct probe of the thermal energy of the intergalactic medium and intracluster medium. Two of the main drawbacks of tSZ studies are that the tSZ has been shown to be sensitive to a wide number of 
astrophysical processes introducing degeneracies and that the tSZ is a measure of the projected electron thermal energy along the line of sight. This smears all redshift information and the contributions of the various astrophysical processes become badly entangled with the projection effects. 

Recent studies have proposed the reconstruction and recovery of redshift information by cross-correlating the tSZ effect with galaxies and their photometric redshift estimates \citep{ZP01, Shao11b}. One of the leading methods proposed in the literature is tomographic reconstruction in which we crudely bin the data into redshift slices and construct the 2D projection for each bin. The auto (single bin) and cross (between bins) correlations can then be used to constrain model parameters and extract information. 

This paper is concerned with extending these studies to a full 3D analysis in which we necessarily avoid binning data and therefore avoid the consequential loss of information. In principle, 3D studies would allow a full sky reconstruction that includes the effects of sky curvature and extended radial coverage. As SZ studies are often followed by photometric or spectroscopic galaxy surveys, we expect that photometric redshifts up to $z \sim 1.3 - 2$ will be readily available in due course. We investigate the cross-correlation of the tSZ with an external tracer given by cosmological weak lensing and photometric redshift surveys. 

Current ongoing and proposed future ground based surveys, such as SZA\footnote{http://astro.uchicago.edu/sza},
ACT\footnote{http://www.physics.princeton.edu/act}, APEX\footnote{http://bolo.berkeley.edu/apexsz},
SPT\footnote{http://pole.uchicago.edu} and the recently completed all sky Planck survey \cite{Planck13y}, have published
a map of the entire y-sky with a great precision (also see \cite{HS14}). The high multipole $\ell \sim 3000$ tSZ power spectrum has been observed by the SPT \citep{Lu10, Saro13,Hanson13,Holder13,Vieira13,Hou14,Story13,High12} collaboration with the ACT \citep{Fw10,Dn10,  Shegal11,Hand11,Sher11,Wil13,Dunk13,Calab13} collaboration reporting an analysis on similar scales. It is expected that ongoing surveys will improve these
measurements due to their improved sky coverage as well as wider frequency range.

It is important to appreciate why
the study of secondaries such as tSZ should be an important aspect of any CMB mission.
In addition to the important physics the secondaries probe,
accurate modeling of the secondary non-Gaussianities is required to
avoid $20\%-30\%$ constraint degradations in
future CMB data-sets such as Planck\footnote{http://www.rssd.esa.int/index.php?project=planck} and CMBPol\footnote{http://cmbpol.uchicago.edu/} \citep{Smidt10}.

While the tSZ surveys described above provide a direct probe of the baryonic Universe, weak lensing
observations \citep{MuPhysRep08} on the other hand can map the dark matter distribution in an unbiased way.
In recent years there has been tremendous progress on the technical front in terms of specification
and control of systematics in weak lensing observables. There are many current ongoing weak lensing surveys such as
CFHT{\footnote{http://www.cfht.hawai.edu/Sciences/CFHLS/}}
legacy survey, Pan-STARRS{\footnote{http://pan-starrs.ifa.hawai.edu/}}
and the Dark Energy survey (DES){\footnote{https://www.darkenergysurvey.org/}}. In the future, the Large Synoptic Survey Telescope (LSST){\footnote{http://www.lsst.org/llst\_home.shtml}},
Joint Dark Energy Mission (JDEM){\footnote{http://jdem.gsfc.nasa.gov/}} and Euclid {\footnote{http://sci.esa.int/euclid/}} 
will map the dark matter and dark energy distribution of the entire sky in unprecedented detail.
Among other things, these surveys hold great promise in shedding light on
the nature of dark energy and the origin of neutrino masses \citep{JK11}, where the
weak lensing signals dominate the others considered by e.g. the Dark Energy Task Force \citep{Al11}.
However, the optimism that has been associated with weak lensing is predicated on first overcoming the vast systematic uncertainties in both
the measurements and the theory \citep{HS04,MHH05,CH02,HS03,Wh04,Hu06,MTC06}.
The statistics of the weak lensing convergence have been studied in great detail using an extension of perturbation theory  \citep{MuJa00,MuJai01,MuVaBa04} and
methods based on the halo model \citep{CH00,TJ02,TJ03}. These studies developed techniques that can be used to predict the lower-order moments (equivalent to the
power spectrum and multi-spectra in the harmonic domain) and the entire PDF
for a given weak lensing survey. The photometric redshifts of source galaxies are useful
for tomographic studies of the dark matter distribution and in establishing a three-dimensional
picture of their distribution \citep{MunshiK11}. Finally, cross correlations with other tracers of large scale structure, such as intensity mapping from future 21cm surveys, could also be considered \citep{Chang08}. 

This paper is primarily motivated by the recent paper \citep{WHM14} where
the CFHTLenS data with Planck tSZ maps was correlated. They measure a non-zero correlation between the two
maps out to one degree angular separation on the sky, with an overall significance of six sigma and use the 
results to conclude a substantial fraction of the ”missing” baryons in the universe may reside in a low density 
warm plasma that traces dark matter. An internal detection of the tSZ effect and CMB
lensing cross-correlation in the Planck nominal mission data has also recently been reported at a significance of 6.2 sigma \citep{HS14}.
While these correlations were computed using 2D projections, we develop techniques for cross-correlation studies in 3D that go beyond the tomographic treatment \citep{HT95,H03,BHT95,Castro05,PM13}. 

This paper is organised as follows. In Section \textsection\ref{sec:Notation} we outline some key notation and cosmological parameters that will be adopted throughout this paper. Section \textsection\ref{sec:tSZ} forms the core of our paper and introduces in more detail the tSZ effect, cosmological weak lensing and photometric redshift surveys. The cross-correlations of the tSZ with the external tracers is detailed and a discussion of realistic survey effects is introduced. We highlight different approaches to cosmological weak lensing, notably the halo model, and understand how redshift space distortions affect our spectra. Finally, section \textsection\ref{sec:conclu} is reserved for concluding remarks as well as a discussion of our results. 

In this paper we focus on the tSZ effect. The corresponding results for the kSZ effect will be presented elsewhere.

\section{Notations}\label{sec:Notation}
\label{Not}
\n
In this section we introduce our notations for tSZ effect as well as for the weak lensing convergence
that will be used later. We will use the following line element:
\be
ds^2 = -c^2 dt^2 + a^2(r)(dr^2 + d^2_{\rm A}(r)(\sin^2\theta d\theta^2 + d\phi^2) )
\ee
Here $d_A(r)$ is the comoving angular diameter distance at a (comoving) radial distance $r$ and can be given in terms of the curvature density parameter $\Omega_K = 1 - \Omega_{\rm M} - \Omega_{\Lambda}$
\begin{align}
 d_A \left( r \right) &=  {\lambda_{\rm H}}\frac{\sin_\rk ( | \Omega_\rk |^{1/2} r / \lambda_{\rm H} )}{ | \Omega_\rk |^{1/2} }; \quad  \lambda_{\rm H} = {c}{\,\rH^{-1}_0};
\end{align}
\n
where $\sin_\rk$ means $\sinh$ if $ \Omega_\rk > 0$ or $\sin$ if $ \Omega_\rk < 0$; if $\Omega_\rk = 0$, then $d_A \left( r \right) = r$. The radial comoving distance from a source at redshift $z$ to an observer at $z = 0$ is given by:
\begin{align}
&r(z) = \lambda_{\rm H}\displaystyle\int\limits^z_0 \frac{dz^{\prime}}{E (z')} \, ; \qquad
E^2 (z) = \frac{\rH^2 (z)}{\rH^2_0} = \Omega_{\Lambda} + \Omega_\rk \left( 1 + z \right) + \Omega_{\rm M} \left( 1 + z \right)^3. 
\label{eqn:Ez}
\end{align}
 \n
We also adopt the notation that $a(z)=1/(1+z)$ is the scale factor at redshift $z$ normalised to unity at redshift $z=0$. 
The particular cosmology that we will adopt for numerical 
studies is specified by the following parameters (to be introduced later):
$\Omega_\Lambda = 0.741,\; h=0.72,\; \Omega_b = 0.044,\; \Omega_{\rm CDM} = 0.215,\;
\Omega_{\rm M} = \Omega_b+\Omega_{\rm CDM},\; n_s = 0.964,\; w_0 = -1,\; w_a = 0,\;
\sigma_8 = 0.803,\; \Omega_\nu = 0$. In such a cosmology, $\Omega_K = 0$ and we can take $d_A (r) = r$ for numerical results. We do, however, keep functions of the curvature in our equations for generality. Throughout this paper $c$ will denote speed of light and will be set to unity.
\begin{figure}
\centering
\textbf{\;\;\;\;\;\;\;\;\;\;\;\; WL Power Spectrum}\par\medskip
  \includegraphics[width=65mm]{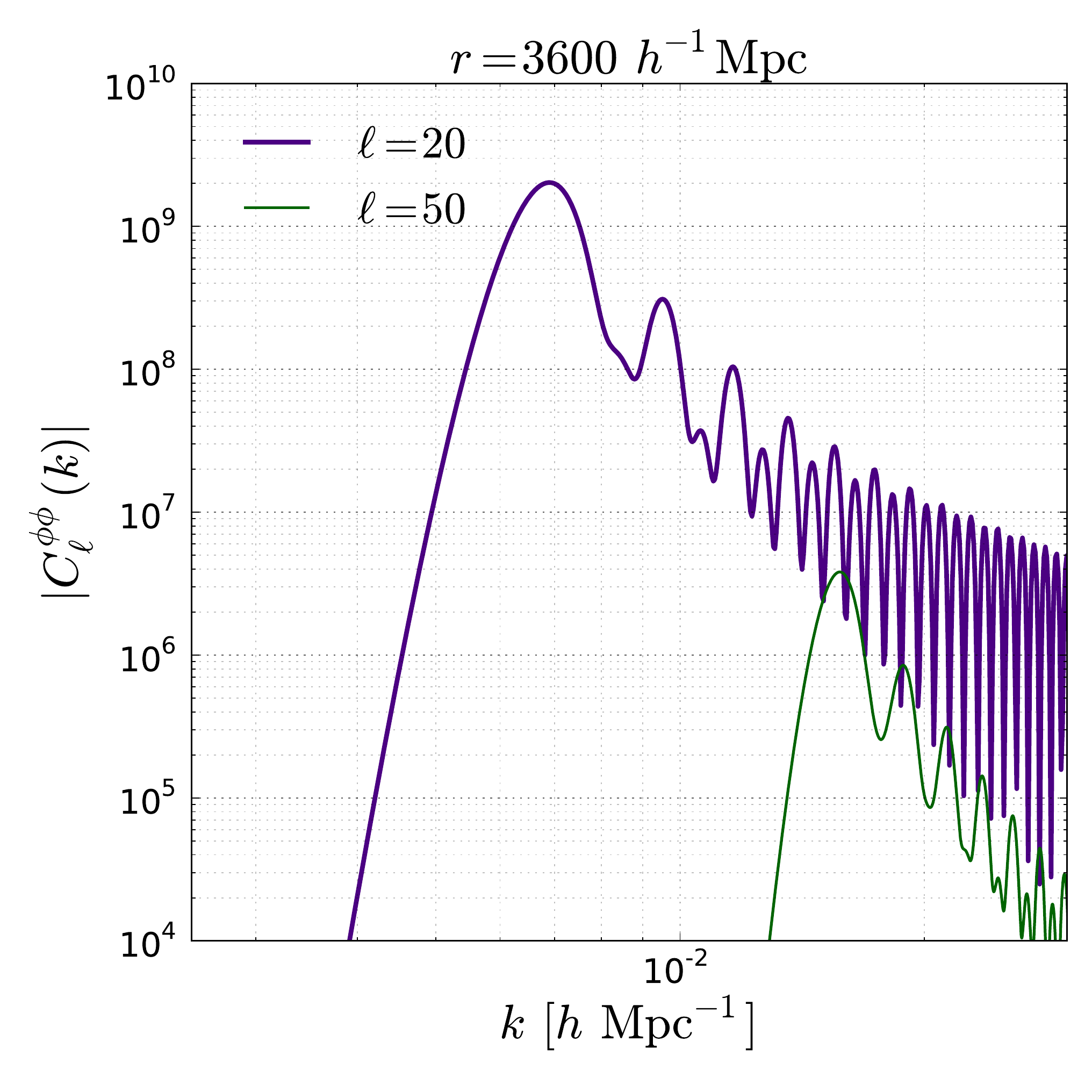}
\caption{Weak lensing potential power spectrum for a survey of depth $r = 3600  h^{-1} \textrm{Mpc}$ and multipoles $\ell = \lbrace 20 , 50 \rbrace$. The other lensing variables, such as convergence and shear, can be reconstructed from this potential. The lensing spectra are a function of $\ell$ and two wavenumbers $k_1$ and $k_2$. In this plot we only consider the diagonal contribution $k_1 = k_2 = k$. We are interested in the low-$k$ behaviour at scales where linear regime predictions are robust. Scales above $k = 1.5 h \rm Mpc^{-1}$ are entering the highly non-linear regime.  }
\label{fig:Cl_phi-phi}
\end{figure}
\section{Reconstructing the Thermal Sunyaev-Zeldovich Effect in 3D}\label{sec:tSZ}
Mapping the thermal Sunyaev-Zel'dovich (tSZ) sky is of particular interest given that the tSZ is a 
powerful probe of galaxy-clusters. This is related to the fact that the efficiency of free-electrons generating the tSZ effect is virtually
independent of redshift. This is in direct contrast to probing high redshift clusters using X-ray emissions. Various groups, including the
ACT group and the SPT group, routinely report blind detections of high redshift clusters using tSZ surveys. The Planck collaboration has already released frequency cleaned all-sky $y$-maps \citep{Planck13y} as well as a catalogue of 861 confirmed galaxy clusters \citep{Planck13cl}.
In addition to detection of galaxy clusters, the tSZ effect can be used to probe the thermal history of the Universe as it directly probes the thermal energy of the intergalactic medium (IGM) and the intercluster medium (ICM) respectively. A number of simulations have confirmed that the tSZ effect is very sensitive to a compendium of astrophysical
processes that include the radiative cooling and feedback in terms of SN energy injections. In addition, the tSZ is the projection of electron pressure
along the line of sight. There exists degeneracies among the various competing processes responsible for tSZ effect. The idea of recovering redshift information using a cross-correlation of tSZ maps with galaxy surveys, both photometric and spectroscopic, using tomographic bins has previously been studied \citep{ZP01, Shao11b}. In this section we extend these results to full 3D analysis using a sFB transformation that aims to utilise redshift information of specific sources from the start \citep{PM13}. First we outline the sFB formalism before providing a more detailed introduction to the tSZ effect. We then proceed to construct cross-correlation spectra for tSZ-WL and tSZ-spectroscopic redshift surveys. This includes a discussion of the halo model approach to large scale clustering as well as a treatment of realistic selection functions and photometric redshift errors. 

\subsection{Spherical Fourier-Bessel Expansion}
Spherical coordinates are a natural choice for the analysis of cosmological data sets as, by an appropriate choice of coordinates, we can place an observer at the origin of the analysis. Future WL and tSZ surveys will provide an unprecedented level of detail by generating extended sky maps with large radial coverage. We therefore require a simultaneous treatment of the extended radial coverage and the spherical sky geometry. For this problem the spherical Fourier-Bessel (sFB) expansion is a natural basis for the analysis of cosmological random fields. The sFB formalism has seen an increased use in the literature over the past few years, for example it has been used in: weak lensing studies \citep{H03,Castro05,HKT06,MuCoHe11,Kitching11,ASW12}, redshift space distortions \citep{HT95,PM13}, relativistic effects \citep{February13,Yoo13} and studies of baryon acoustic oscillations \citep{Rassat12, PM13, Grassi13}. 

We initially consider a homogeneous 3D random field $\psi (\hat{\Omega} , r)$ such that $\hat{\Omega}$ defines a position on the surface of the sphere and $r$ denotes the comoving radial distance. The eigenfunctions of the Laplacian operator in spherical coordinates are constructed from products of the spherical Bessel function $j_{\ell} (kr)$ and the spherical harmonics $Y_{\ell m} (\hat{\Omega})$ with concomitant eigenvalues of $-k^2$. As such, the 3D expansion of a scalar field in the sFB formalism and its inverse transformation are given by a spectral decomposition of the 3D field with respect to these eigenfunctions:
\begin{align}
 \psi \left( {\bf{r}} \right) = \sqrt{\frac{2}{\pi}} \, \displaystyle\int k dk \, \displaystyle\sum_{\lbrace \ell m \rbrace} \psi_{\ell m} (k) \, j_{\ell} (k r) \, Y_{\ell m} (\theta , \varphi ) ; \quad
 \psi_{\ell m} (k) = \sqrt{\frac{2}{\pi}} \, \displaystyle\int d^3 {\bf r} \, \psi ({\bf{r}}) \, k \, j_{\ell} (kr) \, Y^{\ast}_{\ell m} \left( \theta , \varphi \right).
\end{align}
\n
In our notation $\lbrace \ell m \rbrace$ are quantum numbers and $k$ is simply the wavenumber. These equations will be a recurring feature in the 3D analysis presented in this paper. Using this decomposition we can define a 3D power spectra of the field $\psi ({\bf{r}}\, ; r)$ by:
\begin{align}
 \left\langle \psi_{\ell m} \left( k ; r \right) \, \psi^{\ast}_{\ell^{\prime} m^{\prime}} \left( k^{\prime} ; r \right) \right\rangle &= \mathcal{C}^{\psi \psi}_{\ell} (k ; r) \delta_{1D} \left( k - k^{\prime} \right) \, \delta^{\rm K}_{\ell \ell^{\prime}} \, \delta^{\rm K}_{m m^{\prime}} .
\end{align}
\n
Remember, we have implicitly assumed that the field $\psi ({\bf{r}};r)$ is statistically homogeneous and isotropic such that the decomposition in the full sky has been defined at some instant in time corresponding to $r$. The time dependence of the coefficients is necessary to ensure 3D spatial homogeneity. In the 3D homogeneous case, the power spectrum is independent of $\ell$ and collapses to the standard Cartesian Fourier power spectrum $\myC_{\ell} (k ; r) = P(k;r)$. 

Now if we cross-correlate the field at two different radial distances $r$ and $r^{\prime}$, which is effectively calculating the cross-correlation between two different homogeneous and isotropic fields $\psi ({\bf{r}} ; r)$ and $\psi ({\bf{r}}^{\prime} ; r^{\prime})$, then it is possible to argue that 3D homogeneity and isotropy still holds \citep{Castro05}:
\begin{align}
 \left\langle \psi_{\ell m} ( k ; r) \, \psi^{\ast}_{\ell^{\prime} m^{\prime}} (k^{\prime} ; r^{\prime}) \right\rangle &= \myC^{\psi \psi}_{\ell} (k ; r , r^{\prime}) \, \delta_{1D} (k - k^{\prime}) \, \delta^{\rm K}_{\ell \ell^{\prime}} \, \delta^{\rm K}_{m m^{\prime}}.
\end{align}
\n
We can analogously construct the cross-correlation of two different homogeneous 3D fields $\psi ({\bf{r}} ; r)$ and $\phi ({\bf{r}} ; r)$, such as weak gravitational lensing of the CMB and cosmic shear fields, as follows 
\begin{align}
 \left\langle \psi_{\ell m} \left( k ; r \right) \, \phi^{\ast}_{\ell^{\prime} m^{\prime}} \left( k^{\prime} ; r^{\prime} \right) \right\rangle &= \mathcal{C}^{\psi \phi}_{\ell} (k ; r, r^{\prime}) \delta_{1D} \left( k - k^{\prime} \right) \, \delta^{\rm K}_{\ell \ell^{\prime}} \, \delta^{\rm K}_{m m^{\prime}}.
\end{align}

Importantly, for cosmological weak lensing, the 3D lensing potential $\phi$ is not homogeneous and isotropic in 3D space as it is defined as a 2D projection at each source distance $r$ \citep{Castro05}. Consequentially, its 2D projection retains the homogeneous and isotropic characteristics but this does not extend to the radial direction. This is contrary to the gravitational potential which is 3D homogeneous and requires a slight tweak to the equations presented above. The cross-correlation of two 3D inhomogeneous fields can be expressed as follows
\begin{align}
 \left\langle \phi_{\ell m} ( k ) \, \psi^{\ast}_{\ell^{\prime} m^{\prime}} (k^{\prime}) \right\rangle &= \myC^{\phi \psi}_{\ell} (k, k^{\prime}) \, \delta_{1D} (k - k^{\prime}) \, 
\delta^{\rm K}_{\ell \ell^{\prime}} \, \delta^{\rm K}_{m m^{\prime}}
\end{align}
\n
where the expansion coefficients are written without the time dependence explicitly shown as the fields are, by definition, not homogeneous and isotropic in 3D space (please see \citep{Castro05} for an extended discussion of this point). 

In addition to the above, we can also consider the scenario where we have a generic 2D projected field $\psi (\hat{\Omega})$ that samples an underlying 3D field ${{\Psi}} ({\bf{r}})$ according to a weight function $w_{\psi} (r)$: 
\begin{align}
 &\psi (\hat{\Omega}) = \int\limits^{\infty}_0 dr \, w_{\psi} (r) \, \Psi ({\bf{r}}); \qquad
 \psi_{\ell m} = \sqrt{\frac{2}{\pi}} \int\limits^{\infty}_0 dr w_{\psi} (r) \, \int dk \, k \, j_{\ell} (kr) {{\Psi}}_{\ell m} (k ; r).
\end{align}
\n
From this representation we can construct the power spectra and cross-correlated spectra of the 2D projected field with other 2D or 3D fields in the sFB formalism. The cross correlation of 2D projected harmonics $\psi_{\ell m}$ with a 3D field $\phi_{\ell m}$ is given by:
\begin{align}
 \left\langle \phi_{\ell m} (k) \psi^*_{\ell^{\prime} m^{\prime}} \right\rangle &= \myC^{\phi \psi}_{\ell} (k) \, \delta^{\rm K}_{\ell \ell^{\prime}} \, \delta^{\rm K}_{m m^{\prime}}.
\end{align}

Finally, the sFB decomposition may be extended to spin-$s$ fields as until now we have only dealt with scalar fields of spin-$0$. This has been covered in the literature and is conceptually based on constructing eigenfunctions to a spin-$s$ field (see \cite{Castro05,MunshiK11} for recent applications). The decomposition is given by:
\begin{align}
 &_{s}{\psi ({\bf{r}} )} = \displaystyle\int\limits^{\infty}_0 dk \, \displaystyle\sum_{\lbrace \ell m \rbrace} \, \left[ _{s}{\psi_{\ell m}} \right] \, \left[ _{s}{Z}_{k \ell m} ({\bf{r}}) \right];  \qquad
 _{s} \psi_{\ell m} (k) = \displaystyle\int d^3 {\bf{r}} \, \left[ _{s}\psi ({\bf{r}}) \right] \, \left[ _{s}Z_{k \ell m}^{\ast} ({\bf{r}}) \right].
\end{align}
\n
where the generalised spin-$s$ harmonic eigenfunctions in the sFB formalism are given by:
\begin{align}
 _{s}Z_{k \ell m} (\Omega , r) = \sqrt{\frac{2}{\pi}} \, k \, j_{\ell} (kr) \, _{s}Y_{\ell m} (\Omega).
\end{align}
\n
These functions simply reduce to the scalar sFB functions in the limit $s=0$, as expected. The spin-weighted harmonics will be used later to construct higher-spin generalisations of the shear and convergence in weak lensing. 

\subsection{Thermal Sunyaev Zel'dovich Effect}
The Sunyaev-Zel'dovich effect is generated by inverse Compton scattering of CMB photons by intervening electrons in hot, ionised gas. In short, a low energy CMB photon encounters a high energy electron such that the resultant scattering imparts energy onto the photon thereby increasing its frequency. This upsurge in frequency results in a spectral distortion of the CMB in a well understood manner. As energies are relatively low, no more than a few keV in most galaxy clusters, it is often sufficient to restrict ourselves to a non-relativistic treatment. One of the most powerful characteristics of the thermal Sunyaev-Zel'dovich (tSZ) effect is that the main observable, the Compton $y$-parameter, does not seem to have any significant dependence on the redshift. For this reason it is hoped that large-scale SZ maps can probe the redshift evolution of structures, the intra-cluster medium (ICM) and trace out the thermal history of the Universe. As a side note, it is also important to mention that, in addition to 
the above, the 
motion of the hot, ionised gas with respect to the CMB photons produces a spectral distortion arising from Doppler effects, allowing for the estimation of the peculiar velocities of the clusters. As a disclaimer we note that in our analysis we necessarily neglect non-thermal contributions to the tSZ effect. 

The tSZ effect generates a contribution to the CMB temperature fluctuation which is typically expressed as:
\begin{align}
\delta_{\rm T}(\nu,\oh) &= \frac{\delta \rT(\oh)}{\rT} = g(x)y(\oh) . 
\end{align}
\n
In this expression $g(x_{\nu})$ corresponds to the spectral dependence and $y(\oh)$ encodes 
the angular dependence; $x_{\nu}$ represents the dimensionless frequency and $\oh=(\theta,\phi)$ corresponds to a 
unit vector that signifies pixel positions on the sky. A subscript $s$ will be used to denote the smoothed maps e.g. $y_s(\oh)$.  
In the non-relativistic limit $g(x)$ takes the following form:
\begin{align}
g(x) &= x\coth \left ({x \over 2} \right ) -4 = \left ( x {e^{x} +1 \over e^{x} -1} - 4\right ); \quad\quad  x_{} = {h\nu \over k_B\rT}= {\nu \over 56.84 {\rm GHz}} = {5.28 {\rm mm} \over \lambda};
\label{eq:def_g}
\end{align}
Here $k_B$ and $h$ are the Boltzmann and Planck constant respectively; $\nu$ denotes the 
frequency of the photon and $\rT=2.726 \; {\rm K}$ is the mean temperature of the CMB sky. The tSZ effect presents itself as a CMB temperature 
decrement at $\nu \ll 218{\rm GHz}$ and
as an temperature increment at $\nu \gg 218{\rm GHz}$ with a null point at $\nu=218{\rm GHz}$. In the Rayleigh Jeans 
limit, characterized by $x \ll 1$, $g(x) \approx -2$ is roughly independent of frequency. The other limiting situation is for $x \gg 1$,  for which $g(x) \approx (x-4)$. 
Key information on the thermal history of the Universe is encoded in the $y(\oh)$ maps that
are extracted from the frequency maps obtained through multi frequency CMB observations. The $y$ maps
are opacity weighted integrated pressure fluctuations along the line of sight \citep{CHT00} \footnote{Note that \cite{CHT00} does not separate out the spectral distortions $g(x)$ from $y$ whereas we do, i.e. $y^{\rm Cooray} = g(x) y ( \oh)$. This also leads to different definitions for the window function $w_{\rm SZ} (r)$. }
\begin{align}
y(\oh) &\equiv \int ds\; {n_e \sigma_{\rm T}} {k_B T_e \over m_e c^2}= {\sigma_{\rm T} \over m_e c^2} \int_0^{r_{\rm H}} dr\; a(r) \, n_ek_BT_e(\oh,r) =  
{\sigma_{\rm T} \over m_e c^2}\int_0^{\eta_{\rm H}} d\eta \; a(\eta)\; \Pi_e(\eta,\oh) = \int_0^{r_{\rm H}} \; dr\; w_{\rm SZ}(r) \pi_e(r) 
\label{eq:weight_tsz} .
\end{align}
We have introduced various notations that appear in the literature in the context of study of the tSZ effect
here $\Pi_e = n_ek_BT_e$. 
In our notation $m_e$ corresponds to the electron mass, $k_B$ denotes the Boltzmann's constant,
$\sigma_{\rm T}= 6.65 10^{-25} {\rm cm^2}$ represents the Thompson cross-section, $n_e$ denotes the number
density of electrons and $T_e$ to the electron temperature. The conformal time is denoted by $d\eta = dt/a(t)$. 
The line of sight integral 
depends on the comoving radial co-ordinate distance $r$ and the corresponding scale factor of the Universe $a(r)$.
The weight is defined as:
\begin{align}
 w_{\rm SZ}(r) = \dot \tau(r) = \sigma_{\rm T} n_e(r)a(r) ,
\end{align}
where the dot defines the derivative with respect to 
comoving radial distance $r$ and the 3D pressure fluctuation is defined as $\pi_e = k_B T_e /m_e c^2$. To our detriment, however, the redshift information is typically lost due to the projection along the line of sight. These projection effects severely compromise the power of the tSZ effect in probing the thermal history of the Universe. Tomographic and 3D methods aim to recover this otherwise discarded information. As such, we aim to cross-correlate the comptonization map $y(\oh)$ with tomographic and projected maps from weak lensing surveys to constrain the thermal history of the Universe and its evolution with redshift. Throughout  
we will consider the Rayleigh-Jeans part of the spectrum $\delta_T = -2y $; for ACT and SPT operating
at $\nu = 150 {\rm GHz}$ from Eq.(\ref{eq:def_g}) we get $g(x) = -0.95$. 

Detailed modeling of
the bias is only required for the computation of the variance, $\la\delta y^2(\oh)\ra$, which
samples the pressure fluctuation power spectrum $P_{\pi\pi}$ and can be expressed as:
\be
\la \delta y^2(\oh)\ra_c = \inc d {r}
{\omega_{\rm SZ}^2(r) \over d^2_A(r)} \int {d^2 {\bf l} \over (2
\pi)^2}~ {\rm P}_{\pi\pi} { \left [ {\ell\over d_{\rm A}(r)}, r \right ]} b^2_{\ell}(\theta_s).
\label{eq:ps}
\ee
The pressure power spectrum ${\rm P}_{\pi\pi}(k,z)$ at a redshift $z$ is expressed in 
terms of the underlying power spectrum ${\rm P}_{\delta\delta}(k,z)$
using a bias $b_{\pi}(k,z)$ i.e. ${\rm P}_{\pi\pi}(k,z)=b_{\pi}^2(k,z){\rm P}_{\delta\delta}(k,z)$. The bias $b_{\pi}(k,z)$ is assumed
to be independent of length scale or equivalently wave number $k$; i.e. $b_{\pi}(k,z)=b_{\pi}(z)$. The redshift dependent
bias can be expressed as: $b_{\pi}(z)= b_{\pi}(0)/(1+z)$. Here $b_{\pi}(0)$ can be written as 
$b_{\pi}(0)= k_B T_e(0) b_{\delta}/m_e c^2$. Different values of $b_{\delta}$ have been reported by various authors;
e.g. \citep{Ref00} found $b_{\delta} \approx 8-9$ and $T_e(0)\approx 0.3-0.4$. On the other hand 
\citep{Sel01} found $b_{\delta} \approx 3-4$ and $T_e(0)\approx 0.3-0.4$.
 Typical value of $b_{\pi}(0)$ found by \citep{CO99} is $b_{\pi}(0)=0.0039$.
This is a factor of two lower than the value used by \citep{GS99a,GS99b} and \citep{CH01}. A 
Gaussian beam $b_{\ell}(\theta_s)$ with FWHM at $\theta_s$ is assumed. Unless otherwise stated we adopt the value of \citep{CO99} of $b_{\pi} (0) = 0.0039$ in our linear biasing scheme. We will consider a spectrum of values later for $b_{\pi} (0)$ later on in this paper (see Figure \ref{fig:tSZ-WL}). 
\subsection{Cross-correlating tSZ with Weak-Lensing Surveys in 3D}
\subsubsection{Weak Lensing}
Gravitational lensing is the concomitant deflection of light that arises from fluctuations in the gravitational potential. Lensing of a background source refers to the distortions of the source images generated by the deflection of light, along the line of sight between a source and an observer, caused by the fluctuations in the gravitational potential of the intervening mass distribution. The two most notable effects of gravitational lensing are the shearing and magnification of the images of the sources.

The weak lensing potential $\phi ({\bf{r}})$ can be related to the gravitational potential $\Phi$ by the line of sight integral:
\begin{align}
 \phi ({\bf{r}}) &= \frac{2}{c^2} \displaystyle\int\limits^r_{0} d r^{\prime}  F_{\rm K}(r,r')
\Phi (r^{\prime} , \theta , \varphi ); \quad F_{\rm K}(r,r') \equiv \frac{f_K (r - r^{\prime})}{f_K (r) f_K (r^{\prime})}; 
 \label{eqn:phi_WL}
\end{align}
\n
making using the Born approximation in assuming that the path of the photons is unperturbed by the lens. Here the function $f_{\rm K}(r)$ is a distance function depending on the curvature of the Universe: $f_{\rm K} (r)=\sin_{\rm K}(r)$.

Being able to link the 3D lensing potential $\phi$ to the 3D gravitational potential $\Phi$ is one of the most important steps in linking the observables of cosmological weak lensing to theoretical predictions. In a perturbed cosmology, the gravitational potential can be linked to the overdensity field $\delta ({\bf{r}}) = \delta \rho({\bf{r}}) / \rho$ by Poisson's equation in comoving coordinates using the comoving gauge:
\begin{align}
 \triangle \, \Phi ({\bf{r}}) &= \frac{3}{2} \frac{\Omega_{\rm M} H^2_0}{2 a(t)} \, \delta \left( {\bf{r}} \right).
\end{align}
\n
The aim of 3D weak lensing is to use information of the distance to individual source galaxies to avoid averaging over the redshift distribution of source galaxies. This raises the interesting possibility of being able to determine the full 3D mass density in a non-parametric way by estimating the unprojected tidal shear perpendicular to the line of sight direction from the distortion of a source galaxies' ellipticity. This is fundamentally different to estimations arising from tomographic weak-lensing or angular line-of-sight approaches which both necessitate the averaging of weak-lensing observables with the line-of-sight galaxy distribution. Distance information is inferred from the corresponding photometric redshift data. The idea of 3D weak lensing was first introduced by \cite{H03} and it has been developed further by numerous authors \citep{Castro05,Kit07,
Massey07,Kit08,MuCoHe11,ASW12}. As we can see, the lensing potential defined in Eq. (\ref{eqn:phi_WL}) is dependent on the Hubble parameter which is sensitive to the contents of the Universe. This is one of the reasons as to why weak-lensing is a powerful probe of dark energy. In addition, the lensing potential is explicitly related to the gravitational potential which is also dependent on the matter content and non-linear growth of structure through its coupling to $\delta ({\bf{r}})$. Any model that we introduce to describe the non-linear growth of structure will have a direct impact on the behaviour of weak lensing observables. 

Performing a spectral decomposition of the lensing potential in the sFB formalism, we find that the lensing harmonics $\phi_{\ell m}$ can be expressed as:
\begin{align}
 \phi_{\ell m} \left( k \right) &= \frac{4 k}{\pi c^2} \int^{\infty}_{0} dk^{\prime} \, k^{\prime} \int^{\infty}_0 dr \, r \, j_{\ell} (kr) 
\int^r_0 F_\rk (r , r^{\prime}) j_{\ell} (k^{\prime} r^{\prime}) \, \Phi_{\ell m} (k^{\prime} , r^{\prime}).
 \label{eqn:phi_lm}
\end{align}
We have introduced a few features in the above. The dependence on $r$ appearing after a semi-colon, such as $\Phi_{\ell m} \left( k ; r \right)$ is just an expression of the time-dependence of the potential. This naturally translates into a dependence on the comoving distance as this intrinsically depends on the look-back time, in a rather circular manner, see \citep{Castro05} for further details. Lastly, we have introduced the harmonic decomposition of the gravitational potential $\Phi_{\ell m} (k ; r)$ which an be related to the overdensity via Poisson's equation:
\begin{align}
 \Phi_{\ell m} \left( k ; r \right) = - \frac{3}{2} \frac{\Omega_{\rm M} H^2_0}{k^2 a(r)} \delta_{\ell m} \left( k ; r \right) . 
\end{align}
The weak lensing power spectrum will be given by: $\left\langle \phi_{\ell m} (k) \phi^{\ast}_{\ell^{\prime} m^{\prime}} (k^{\prime}) \right\rangle = \mathcal{C}^{\phi \phi}_{\ell m} (k,k^{\prime}) \delta^{\rm K}_{\ell \ell^{\prime}} \delta^{\rm K}_{m m^{\prime}}$;
where we must remember that, due to the nature of look-back time, the 3D lensing potential is not homogeneous and isotropic in 3D space but homogeneous and isotropic on the 2D sky. Expanding the harmonics in the sFB formalism as per equation (\ref{eqn:phi_lm}) we see that the power spectrum can be written as (Figure \ref{fig:Cl_phi-phi}):
\ben
&& \mathcal{C}_{\ell}^{\phi \phi} (k_1 , k_2) = \frac{16}{\pi^2 c^4} \int_0^{\infty} dk^{\prime} k^{\prime 2} \, \mathcal{I}_{\ell}^{\phi} (k_1,k^{\prime}) \,
 \mathcal{I}_{\ell}^{\phi} (k_2,k^{\prime}); \label{eqn:I_phi1} \\
&& \mathcal{I}^{\phi}_{\ell} (k_i , k^{\prime} ) = k_i \int\limits^{\infty}_0 dr \, r^2 \, j_{\ell} (k_i r) \, \int\limits^r_0 d r^{\prime} \, F_K (r , r^{\prime}) \, j_{\ell} (k^{\prime} r^{\prime}) \, \sqrt{ P^{\Phi \Phi} (k^{\prime} ; r^{\prime}) } . \label{eqn:I_phi2} 
\een
\n
Typically, however, we will only look at diagonal cuts in the $(k_1 , k_2)$ plane at a given $\ell$, though we could certainly consider the full $(k_1 , k_2 , \ell)$ space should we need to. In Figure \ref{fig:Cl_phi-phi} we show typical power spectra for weak-lensing at configurations $\ell = \lbrace 20 , 50 \rbrace$ for $r = 3600 h^{-1} \rm Mpc$. These spectra are in agreement with the results presented in \cite{Castro05} and provide both a useful consistency check as well as a useful guide to the phenomenology of weak lensing cross-correlations.

We can reduce the dimensionality of the integrals by using the Limber approximation \citep{Limber54} which is valid for small angular separations and hence for large multipole moments $\ell$ in the harmonic domain. Employing the Limber approximation we can simplify Eq.(\ref{eqn:I_phi1}) - Eq.(\ref{eqn:I_phi2}) as follows:
\ben
&& \mathcal{I}^{\phi}_{\ell} (k_i , k^{\prime} ) = \sqrt{\pi \over 2} {\nu \over k^2} F_{\rm K}\left ({\nu \over k_i}, {\nu \over k'}\right )
\sqrt{ P^{\Phi\Phi}\left (k^{\prime}, {\nu \over k^{\prime}}\right)};\\
&& \myC^{\phi\phi}_{\ell}(k_1,k_2) = {4 \over c^2} {\nu^2 \over k_1^2k_2^2} \;\;\int_0^{\infty} F_K\left({\nu\over k^{\prime}},{\nu\over k_1}\right ) 
F_K\left ({\nu\over k^{\prime}},{\nu\over k_2}\right ) 
P^{\Phi\Phi} \left ( k^{\prime}, {\nu \over k^{\prime}} \right ) {k^{\prime}}^2 dk^{\prime}.
\een
\subsubsection{Shear and Convergence}
Weak lensing on the full sky can be aptly described by using spin-weighted spherical harmonics, the weak lensing shear is a spin-2 object after all. The 2D distortion of a source located a given 3D comoving position by intervening matter is given by:
\begin{align}
 \left[ \gamma ({\bf{r}} ) \right]_{i j} &= \left[ \begin{array}{cc}
        \gamma_1({\bf{r}})                 &    {\sin \theta}\, \gamma_2({\bf{r}})\\
        {\sin \theta}\,\gamma_2({\bf{r}})  &   -  {\sin^2 \theta}\,\gamma_1({\bf{r}})
\end{array}  \right] = \left[ \nabla_i \nabla_j - \frac{1}{2} g_{ij} \nabla^2 \right] \phi ({\bf{r}} );
\end{align}
\n
where $\gamma_1$ and $\gamma_2$ are components of the weak lensing shear induced by the gravitational tidal field. These can be encapsulated in a complex shear $\gamma ({\bf{r}}) = \gamma_1 ({\bf{r}}) + i \gamma_2 ({\bf{r}})$ and represent orthogonal modes of the distortion. Additionally we can construct the convergence field tensor that probes the magnification via the isotropic convergence scalar field $\kappa$:
\begin{align}
 \left[ \kappa ({\bf{r}}) \right]_{ij} &= \left[ \begin{array}{cc}
        1                 &    0\\
        0  &  \sin^2 \theta
\end{array} \right] \kappa ({\bf{r}}) = \frac{1}{2} g_{ij} \nabla^2 \phi ({\bf{r}}) .
\end{align}
\n
As given in \cite{Castro05}, the complex shear may be written in terms of the edth-derivative $\edth$ and its complex conjugate $\edthbar$. These two derivative operators were first introduced by \cite{Newman66,Goldberg66} as a generalisation of the covariant derivative to an operator acting on the surface of a sphere. The operator $\edth$ acts as a spin raising operator and $\edthbar$ acts a spin lowering operator on the quantum numbers $s$ of the spin weighted spherical harmonics $_s Y_{\ell m}$. The power of this approach is that we can relate spin-$s$ objects that are not invariant under rotations of the coordinate frame to scalar quantities that are invariant under rotations of the coordinate frame. The complex 3D shear 
$\gamma({\bf r}) =\gamma_1({\bf r})+i\gamma_2({\bf r}) $ itself is a spin-2 object but we can now relate it to an $\edth$ derivative of scalar functions.
The lensing potential is split into even and odd parity components such that:
\begin{align}
\gamma_1 ({\bf{r}}) \pm i \gamma_2 ({\bf{r}}) = \frac{1}{2} \edth \edth \left[ \phi_E ({\bf{r}}) \pm i \phi_B ( {\bf{r}}) \right].
\end{align}
In cosmological weak lensing, these equations are simplified as the shear field induced by gravitational tidal fields only has an even parity contribution, i.e. $\phi_B ({\bf{r}}) = 0$. This allows us to recover the orthogonal components of the shear $\gamma_1$,$\gamma_2$
and the convergence $\kappa$:
\begin{align}
\gamma_1 ({\bf{r}}) = \frac{1}{4} \left( \edth \edth + \edthbar \edthbar \right) \phi ({\bf{r}}); \qquad
\gamma_2 ({\bf{r}}) = - \frac{i}{4} \left( \edth \edth - \edthbar \edthbar \right) \phi ({\bf{r}}); \quad
 \kappa ({\bf{r}}) = \frac{1}{4} \left[ \edth \edthbar + \edthbar \edth \right] \, \phi ({\bf{r}}).
\end{align}
\n
Performing a 3D expansion allows us to relate the above equations for the shear and convergence to the lensing potential. This is made possible by knowing the effects of the $\edth$ and $\edthbar$ derivatives on spin weighted spherical harmonics. In particular it can be shown that:
\begin{align}
 _2\gamma_{\ell m} (k) = {_{-2}}\gamma_{\ell m} (k) = \frac{1}{2} \sqrt{ \frac{( \ell + 2) !}{(\ell - 2)!}}  \phi_{\ell m} (k); \quad
 \kappa_{\ell m} (k) = - \frac{\ell ( \ell + 1)}{2} \phi_{\ell m} (k).
 \label{eqn:gammakappa}
\end{align}
\n
Hence we can construct the power spectra of the convergence and shear as follows \citep{Castro05}:
\begin{align}
 &\myC_{\ell}^{\kappa \kappa} (k_1, k_2) = \left[ \frac{\ell \left( \ell + 1 \right)}{2} \right]^2 \myC^{\phi \phi}_{\ell} (k_1 , k_2) ; \qquad \myC_{\ell}^{\gamma \gamma} (k_1 , k_2) = \left[ \frac{1}{4} \frac{\left( \ell + 2 \right) !}{\left( \ell - 2 \right) !} \right] \, \myC_{\ell}^{\phi \phi} (k_1 , k_2) .
\end{align}
\subsubsection{tSZ-Weak Lensing Convergence}
In this section we construct the cross-correlation spectra between the thermal Sunyaev-Zel'dovich effect and 3D weak lensing. As noted, the tSZ effect directly probes the integrated thermal pressure of free electrons along the line of sight giving us valuable information on the thermal history of our Universe and hence the baryonic Universe. Weak lensing observations probe the dark matter Universe in a relatively unbiased way. Due to the integrated nature of the tSZ effect, redshift information can be completely lost diminishing the ability of tSZ observations in distinguishing between different thermal histories. By cross-correlating the tSZ effect with external tracers, such as WL, we hope to recover some of the information that has been lost. The unprojected nature of 3D WL makes this a very interesting candidate for an external tracer, especially as many of the tSZ surveys will have sky coverage that overlaps with upcoming weak lensing surveys. A key idea here, much as in tomography, is that the dark 
matter distribution up to a given redshift will be correlated with the tSZ effect. 

\begin{figure}
\centering
\textbf{tSZ-WL Cross-Correlation}\par\medskip
{\includegraphics[width=60mm]{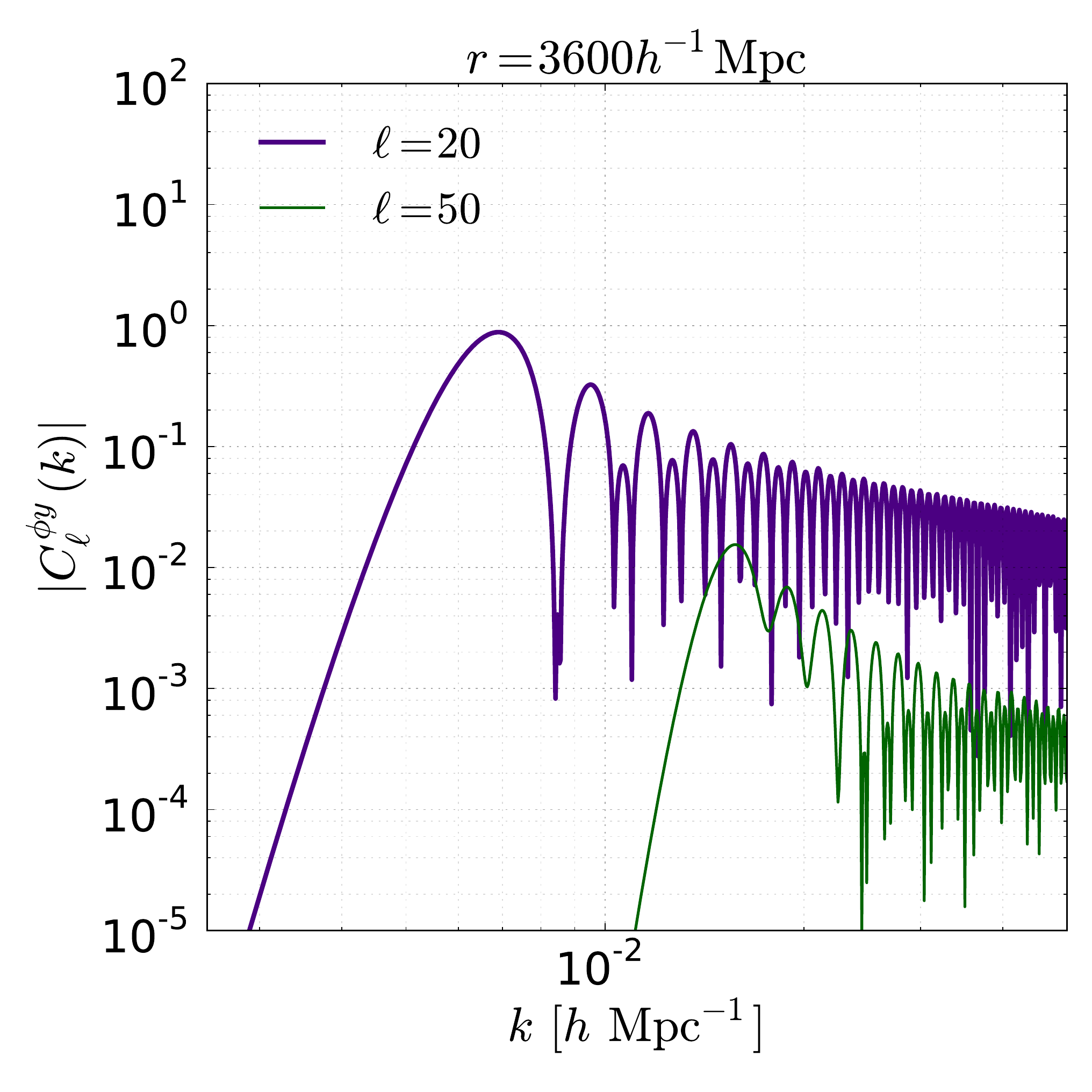}}
{\includegraphics[width=60mm]{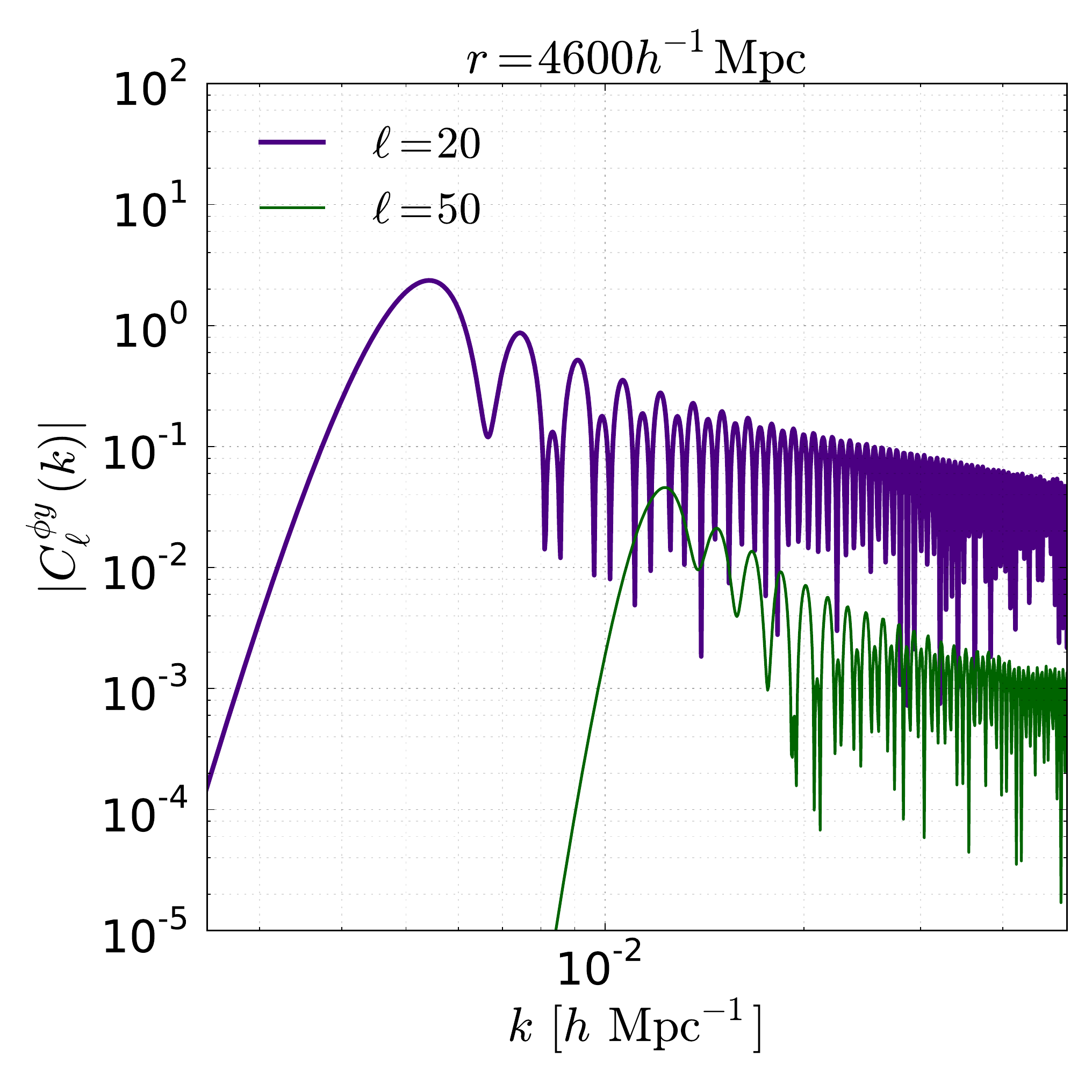}}
{\includegraphics[width=60mm]{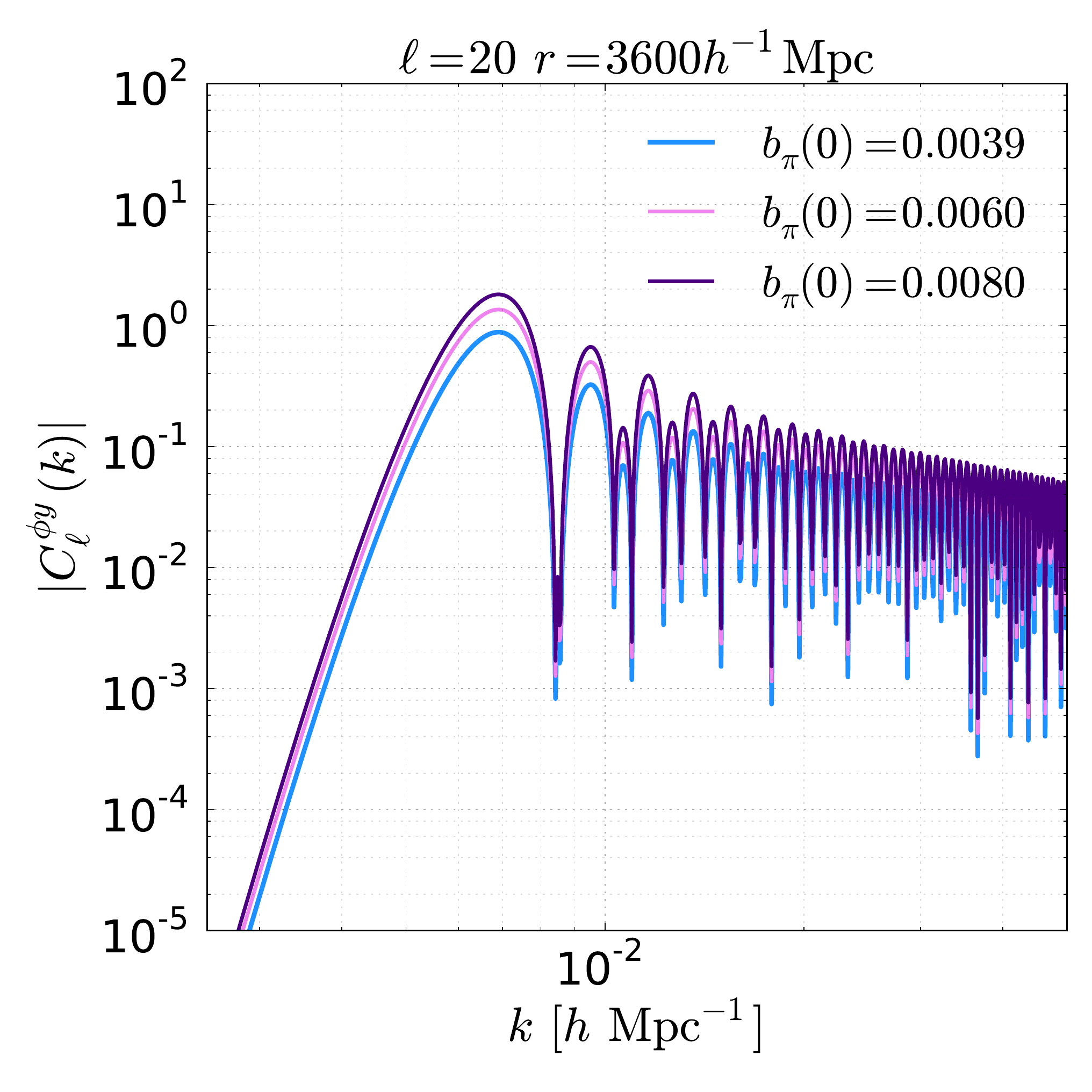}}
\caption{The left panel shows the tSZ-WL cross-correlation for a survey of depth $r = 3600 h^{-1} \rm Mpc$ and the centre panel is for a survey of depth $r = 4600 h^{-1} \rm Mpc$. These plots show the diagonal contribution in the $(k_1 , k_2)$ plane for $\ell \in \lbrace 20, 50 \rbrace$. Note that the approximate Bessel function inequality comes into play around $k r \geq \ell$ after which we have the decaying oscillatory behaviour. For increasing $\ell$ the terms become more sharply peaked. The right panel shows the variation of the tSZ-WL cross-correlation with regards to the linear bias scheme discussed earlier. We take three values for $b_{\pi} (0)$ corresponding to a range of values found in the literature \citep{CO99,GS99a,GS99b,CH01}. }
\label{fig:tSZ-WL}
\end{figure}

In order to construct the cross-correlation spectra we need an expression for the projected 2D tSZ field $y ({\hat{\Omega}})$ sampling the underlying 3D pressure fluctuation field $\pi_e (r)$ decomposed into harmonics:
\begin{align}
 &y ( \hat{\Omega} ) = \int\limits^{\infty}_0 \, dr \, w_{\rm SZ} (r) \, \pi_e (r) \, ; \qquad
 y_{\ell m} = \sqrt{\frac{2}{\pi}} \int\limits^{\infty}_0 dr \, w_{\rm SZ} (r) \, \int_0^{\infty} dk \, k \, j_{\ell} (kr) \, \left[\pi_e \right]_{\ell m} (k ; r) .
\end{align}
\n
We can now cross-correlate the harmonics $y_{\ell m}$ with the lensing potential harmonics $\phi_{\ell m}$ to obtain the following cross-correlation spectra:
$ \left\langle \phi_{\ell m} (k) \, y^*_{\ell^{\prime} m^{\prime}} \right\rangle = \mathcal{C}^{\phi y}_{\ell} (k) \, \delta_{\ell \ell^{\prime}} \delta_{m m^{\prime}};$
where the power spectrum $\myC_{\ell}^{\phi y} (k)$ is explicitly given by (Figure \ref{fig:tSZ-WL}):
\ben
&& \mathcal{C}^{\phi y}_{\ell} (k) = \frac{4}{\pi c^2} \int\limits^{\infty}_0 \, dk^{\prime} \, k^{\prime 2} \,  
\mathcal{I}^y_{\ell} (k^{\prime}) \,\mathcal{I}^{\phi}_{\ell} (k,k^{\prime}); \label{eqn:Cl_phi-y} \quad\quad \\
&& \mathcal{I}^y_{\ell} (k) = k^2 \sqrt{\frac{2}{\pi}} \, \int\limits^{\infty}_0 dr \, w_{\rm SZ} (r) \, j_{\ell} (kr) \, b_{\pi} (k ; r) \, \sqrt{ P^{\Phi \Phi} (k ; r) }; \label{eqn:I_y} \\
&& \mathcal{I}^{\phi}_{\ell} (k , k^{\prime} ) = k \int\limits^{\infty}_0 dr \, r^2 \, j_{\ell} (k r) \, \int\limits^r_0 d r^{\prime} \, F_K (r , r^{\prime}) \, j_{\ell} (k^{\prime} r^{\prime}) \, \sqrt{ P^{\Phi \Phi} (k^{\prime} ; r^{\prime}) }.  \label{eqn:I_phi} 
\een
The bias coefficient, $b_{\pi}(k;r)$, encodes the scale dependent biasing scheme described above. We have assumed that the power spectrum can be well approximated by $P^{\Phi\Phi}(k; r,r')=\sqrt{P^{\Phi\Phi}(k;r)}\sqrt{P^{\Phi\Phi}(k;r')}$. This is tantamount to stating that we are only interested in correlations in the potential field over small distances for which the look back time is negligible and hence $r \simeq r^{\prime}$ \citep{Castro05}. We have also used the result $\myC_\ell(k,k')=P(k)$, where the 3D power-spectrum is defined in terms of the Cartesian Fourier transform:
\ben
&& \Psi({\bf k}) = {1 \over (2\pi)^{3/2}} \int d^3{\bf k}\,\Psi({\bf k})\,e^{i{\bf k}\cdot{\bf r}}; \quad
\Psi({\bf r}) = {1 \over (2\pi)^{3/2}} \int d^3{\bf k}\,\Psi({\bf k})\,e^{-i{\bf k}\cdot{\bf r}}; \\
&& \la \Psi({\bf k})\Psi^*({\bf k}') \ra \equiv (2\pi)^3 \,P_{\Psi\Psi}(k)\,\delta_{3\rm D}({\bf k}-{\bf k}').
\een
In Figure \ref{fig:tSZ-WL} we restrict ourselves to a limited set of cases for numerical calculations. We choose four configurations corresponding to $r_{\textrm{max}} \in \lbrace 3600 , 4600 \rbrace \, h^{-1} \, \textrm{Mpc}$ and $\ell \in \lbrace 20, 50 \rbrace$. The results are well-sampled at low $k$ but the resolution of the numerical integrals drops above $k \sim 10^{-1}$ as the approximate Bessel function inequality $k r \geq \ell$ dominates resulting in a highly oscillatory tail that we do not consider of prime importance for this work. As noted in \cite{Castro05}, the differences between individual spectra are only slight but there are a wider range of useful $\ell$ modes that a full 3D study has access to. This increases the sensitivity of the tSZ-WL cross-correlation to cosmological models. 
In addition to the 3D cross-correlation presented here, recent work has focused on higher order correlations of tSZ and WL using
tomographic bins \citep{MuJoCoSm11}, analytical schemes to describe statistical aspects of the projected $y$-sky using moment-based methods \citep{Mu13} and topological estimators \citep{MuSmJoCo12}. 
\subsubsection{Extended Limber Approximation}
The computations of higher-order multispectra are often difficult due to the presence of complex multidimensional integrals that make numerical computations expensive if not prohibitive. The Limber approximation \citep{Limber54}, and its Fourier space generalisation, are often used to simplify the numerical calculations by reducing the dimensionality of the integrals. The use of the Limber approximation is valid on small angular separations and hence for large multipole moments $\ell$ in the harmonic domain. It requires smooth variations of the integrand compared to the Bessel functions or relevant $\ell$. A framework for calculating higher order corrections to the Limber approximation was presented in \citep{LoVerde08}. Starting with the expression for the angular spectra in Eqn. (\ref{eqn:Cl_phi-y}) and the corresponding kernels in Eqns. (\ref{eqn:I_y}) and (\ref{eqn:I_phi}) we re-write the spectra as follows
\begin{align}
 \myC_{\ell}^{\phi y} (k) &= \frac{4 k}{\pi c^2} \sqrt{\frac{2}{\pi}} \int d r_a w_{\rm SZ} (r_a) \int d r_b r_b^2 j_{\ell} (k r_b) \int d r^{\prime}_b F_K (r_b , r^{\prime}_b );\nn  \\
&\quad \times \int dk^{\prime} \, {k^{\prime}}^4 \, j_{\ell} (k^{\prime} r_a) \, j_{\ell} (k_2 r^{\prime}_b) b_{\pi} (k^{\prime} ; r_a) \sqrt{ P_{\Phi \Phi} (k^{\prime} ; r_a) P_{\Phi \Phi} (k^{\prime} ; r^{\prime}_b) }.
\end{align}
\n
Applying the extended Limber approximation to the $k_2$ integral, we find the expression collapses to the following:
\ben
&& I_{\ell}(k) = w_{\rm SZ} \left( {\nu \over k} \right ) b_{\pi}\left (k^{\prime}; {\nu \over k^{\prime}} \right )
\sqrt{P^{\Phi\Phi}\left (k^{\prime}; {\nu \over k^{\prime}}\right )}; \quad
I_{\ell}(k,k^{\prime}) = {\pi \over 2} {\nu \over k^2} F_{\rm K}\left({\nu \over k},{\nu \over k^{\prime}}\right) 
\sqrt{P^{\Phi\Phi}(k^{\prime}; {\nu \over k^{\prime}})}; \\
&& \myC^{\phi y}_{\ell}(k) = {2 \over c^2}{\nu \over k^2}\int_0^{\infty} {k^{\prime}}^2 dk' w_{\rm SZ}\left ({\nu \over k^{\prime}}\right ) 
b_{\pi}\left (k^{\prime}; {\nu \over k^{\prime}} \right )P^{\Phi\Phi}\left ( k^{\prime}; {\nu \over k^{\prime}} \right ); 
\quad\quad \nu = \ell+{1 \over 2}.
\een
\subsubsection{tSZ-Weak Lensing Shear and Flexions}
Alternatively, it is possible to express the power-spectrum of weak lensing observables, such as the convergence, shear, flexions
and the Compton $y$-parameter maps, in terms of $\myC_{\ell}^{\phi y}$. As we saw, the components of the 3D shear $\gamma_1(\br)$ and $\gamma_2(\br)$ and the convergence $\kappa(\br)$ can be expressed in terms of the complex 3D lensing potential $\phi ({\bf{r}})$ using the spin-raising $\edth$ and spin-lowering operators $\bar\edth$ \citep{Castro05}. We will introduce  
${}_{\pm 2}\Gamma(\br) = \gamma_1(\br)\pm i\gamma_2(\br)$ to denote the complex shear $\gamma(\br)$ and its conjugate $\gamma^*(\br)$.
The harmonics of $\Gamma$ can be decomposed in terms of Electric ``E''  and Magnetic ``B'' mode polarizations. The $\Gamma$''s are spin-2 objects and can be decomposed using the spin-2 spherical harmonics \cite{MuSmHeCoCoo11}:
\begin{align}
 {}_{\pm 2}\Gamma_{\ell m}  = -[E_{\ell m} \pm i B_{\ell m}] .
\end{align}
\n
Ignoring the B-mode contribution, as gravitational tidal fields only generate an electric contribution, we have ${}_{\pm 2}\Gamma_{\ell m}  = -E_{\ell m} $ which is related to the lensing harmonics via
\begin{align}
E_{\ell m}(k) &= -{1 \over 2} \sqrt{(\ell+2)!\over (\ell-2)!} \phi_{\ell m}(k) .
\label{eq:shearkappa}
\end{align}
\n
In addition we have the $\kappa_{\ell m}$ harmonics defined in Eqn~\ref{eqn:gammakappa}. 

Higher order spin objects that can be generated from the lensing potential, known as \textit{flexions}, have often been used to study weak lensing \citep{GN02,GN05,BRE00,BG05,Schneider08}. They are related to the derivatives of the shear or convergence and are sensitive to information about substructures beyond that which can be
studied using just the shear or convergence alone. The most commonly used flexions are the spin-1 or {\em first} flexion $\cal F$ and the spin-3 or {\em second} flexion $\cal G$. Their relationship with the shapelet formalism
have been discussed at length in the literature \citep{Ref03,BJ02,RefBeac03}. Both of these 
flexions have been used extensively in the literature for individual
halo profiles and as well as the study of substructures \citep{Beacon06}. These flexions can be used to study weak 'arciness' in images of lensed galaxies. The flexions are defined as follows
\begin{align}
&{\cal F}(\br) = {1 \over 6} \left (\bar\edth\edth\edth+\edth\bar\edth\edth + \edth\edth\bar\edth \right)\phi(\br); \qquad
{\cal G}(\br) = {1 \over 2} \left( \bar\edth\bar\edth\bar\edth \right) \phi(\br).
\end{align}
\n
The harmonic decomposition of these objects is obtained by expanding in a spin weighted spherical harmonic basis ${}_sY_{\ell m}(\oh)$ (see \cite{MuSmHeCoCoo11} for
a detailed derivation and discussion) and evaluating the $\edth$ and $\edthbar$ derivatives on the spin weighted spherical harmonics. 
\begin{align}
&{\cal F}_{\ell m} = {1 \over 6} \left[ \ell (\ell+1) \right]^{1/2} \left[ 3\ell^2 + 3\ell-2 \right] \phi_{\ell m}; \qquad
{\cal G}_{\ell m} = {1 \over 2} \sqrt {(\ell+3)! \over (\ell-3)!} \, \phi_{\ell m}.
\label{eq:flex}
\end{align}
\begin{figure}
 \centering
 \textbf{Conditional Probability Function}\par\medskip
{ \includegraphics[width=0.328\textwidth]{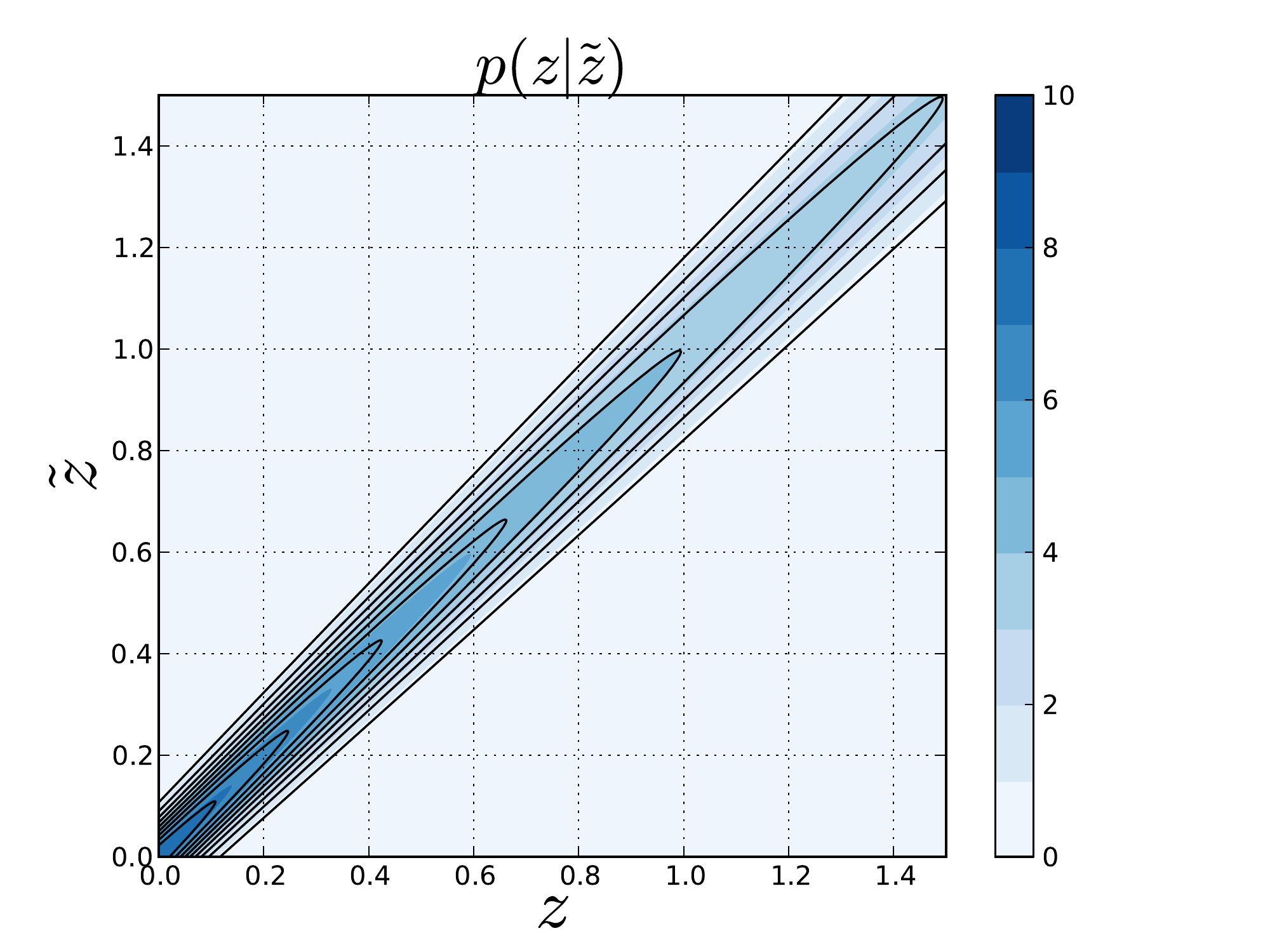}}
{ \includegraphics[width=0.328\textwidth]{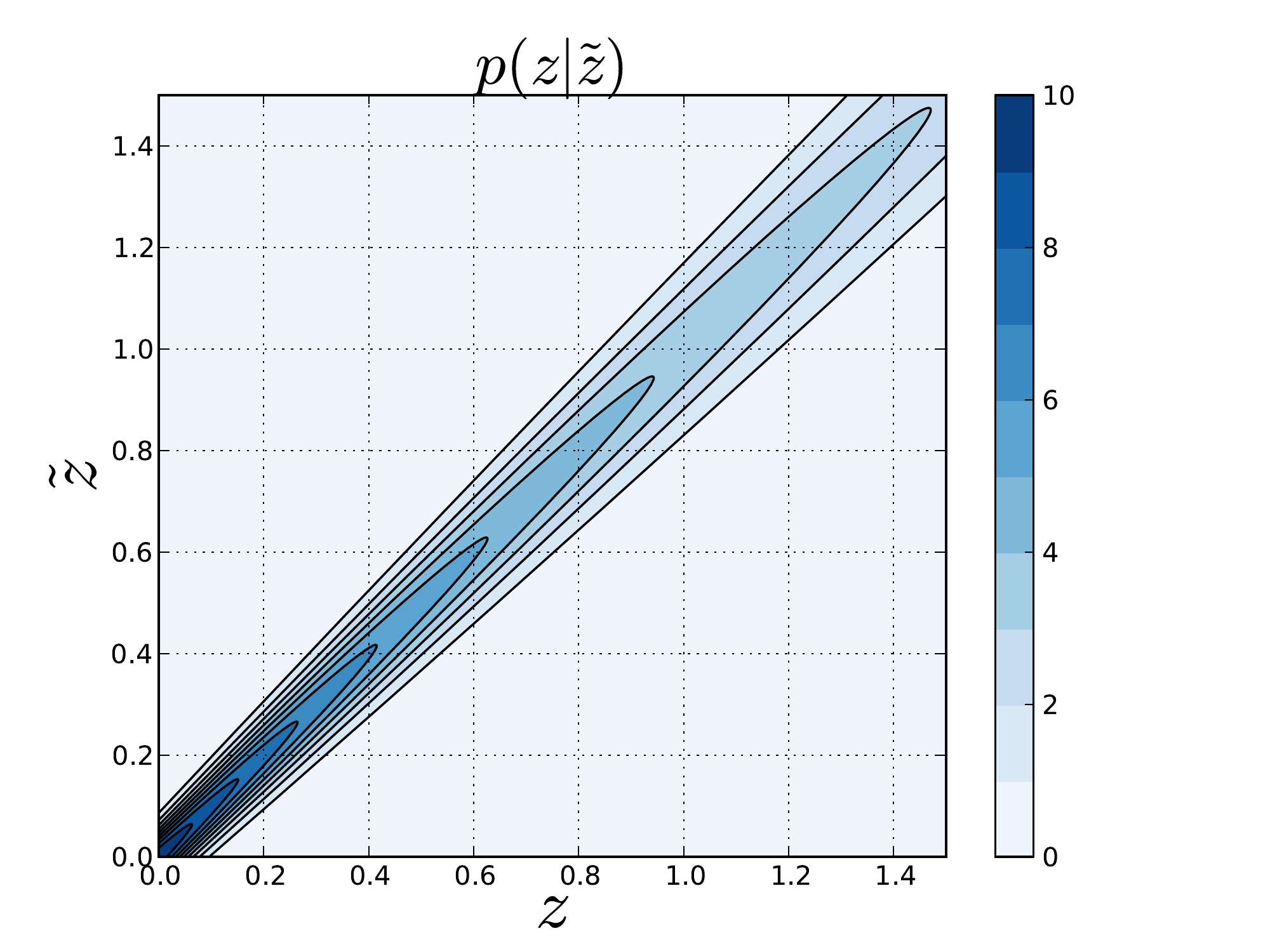}}
{ \includegraphics[width=0.328\textwidth]{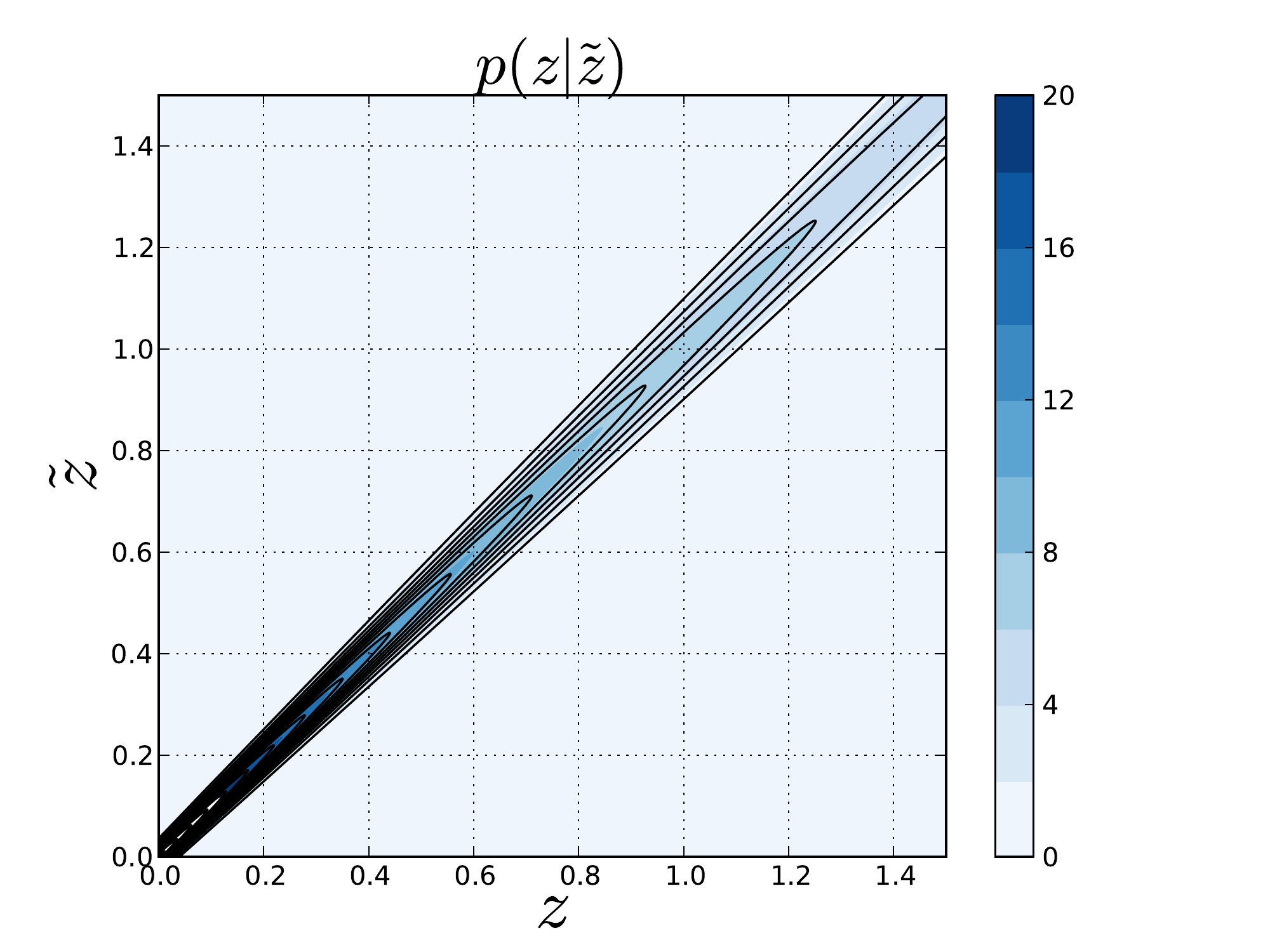}}
 \caption{The conditional probability function up to a redshift of $z = 1.5$ with a redshift dispersion error given by $\sigma_z (z)$. The fiducial model that we adopt in this paper is $\sigma_z (z) = 0.05 (1+z)$ on the left most plot. The middle plot is an approximate fitting formula for a 5-band survey and on the far right for a 17-band survey \citep{Taylor07}. Note that we have adopted a modified color scale in the final plot for convenience. }
 \label{Fig:prob}
\end{figure}
\n
Using Eq.(\ref{eq:shearkappa}), Eq.(\ref{eq:flex}), Eq.(\ref{eqn:gammakappa}) and Eq.(\ref{eqn:Cl_phi-y}) we arrive at the following expressions for the cross-spectra:

{\begin{centering}
\begin{equation}
\myC^{{\Gamma}y}_{\ell}= \sqrt{(\ell+2)!\over (\ell-2)!} \, \myC^{\phi y}_{\ell}; \quad 
\myC^{{\kappa}y}_{\ell}= {\ell(\ell+1) \over 2} \, \myC^{{\phi}y}_{\ell}; \quad
\myC^{{\cal F}y}_{\ell}= {1 \over 6} \left[ \ell (\ell+1) \right]^{1/2} \left[ 3\ell^2 + 3\ell-2 \right] \, \myC^{{\phi}y}_{\ell}; \quad
\myC^{{\cal G}y}_{\ell}= {1 \over 2} \sqrt {(\ell+3)! \over (\ell-3)!} \, \myC^{{\phi}y}_{\ell}.
\end{equation}
\end{centering}
}
\n
Note that for both shear components $_{\pm 2}\Gamma$ we recover the same power spectrum $\myC^{_{\pm 2}{\Gamma}y}_{\ell} \equiv\myC^{{\Gamma}y}_{\ell}$. 
\subsubsection{Results}
In this section we adopt the fiducial $\Lambda \textrm{CDM}$ model described above and consider a survey to $r_{\textrm{max}} = 1800, 3600 h^{-1} \, \textrm{Mpc}$ for the multipoles $\ell = 5, 20, 50, 200$. Note that we can see the effect of the approximate Bessel function inequality $k r \geq \ell$ with increasing $\ell$. At higher multipoles the diagonal terms of the cross-spectra do not become important until $\ell \approx k r_{\textrm{max}}$. Similar results can be seen in the weak lensing 3D spectra \cite{Castro05}. 

The results obtained so far are simplified, for clarity, as we ignore the fact that distance estimates from photometry contain errors and also that the number density of sources will decrease with redshift. The errors in distances will simply be radial errors in the sFB formalism. In the next section we outline some of the complications that a more realistic survey configuration gives rise to. 

\subsection{Realistic Selection Function}
For the redshift distribution of source galaxies for the surveys we will adopt following analytical fit \citep{Hut02,Shao11b}:
\begin{centering}
\begin{align}
n(z) &= {\bar n} \, {z^2 \over 2 z_0^3} \exp \left ( -{z \over z_0} \right ); \quad\quad \int_0^{\infty}\, dz\, n(z) = {\bar n}.
\end{align}
\end{centering}
\n
We will consider two different surveys (1) Dark Energy Survey (DES)\footnote{http://www.darkenergysurvey.org/}
and (2) Large Synoptic Survey Telescope (LSST)\footnote{http://www.lsst.org/lsst/}.
We will adopt $z_0 =0.3$ for the DES and $z_0=0.4$ for LSST. The galaxy number density per steradian is denoted above
as ${\bar N}_g=1.2\times 10^7 {\bar n}_g$, with $\bar n$ being the galaxy number density per square arcmin.
We will adopt ${\bar n}=15$ for DES and ${\bar n}=40$ for LSST.
\begin{figure}
\captionsetup[subfigure]{labelformat=empty}
\centering
\textbf{$\kappa$-$y$ Cross-Correlation: Photometric Redshift Uncertainty}\par\medskip
\includegraphics[width=65mm]{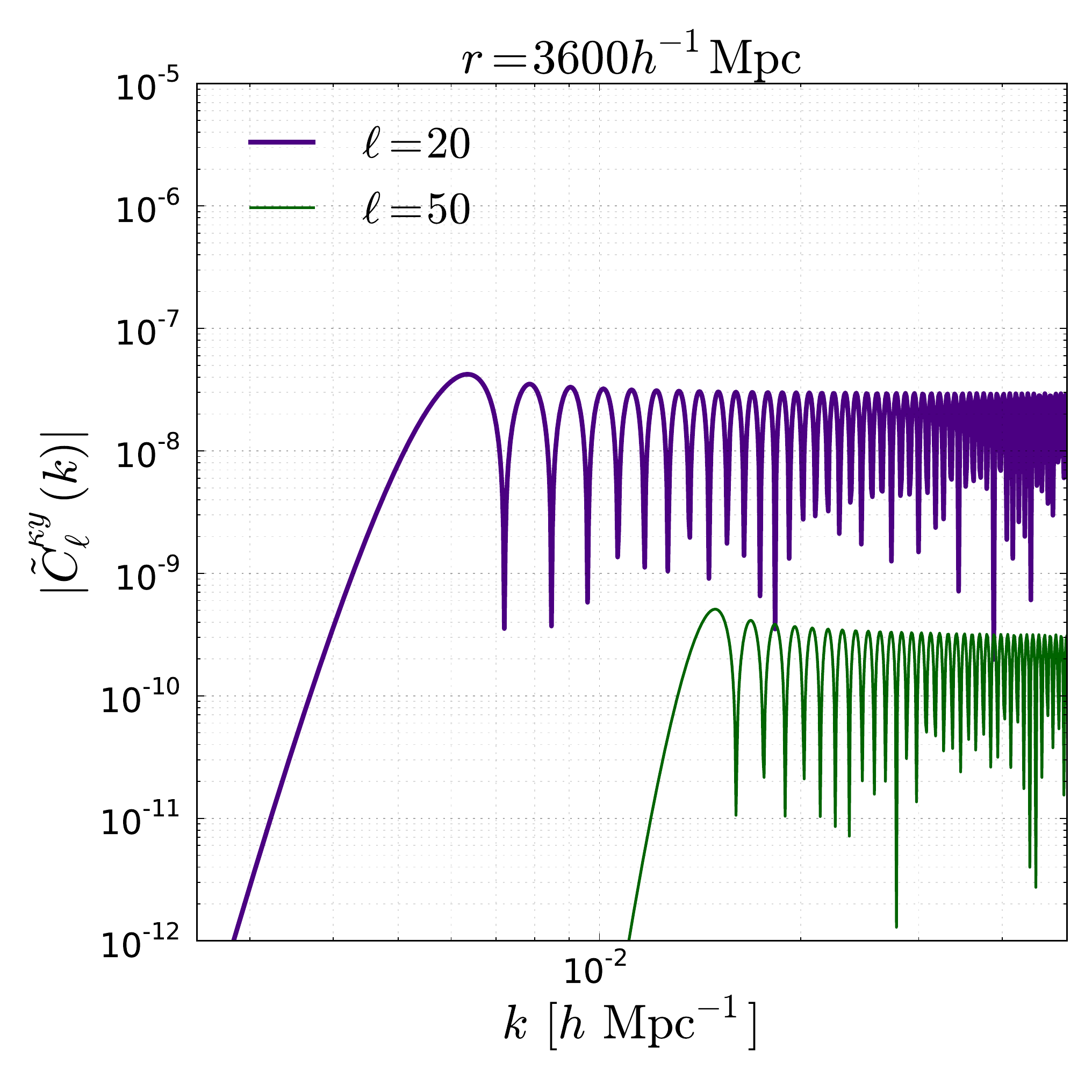}
\caption{In this figure we plot the observable convergence power spectra that have been smoothed due to the effects of photometric redshift uncertainty. Remember, photometric redshift errors are radial errors in the spherical Fourier-Bessel formalism and this means that the observations are smoothed along the line of sight. The adopted fiducial model for redshift error dispersion was $\sigma_z (z) = 0.05 (1+z)$. The left panel corresponds to $\ell = 20$ and the right hand panel to $\ell = 50$. Both have been calculated for a survey size of $r = 3600 h^{-1} \rm Mpc$.}
\label{fig:kappay}
\end{figure}
\subsection{Photometric Redshift Error}
For the depths of surveys proposed in upcoming WL studies, it is often impractical to obtain the spectroscopic redshifts and instead photometric redshifts $z_p$ are obtained form broad band photometry. In order to incorporate photometric redshift errors into our calculations we will need to integrate over the posterior redshift distribution $p(z|\tilde{z})$. 
In this section we wish to relate the observed convergence $\tilde{\kappa}_{\ell m} (k;r)$ that incorporates photometric redshift errors to the \textit{true} underlying convergence $\kappa_{\ell m} (k;r)$. One of the key advantages to the sFB formalism is that errors in distance, such as photometric redshift errors, will simply be radial errors. As such, the angular direction $\oh$ is not affected. The observed convergence harmonics will be given by:
\begin{align}
{\tilde \kappa}_{\ell m}(k;r) &= \sqrt{2 \over \pi } \int d^3\tilde{\br} ~n(\tilde r)~  \kappa(\br) \; k \;  j_\ell(k{\tilde r})\; Y_{lm}(\ho) 
\, w({\tilde r}) 
\end{align}
\n
where $\tilde{r}$ is the observed radial distance that is inferred from a photometric redshift of $\tilde{z}$. The true convergence naturally depends on the correct distance in the true cosmology $r$. In the following we neglect uncertainties in the photometric redshift distribution of sources and we ignore the effects of source clustering which will have a sub dominant contribution to the overall error budget. We can therefore relate the observed radial coordinate to a given photometric redshift as follows: $n({\tilde r}) d^3{\tilde r} = {\bar n_z(\tilde z)}d\tilde z\; d\oh /4\pi$. Using the above expression, we can re-write the observed convergence as
\begin{align}
{\tilde \kappa}_{\ell m}(k;r) &= \sqrt{\frac{1}{8 \pi^3}} \int dz , d \oh \, \bar{n}_z (\tilde{z}) \, \kappa ({\bf{r}}) \, k \, j_{\ell} (k \tilde{r}) \, Y_{\ell m} (\oh) \, w(\tilde{z}) .
\end{align}
\n
The dominant effect of photometric redshift errors are to smooth the source distribution $\bar{n}(\tilde{z})$ along the line-of-sight. If we introduce $p (z|\tilde{z})$ to denote the conditional probability of the true redshift being $z$ given the photometric redshift $\tilde{z}$, then the above can be re-written as 
\begin{align}
{\tilde \kappa}_{\ell m}(k;r) &= \sqrt{1 \over 8\pi^3 } \int d{\tilde z}\; \int dz\; \int d\oh \; {\bar n(\tilde z)} \; p(z|\tilde z)\; \kappa(\br)\; k \; j_l(k{\tilde r}) \; Y_{lm}(\oh) \; w(\tilde z); \label{eq:photoz1} .
\end{align}
\n
Now we just need to expand out $\kappa ({\bf{r}})$ with respect to the true radial distance $r$ and use the spherical harmonic relations to perform the angular integration. This eventually reduces to the following expression for the observed convergence harmonics
\begin{align}
{\tilde \kappa}_{\ell m}(k;r) &= \sqrt{1 \over 8\pi^3 } \int d{\tilde z}\; \int dz \; {\bar n(\tilde z)} \; p(z|\tilde z) \; k \; j_l( k{\tilde r}) \int dk^{\prime} k^{\prime} j_{\ell}(k^{\prime} \tilde{r}) w(\tilde{z}) \kappa_{\ell m} (k^{\prime} ; r)  \label{eq:photoz1} . 
\end{align}
\n
It is now a laborious procedure to construct the cross correlation of $\tilde{\kappa}_{\ell m} (k;r)$ with the tSZ harmonics $y_{\ell m}$ such that $\tilde\myC^{\kappa y}_{\ell}(k;r) = \la \tilde \kappa_{\ell m}(k;r)y^*_{\ell m}\ra$. The power spectrum is given by the following expression (Figure \ref{fig:kappay})
\begin{align}
{\tilde \myC}^{\kappa y}_{\ell}(k;r) &= \sqrt{1 \over 8\pi^3 }\left [\int d{\tilde z}\int dz \; {\bar n}(\tilde z) w(\tilde z) p(z|{\tilde z})\right ] k j_{\ell}(k\;{\tilde r})
\int dk' k' j_{\ell}(k'{\tilde r})\myC^{\kappa y}_{\ell}(k';{\tilde r}) \label{eq:photoz2}.
\end{align}
\n
Typically, the conditional probability associated with photometric redshift errors, $p(z|\tilde z)$, is modelled as a Gaussian for simplicity. This assumption may have catastrophic failures but provides a simple and intuitive starting point for error analysis. The functional form of the conditional probability that we adopt is given by (Figure \ref{Fig:prob})
\begin{align}
p(z|\tilde z) &= { 1 \over \sqrt {2 \pi} \sigma_z(z) } \exp \left [ { -({\tilde z}- z + \beta )^2 \over 2 \;\sigma_z^2(z)}\right ].
\label{eqn:Probability}
\end{align}
\n
In this expression $\beta$ is the possible bias in the photometric redshift calibration and $\sigma_z(z)$ is the redshift dependent dispersion in error. For our fiducial model we adopt $\sigma_z (z) = 0.05 (1+z)$ and assume that we can neglect the redshift calibration term, $\beta = 0$. See Figure \ref{Fig:prob} for the probability kernel defined in Eqn. (\ref{eqn:Probability}) for our fiducial model. As we can see, redshift errors simply translate into radial errors and this induces mode-mode couplings, as can be seen by the integral over the true spectra. The result of photometric redshift errors is that the observations are smoothed along our line of sight, this can be seen in Figure \ref{fig:kappay} where the structure in the oscillatory tail has been smoothed out. See \citep{Kitching11} for a more detailed study of photometric redshift errors in 3DWL. In this approach the authors integrate over the posterior redshift distribution for each galaxy $p_g (z|\tilde{z})$ creating a more accurate covariant 
matrix than the approach taken here of reducing the redshift distributions to a simpler form in which the 
distributions are assumed to be the same for each galaxy at a given redshift. 
\subsection{Signal To Noise}
\begin{figure}
 \centering
 \textbf{$\kappa$-$y$ Cross-Correlation: Signal-To-Noise}\par\medskip
{\includegraphics[width=60mm]{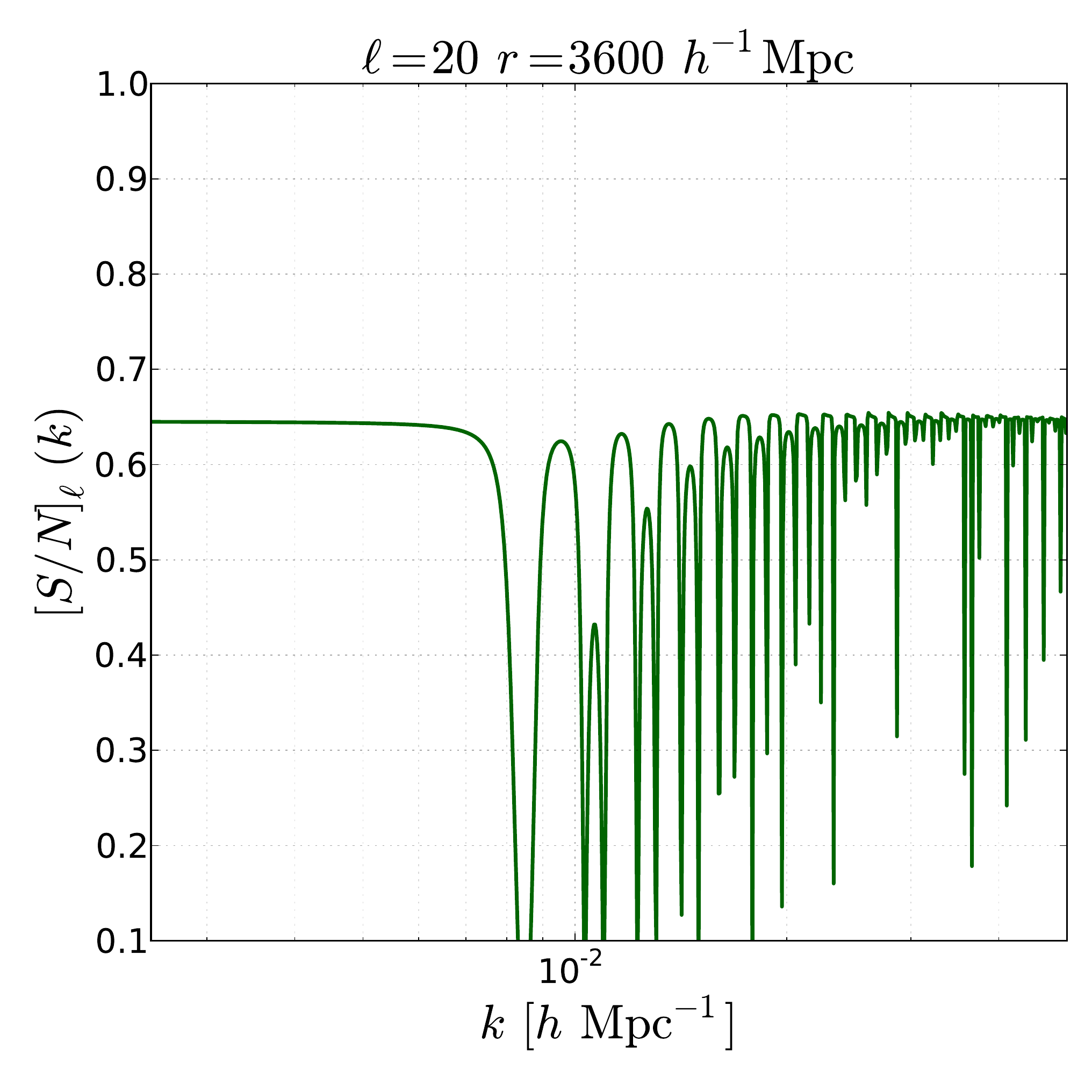}}
{\includegraphics[width=60mm]{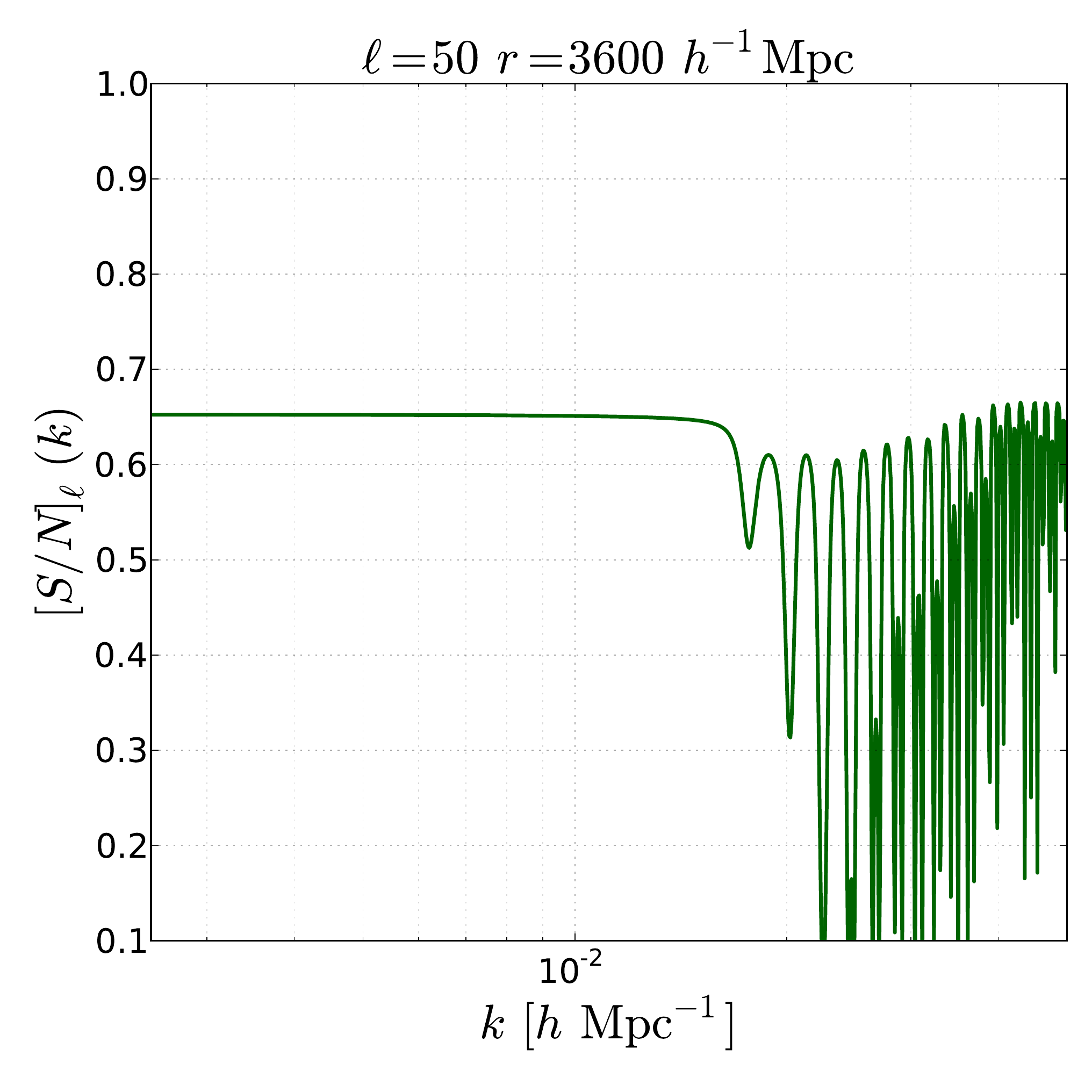}}
{\includegraphics[width=60mm]{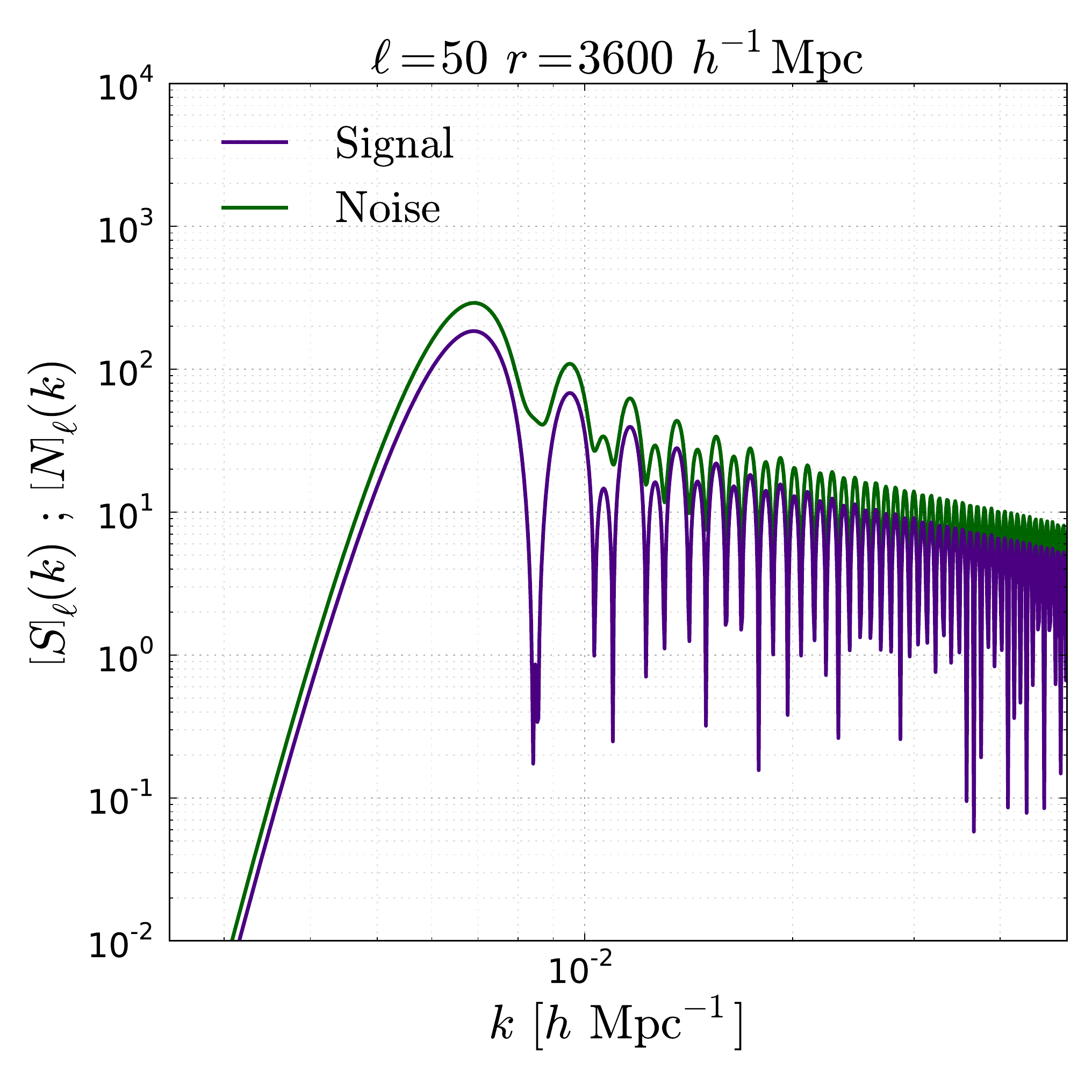}}
\caption{We plot the signal-to-noise as defined in Eq.(\ref{eq:SNR}) for a tSZ-Convergence cross correlation at $\ell = 20, 50$ and $r = 3600 h^{-1} \rm Mpc$. The right most panel is a plot of the signal and noise contributions individually for comparison.}
\label{fig:SNR}
\end{figure}

For two arbitrary data sets $X$ and $Y$ the signal to noise ratio (SNR) of the cross-spectra $\myC_{\ell}^{XY}(k)$ depends on the individual spectra
$\myC_{\ell}^{XX}(k)$ and $\myC_{\ell}^{XX}(k)$  as well as the cross-spectra $\myC_{\ell}^{XY}(k)$ itself. The signal to noise for the $XY$ cross-spectra is given as follows
\begin{align}
 [S/N]_{\ell} (k) &= \frac{ \myC_{\ell}^{X Y} (k) }{\sqrt{ \myC^{XX}_{\ell} (k) \myC^{YY}_{\ell} (k) + \left[ \myC^{X Y}_{\ell} (k) \right]^2 }} .
\end{align}
\n
As an example, we consider the SNR for weak lensing convergence-tSZ cross correlations. From the equation above this simply reduces to (Figure \ref{fig:SNR})
\begin{align}
[S/N]_{\ell} (k) &= \frac{ \myC_{\ell}^{\kappa y} (k) }{\sqrt{ \myC^{\kappa \kappa}_{\ell} (k) \myC^{yy}_{\ell} + \left[ \myC^{\kappa y}_{\ell} (k) \right]^2 }} .
\label{eq:SNR}
\end{align}
\n
We plot a representative signal to noise for our power spectra in Figure \ref{fig:SNR} for $r = 3600 h^{-1} \rm Mpc$ and $\ell = \lbrace 20 , 50 \rbrace$.  At low $k$ the SNR is dominated by the contributions from  the tSZ power spectrum via an offset. Remember, both the WL and tSZ-WL power spectra have negligible contributions at very low $k$ as the spectra fall off relatively sharply. As $k$ approaches the peak of the Bessel function at $k r \sim \ell$ the noise induced by the weak lensing power spectra becomes more prominent as does the signal from the tSZ-WL cross-correlation. 

\subsection{Halo Model for Large Scale Clustering Properties of Pressure}
\subsubsection{The Halo Model: Overview}
Simulations suggest that most of the baryons are at low overdensity regions $\delta < 10$ at high redshifts $z$ but 
are within collapsed halos at low redshifts $z \lesssim 1$. This means that baryons with overdensities $\lesssim 10$ track the underlying dark matter distribution. These baryons have temperatures on order that of the photoionisation temperature of hydrogen and helium. We follow \citep{CHT00,AC1,AC2,Ref02} and calculate the SZ effect due to the baryons within collapsed virialised halos at low redshifts. This approach uses the halo model for large scale clustering and is expected to be valid beyond the range of validity of the linear biasing scheme that we implemented in the previous sections. We compute the 3D cross-spectra in this formalism and compare to the results from the linear biasing scheme above. 

The halo model has been introduced and extended by numerous authors \citep{Sel00,MF00,KomSpe01,KomSel02,CH01,AC1,AC2,Cooray01}. A detailed review article of this approach may be found in \cite{{CooSeth02}} which we direct the interested reader towards for details beyond that presented here. The basic idea in the halo model is that we can relate the complex dark matter distribution by a simpler population of dark matter halos that accurately reconstructs the relevant statistical properties of the dark matter distribution. The core building block in the halo model is therefore the assertion that the dark matter halos will be locally biased tracers of linear density perturbations. This permits a Taylor series expansion allowing us to relate the overdensity of halos to the linear density fluctuations:
\begin{align}
 \delta_h \left( {\bf{x}} , M ; z \right) &= b_0 + b_1 (M;z) \delta_{\textrm{lin}} ({\bf{x}} ; z) + \frac{1}{2} b_2 (M;z) \delta^2_{\textrm{lin}} ({\bf{x}};z) + \dots 
\end{align}
\n
where $b_{\beta} (M;z)$ constitute the bias parameters that will be dependent on the formalism used to generate the halo mass function and any related assumptions. Given that the halos are biased tracers of the density perturbations, the number density of halos will also fluctuate according to:
\begin{align}
 \frac{d n}{d M} ({\bf{x}} ; z) &= \frac{d \bar{n}}{d M} (M;z) \, \delta_h ({\bf{x}} , M ; z). 
\end{align}
\n
and the density field in Fourier space will be given by:
\begin{align}
 \delta ({\bf{k}}) &= \sum_i \textrm{e}^{i {\bf{k}} \cdot {\bf{x}}} \, \delta_h ({\bf{k}} ; M_i ).
\end{align}
\n
Once we have a prescription for the dark matter density profile $\delta$, we can construct related quantities for the gas density profiles $\Pi$ (and hence baryon pressures) and galaxy distributions $g$. The power spectra is constructed by analogy to the normal Fourier space power spectrum:
\begin{align}
 \left\langle \delta_i ({\bf{k}}) \delta^{\ast}_i ({\bf{k}}^{\prime})  \right\rangle &= \left( 2 \pi \right)^3 \delta( {\bf{k}} - {\bf{k}}^{\prime}) P^t_i (k).
\end{align}
\n
with $i$ denoting $\delta, \Pi$ or $g$. The halo power spectrum can be broken down into two contributions at leading orders, a perturbative correction arising from correlation between halos and a nonlinear contribution arising from the correlation within halos described by the halo profile. As well as considering auto-correlations we can also construct the cross-correlated spectra $P^t_{i j} (k)$ between the appropriate regimes. Although inflationary cosmology generically predicts a highly Gaussian spectrum of initial perturbations, the concomitant non-linear growth of structure under gravitational collapse and the formation of virialised halos will lead to non-Gaussian features in the statistical properties of the dark matter distribution. Hence the same will be true in the underlying pressure field that the tSZ effect probes at low redshifts. A full study of the tSZ spectrum would construct measurements beyond the 2-point correlation function and consider an extended hierarchy of correlation functions (see,
for example, \citep{MuSmJoCo12,Mu13}). 

\begin{figure}
\centering
\textbf{Halo Model Power Spectra: Contributions from Halo Terms}\par\medskip
{\includegraphics[width=45mm]{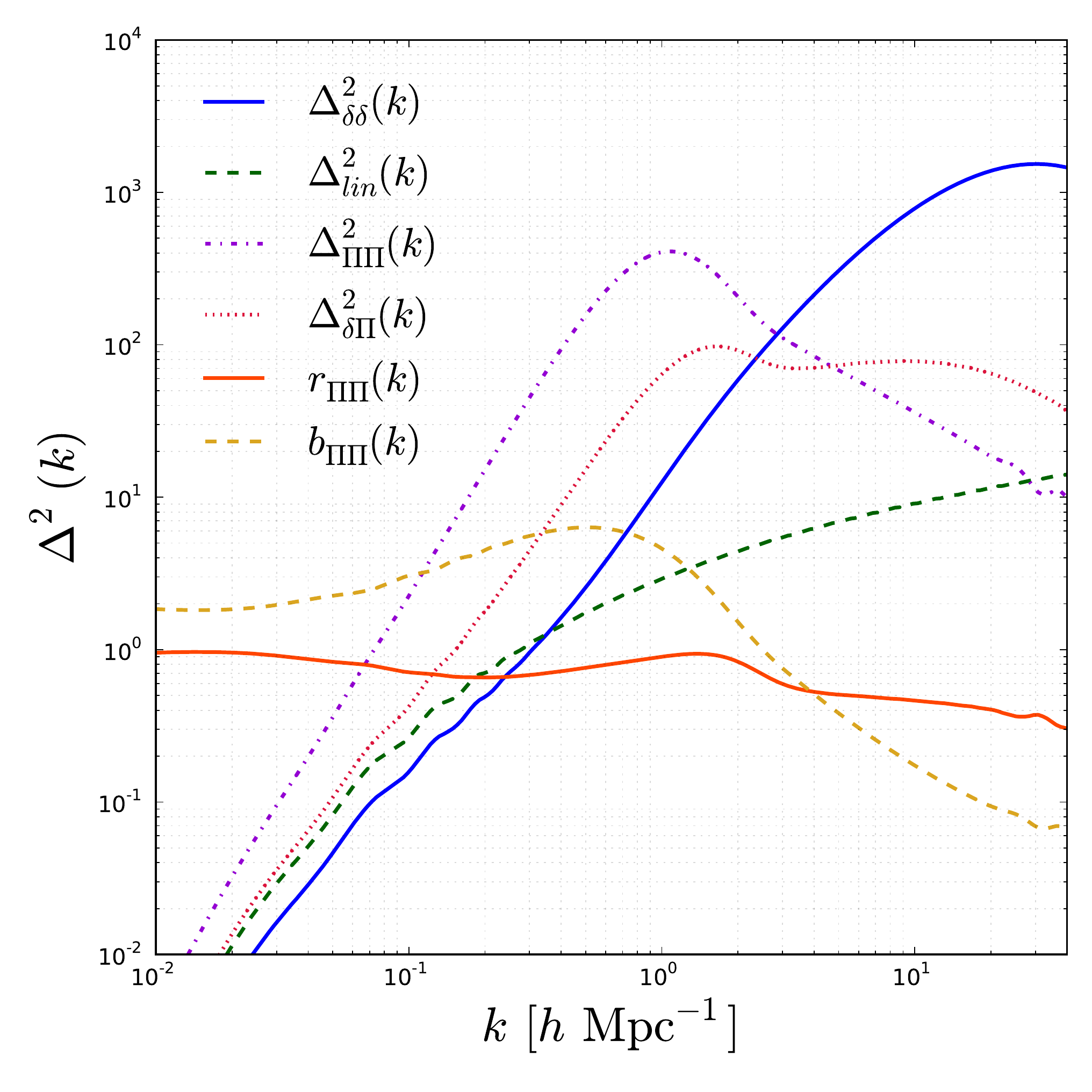}}
{\includegraphics[width=45mm]{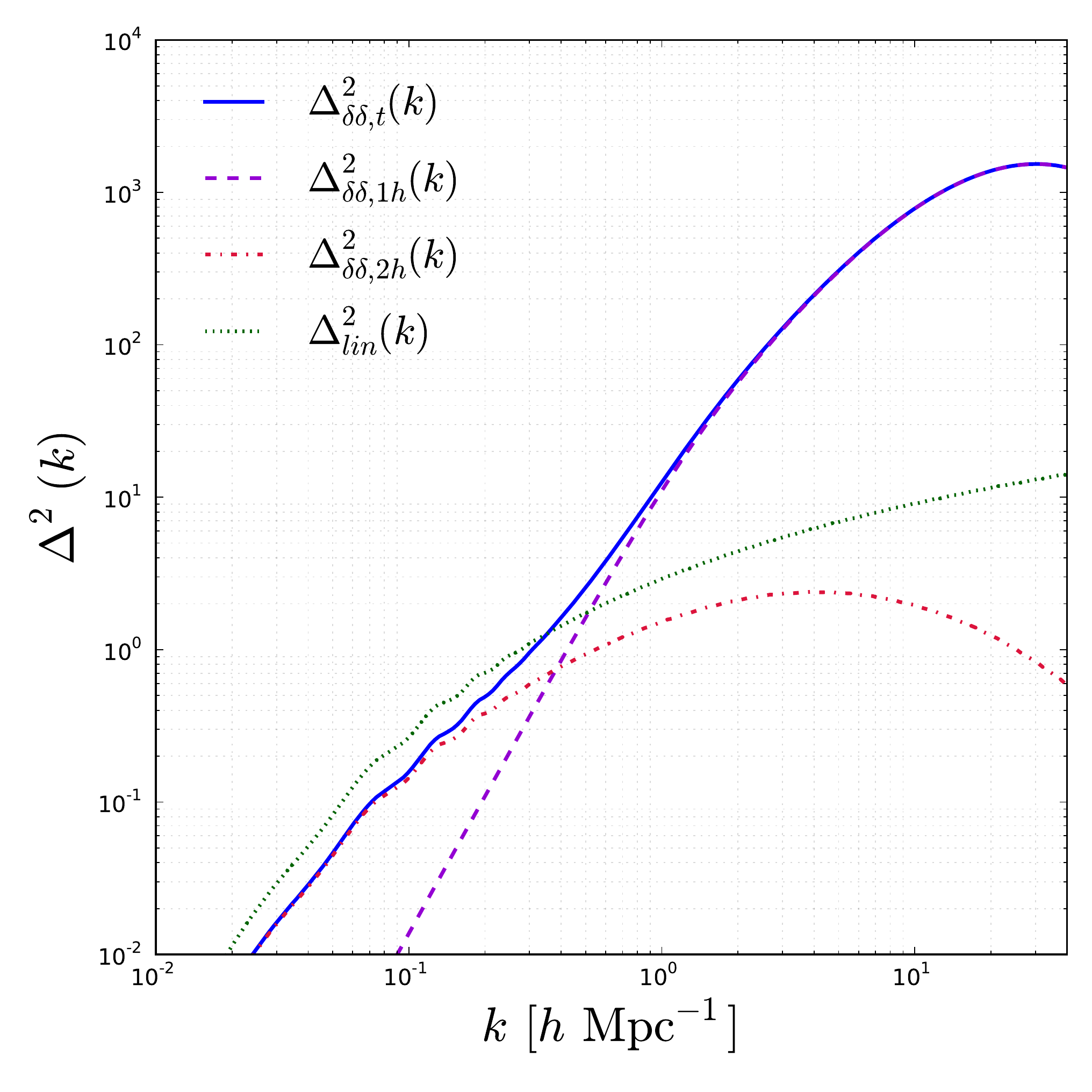}}\\
\hspace{0mm}
{\includegraphics[width=45mm]{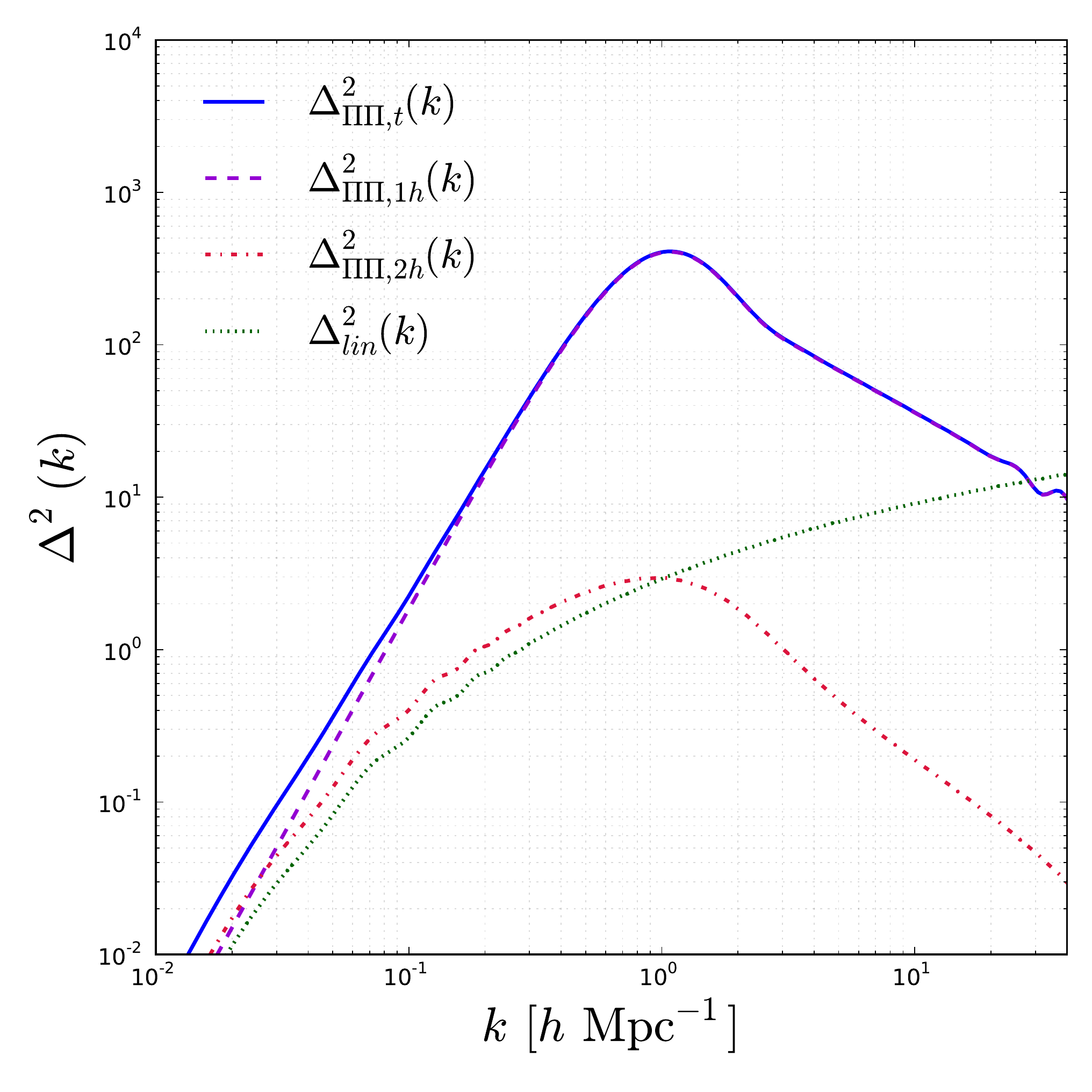}}
{\includegraphics[width=45mm]{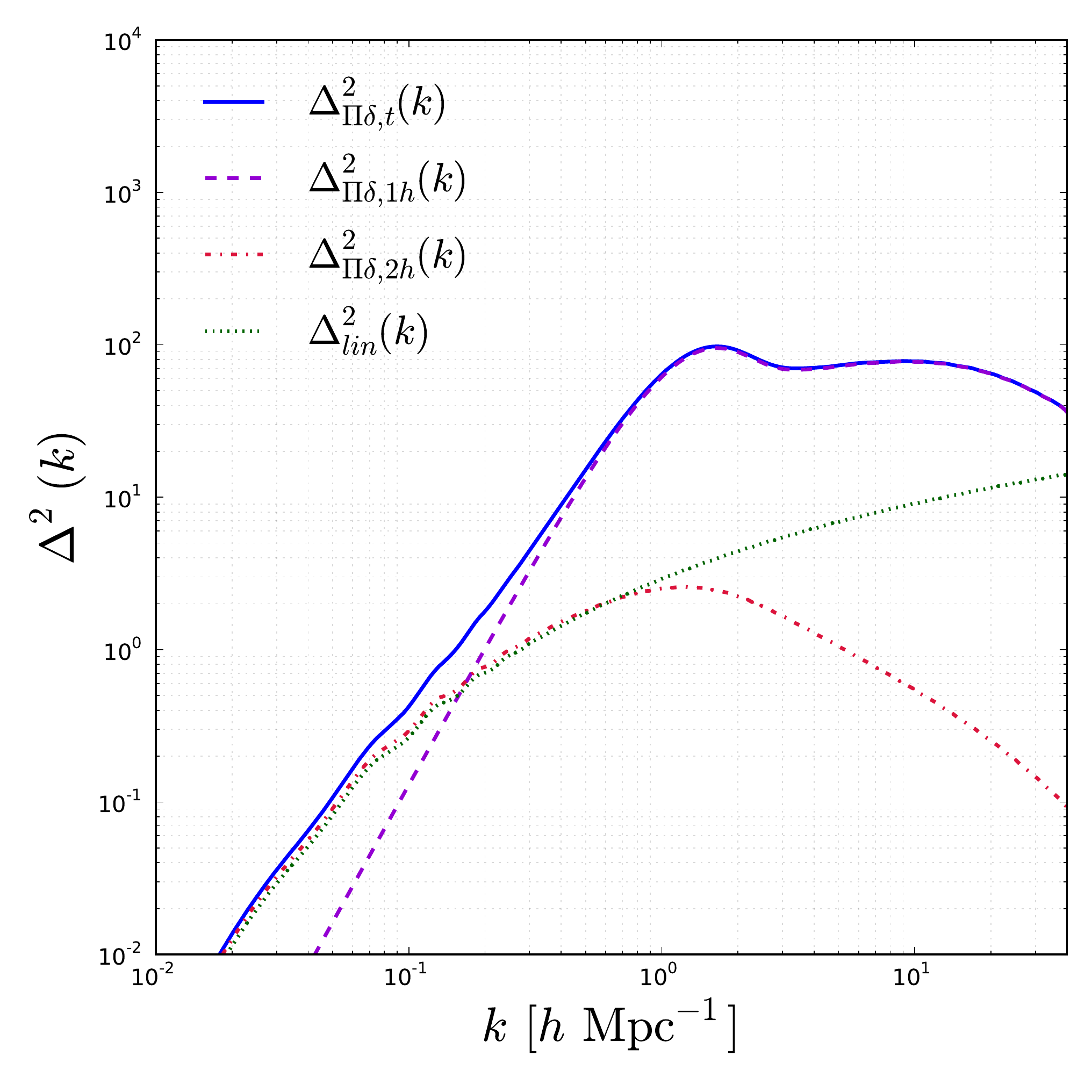}}
\caption{We show the various power spectra with single halo (1h), double halo (2h) and total (t) contributions. As can be seen, the 1-halo contribution dominates the various spectra. The virial temperature describes the electrons allowing us to construct the pressure bias $b_{\Pi} (k)$ and the correlation coefficient $r_{\Pi} (k)$ between the dark matter and baryonic distributions. Most of the contribution to the SZ effect arises from massive clusters of galaxies whereas the smaller mass halos and structures at low electron temperature do not contribute as significantly. }
\label{fig:HaloModel}
\end{figure}

The halo model depends on two crucial ingredients, a prescription for the halo mass function \citep{PS74,ST99} and the halo density profile \citep{NFW96}. The halo model has been extended to SZ calculations by a number of authors by relating the baryonic gas profiles to the underlying dark matter profiles under reasonable physical assumptions \citep{AC1,KomSel02}.
However, it is important to note that halo model does not incorporate the non-thermal pressures, such as that caused by turbulence, gas cooling or star formation. This is a topical area and has been the focus of study in the literature of recent (e.g. \cite{Shi14}). Following the literature \citep{CH01,AC1,AC2}, we introduce a general integral over the halo mass function that contains terms related to the dark matter, gas pressure and baryon density fields:
\begin{align}
&& I^{\beta,\eta,\gamma}_{\mu,i_1,i_2\dots i_{\mu}}(k_1,\dots,k_{\mu};z) &\equiv \int d \, \textrm{ln} M \left (M \over \bar\rho \right)^{\mu} {dn \over d \, \textrm{ln} M}(M,z) \, 
b_{\beta}(M) \, \left ({\bar \rho \over M}{\la N_g\ra \over \bar n_g} \right )^\gamma \left[ T_e(M,z) \right]^{\eta} \, y_{i_1}(k_1,M) \dots y_{i_\mu}(k_\mu,M).
\label{eq:master}
\end{align}
\n
The above integral contains the bias parameters $b_{\beta} (M,z)$, telling us how the halos trace the overdensity field, the halo mass function $[ d \bar{n} / d M ] (M, z)$ giving the number density of halos at a given virialised mass, the electron temperature $T_e (M;z)$ to account for clustering properties associated with baryons and the 3D Fourier transform of the density profiles $y_{i,\mu} (k_{\mu} , M)$. The Fourier transform is explicitly given by:
\begin{align}
 y_i (k,M) &= \frac{1}{M_i} \int\limits^{r_v}_0 dx \, x^2 \, 4 \pi \, \rho_i (x,M) \, j_0 (k x).
\end{align}
\n
where the subscript $i$ represents the density $\delta$ or gas $g$. The Fourier transform has been weighted by the mass such that it is unity at $k=0$. The power spectra are decomposed into contributions from the single halos $P^{PP}_{i j} (k)$ and contributions from the halo-halo correlations $P^{hh}_{i j} (k)$. Here we explicitly write out the various definitions for the power spectra and cross-spectra that we consider in this paper (Figure \ref{fig:HaloModel}):
\ben
 && (i)\;\quad {\delta\delta\ : \,} \quad P^t_{\delta\delta} = P_{\delta\delta}^{\rm PP}+P_{\delta\delta}^{hh} \quad
P^{\rm PP}_{\delta\delta}(k) = I_{2,\delta\delta}^{0,0,0}(k,k); \quad P^{hh}_{\delta\delta}(k) = [I_{1,\delta}^{0,0,0}]^2P_{\rm lin}(k); \label{eq:eq1}\\
 && (ii)\;\quad {\Pi\Pi\; : \,} \quad P^t_{\Pi\Pi} = P_{\Pi\Pi}^{PP}+P_{\Pi\Pi}^{hh} \quad
P^{\rm PP}_{\delta\delta}(k) = I_{2,gg}^{0,2,0}(k,k); \quad P^{hh}_{\delta\delta}(k) = [I_{1,g}^{1,1,0}]^2P_{\rm lin}(k); \\
 && (iii)\;\quad {\delta\Pi\; : \,} \quad P^t_{\Pi\delta} = P_{\Pi\delta}^{PP}+P_{\Pi\delta}^{hh} \quad
P^{\rm PP}_{\Pi\delta}(k) = I_{2,g\delta}^{0,1,0}(k,k); \quad P^{hh}_{\Pi\delta}(k) = [I_{1,g}^{1,1,0}][I_{1,\delta}^{1,0,0}] P_{\rm lin}(k); \\
&&  (iv)\;\quad {g\Pi\; : \,} \quad P^t_{\Pi\delta} = P_{\Pi\delta}^{PP}+P_{\Pi\delta}^{hh} \quad
P^{\rm PP}_{\Pi\delta}(k) = I_{2,g\delta}^{0,1,1}(k,k); \quad P^{hh}_{\Pi\delta}(k) = [I_{1,g}^{1,1,0}][I_{1,\delta}^{1,0,1}] P_{\rm lin}(k).
\label{eqn:HaloPowerSpectra}
\een
\subsubsection{The Halo Model: Ingredients}
The first ingredient we need in the halo model is a prescription for the halo mass function and related quantities. In order to do this we adopt the Press-Schechter (PS) formalism to generate the mass functions $(dn/dM)$ and the corresponding bias parameters $b_{\beta}(M;z)$ that appear in Eq.(\ref{eq:master}) \citep{PS74}. For halos of mass $M$, these quantities are defined as follows \citep{MJW97}: 
\ben
&&{dn\over dM} = \sqrt{2 \over \pi}{\bar \rho \over M}{\delta_c\over \sigma^2}{d\sigma \over dM}\exp \left [-{\delta_c^2 \over 2\sigma^2} \right ]; \quad
b_0(M,z) =1; \quad
b_1(M;z) = 1 + {\nu^2(M;z)-1 \over \delta_c}; \quad \\
&& b_2(M;z) = {8\over 21}\left [b_1(M;z)-1 \right ] + {\nu^2(M;z)-3 \over \sigma^2(M;z)} ; \quad
\nu = {\delta_c \over \sigma(M;z)}.
\een
\n
Here, $\sigma(M,z)$ is the r.m.s. fluctuation of mass $M$ within a top-hat filter at a virial radius $R$ at a redshift $z$, 
i.e. $M=4\pi/3R^3\bar \rho$. The threshold overdensity for spherical collapse is approximately $\delta_c=1.69$ and the background mean density of dark matter in the Universe is ${\bar\rho}=2.775 \times 10^{11} \Omega_{\rm M} h^2 M_{\sun} {\rm Mpc}^{-3}$.

We will assume a spherically averaged dark matter halo profile for collapsed halos, $\rho_{\delta}(x)$ \citep{NFW96}. The Navarro-Frenk-White profile assumes that the profile shape of halos is universal and can be characterised by a scaling radius $x_s$ and scaling density $\rho_s$:
\begin{align}
 \rho_{\delta} (x) &= \frac{\rho_s}{(x/x_s) (1 + x/x_s)^2} .
\end{align}
\n
The mass of the profile within the virial radius $x_v$ can be calculated integrating the NFW profile to yield:
\begin{align}
 M_{\textrm{vir}} &= 4 \pi \rho_s \, x^3_s \left[ \log (1+c_s) - \frac{c_s}{1+c_s} \right] .
 \label{eqn:NFW}
\end{align}
\n
Here we have introduced the concentration parameter $c_s = x_v/x_s$ telling us how centrally peaked the profile is. 
However, assuming spherical collapse, the virial mass $M_{\rm vir}$ within the virial radius $x_v$ can also be expressed as:

\begin{align}
M_{\rm vir}&= 4\pi x_{v}^3 \Delta_c(z)\bar\rho(z)/3; \quad\quad
\bar\rho(z) = 2.775 \times 10^{11} E^2(z) \, \Omega_{\rm M} \, h^2 M_{\sun} {\rm Mpc}^{-3}; \label{eq:virial_mass1}\\
\Delta_c(z) &= 18\pi^2 + 82[\Omega_{\rm M}(z)-1] -39 [\Omega_{\rm M}(z)-1]^2; \quad\quad
\Omega_{\rm M}(z) = \Omega_{\rm M}{(1+z)^3 \over E^2(z)};
\end{align}
\n
where $\Delta_c (z)$ is the overdensity of collapse and $E^2 (z)$ is the function introduced in Eq.~(\ref{eqn:Ez}). By evaluating the overdensity of collapse we can obtain the virial radius $x_v$ for a given mass $M_{\rm vir}$. The concentration parameter $c_s$ for a halo of mass $M$ can be re-expressed in terms of a characteristic mass scale $M_*$ defined by $\sigma (M_{*} ; z) = \delta_c$ or, equivalently, $\nu = 1$. Using fitting formula calibrated on $\Lambda \rm{CDM}$ simulations, the concentration-mass relationship is taken to be:

\begin{align}
c_s (M,z) &= a(z)\left [ {M \over M_*(z)}  \right ]^{-b(z)};\quad\quad a(z)= 10.3(1+z)^{-0.3}; \quad\quad b(z)= 0.24(1+z)^{-0.3}.
\end{align}

\n
This correspondence between the mass definitions allows us to eliminate the scaling density $\rho_s$ by equating the virial mass $M_{\rm vir}$ computed in Eq.(\ref{eq:virial_mass1}) with the virial mass computed using the NFW density profile in Eq.(\ref{eqn:NFW}).

\begin{align}
 \rho_s &= \frac{c^3_s}{x^3_v} \, \frac{M_{\rm vir}}{4 \pi} \left[ \rm{log} ( 1 + c_s ) - \frac{c_s}{1 + c_s}  \right]^{-1}
\end{align}

\n
Once we have determined the concentration parameter for the given virial mass, we can explicitly evaluate $\rho_s$ and hence explicitly determine the NFW halo profile. The halos in a given cosmological background can therefore be characterised with just two parameters: the halo mass $M$ and the concentration parameter $c_s$. Remember, the virial radius is determined by the halo mass and the background cosmology dependent overdensity of collapse and is therefore not an independent parameter.

In order to describe the large scale pressure fluctuation, we will assume a hydrostatic equilibrium between the
gas distribution and the dark matter distribution in halos \citep{Makino97}. This allows us to relate the gas density profile with the dark matter halo profile in a reasonable and physical manner. The equation of state is assumed to be well modelled by a polytropic fluid with polytropic index $\gamma$. Hydrostatic equilibrium implies that we have
\begin{align}
{k_B T_e \over \mu\, m_p} {d\log \rho_g(x) \over dx} &= -{G M_{}(<x) \over x^2}.
\end{align}
\n
where $M (<x)$ is the mass inside a radius $x$. The above relationship implicitly assumes that the gas distribution obeys an isothermal temperature distribution. As the dark matter profile is assumed to obey a scaling relationship, the gas density profile will also obey a scaling relationship in terms of a physical parameter $b$ and scale radius $x_s$:
\begin{align}
\rho_g(x) &= \rho_{g0}\; e^{-b} \left ( 1+ {x \over x_s} \right )^{bx_s/x}; \quad\quad  b = {4\pi G \mu m_p \rho_s x_s^2 \over k_B T_e}.
\end{align}
\n
The physical parameter is intrinsically related to the virial temperature of the gas $T_e$
\begin{align}
k_BT_e &= {1 \over 3 r_v}{\gamma G \mu m_p M_{\delta}(x_v)} . 
\end{align}
\n
In this work we adopt a polytropic index of $\gamma = 3/2$ and a mean molecular weight of $\mu = 0.59$ to accompany the proton mass $m_p$. The total mass of the gas in a dark matter halo within a virial radius $x_v$ will be given by:
\begin{align}
M_g(x_v) &= 4\pi \rho_{g0}\;e^{-b}\; x_s^3\int_0^c \;dq\; q^2 (1+q)^{b/q}.
\end{align}
The final ingredient we need is a prescription for the calculation of the galaxy-pressure power spectrum. In order to do this we need to specify an average occupancy of galaxies in halos. This is simply assumed to have a general form given by
\begin{align}
 \la N_g \ra &= \left ( {M \over M_{\rm min}} \right )^{0.6} {\rm for} \;\; M \ge M_{\rm min}; \nn \\
 &= 0 \;\; \quad\quad\quad\quad\;\, {\rm for}\; M < M_{\rm min}.
\end{align}
\n
The minimum dark matter halo mass is taken to be $M_{\rm min} = 10^9 h^{-1} M_{\sun}$. Consequentially, the mean number density of galaxies ${\bar n}_g$ and the average density weighted temperature ${\bar T}_e$ which appear in Eq.(\ref{eq:master}) can be expressed as:
\begin{align}
{\bar n}_g = \int dM \la N_g \ra {dn \over dM}(M,z); \quad
{\bar T}_e = \int dM {M \over \rho_p} {dn \over dM}(M,z) T_e(M,z).
\end{align}
When studying the correlations between power spectra and cross-spectra it is useful to introduce the bias $b_{\Pi}(k,r)$ and cross-spectral coefficient $r_{\Pi \Pi}(k,r)$ 
\begin{align}
b_{\Pi} (k) = \frac{1}{\bar{T}_e} \sqrt{\frac{P_{\Pi \Pi} (k)}{P_{\delta \delta} (k)}}; \quad r_{\Pi \Pi} (k) = \frac{P_{\Pi \delta} (k)}{\sqrt{ P_{\delta \delta} (k) P_{\Pi \Pi} (k)}} .
\end{align}
\begin{figure}
\centering
\textbf{tSZ-WL Cross-Correlation in the Halo Model: Differential Contributions from Halo Terms}\par\medskip
{\includegraphics[width=45mm]{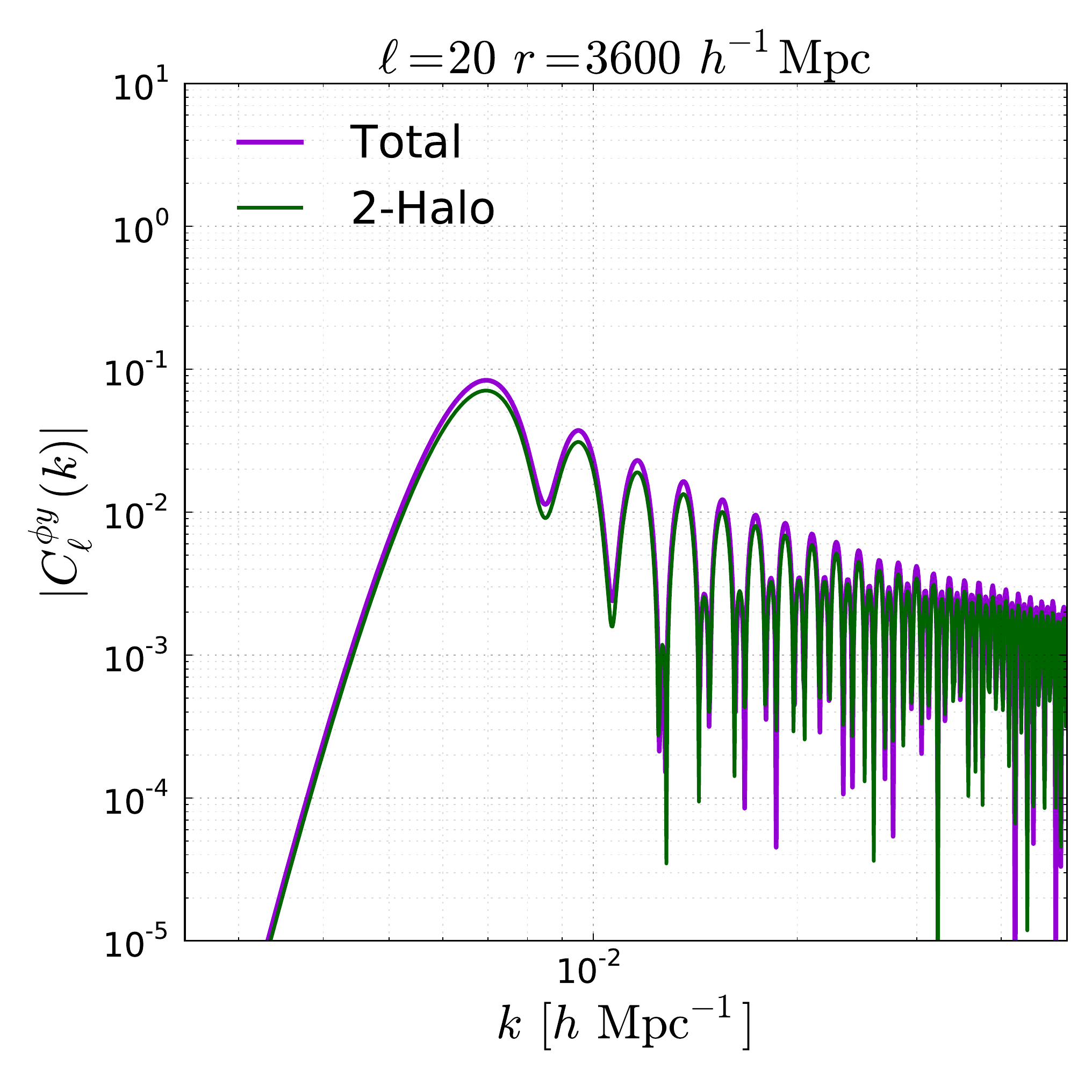}}
{\includegraphics[width=45mm]{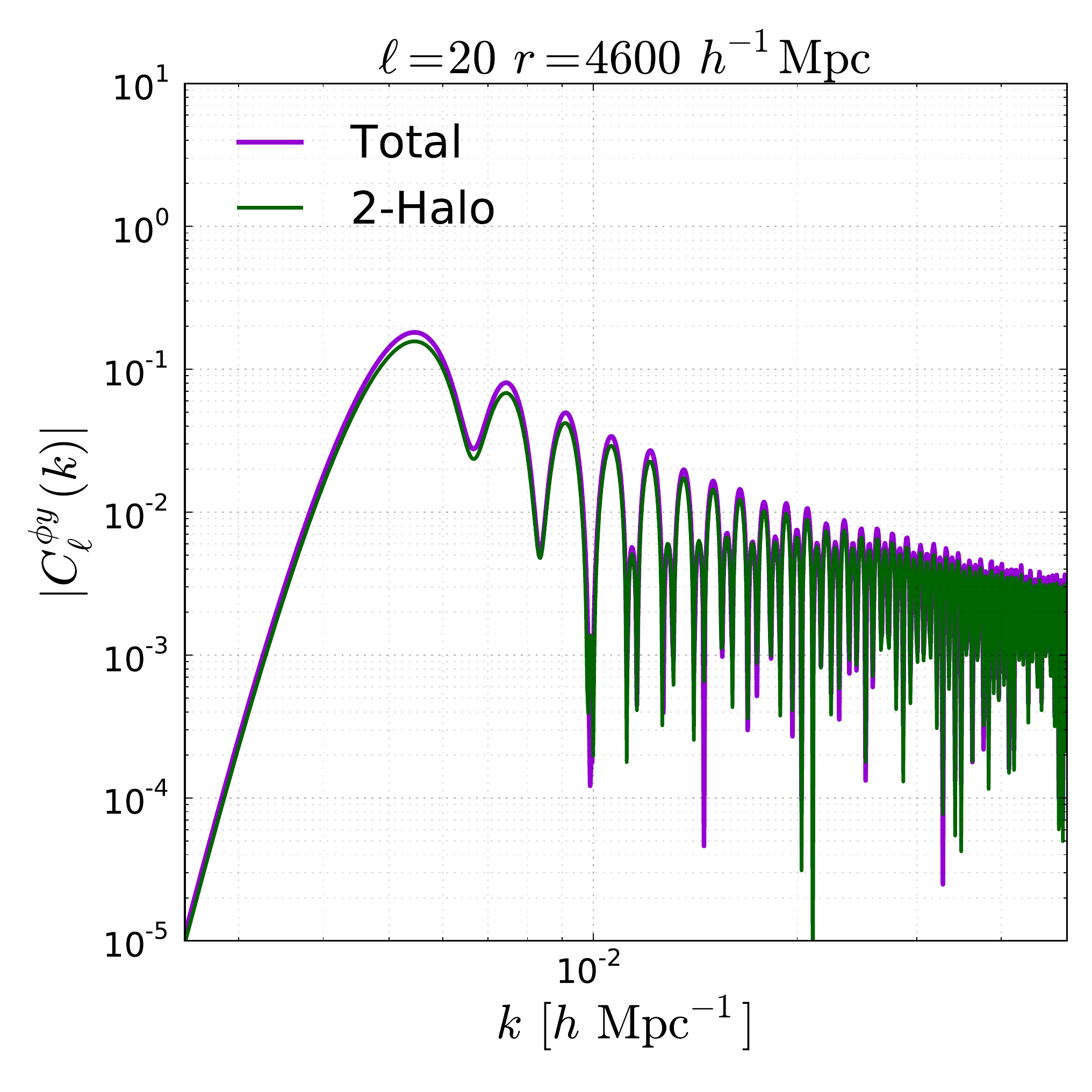}}
{\includegraphics[width=45mm]{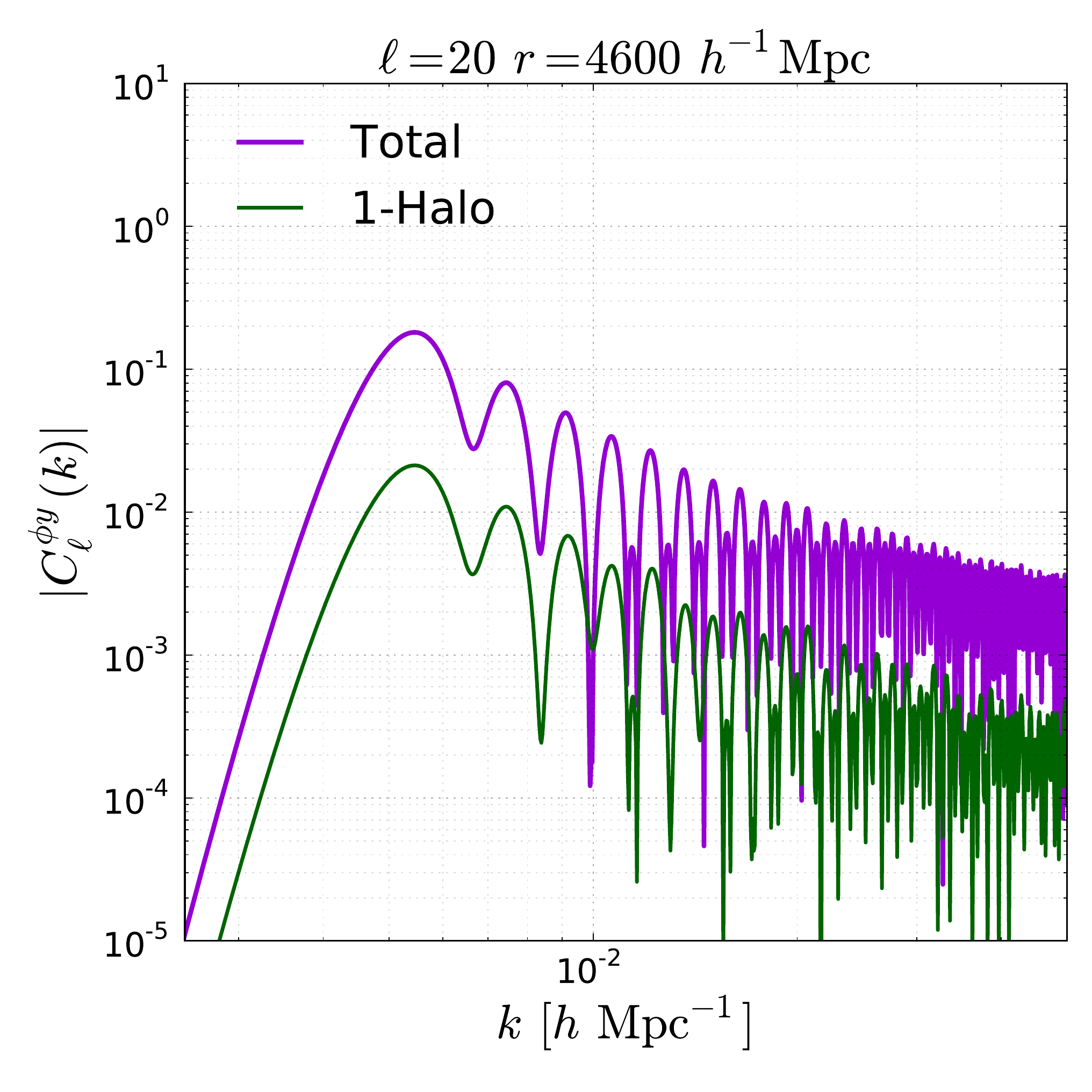}} \\
\caption{The halo model can be used to predict WL-tSZ cross correlation. The figures show the total spectra and the spectra that arises from the 1-halo terms. As expected at these scales, the 2-halo term is dominant and recovers the total spectra to a good degree. The 1-halo terms contribute small but non-negligible corrections to the spectra. This can be seen by inspection of the various power spectra in Figure \ref{fig:HaloModel}, the 1-halo terms fall off too quickly at such low $k$ leaving the 2-halo term dominate.} 
\label{fig:tSZ-WL-HaloModel}
\end{figure}
\subsubsection{Power and Cross Spectra}
We now gather all the various ingredients to construct the halo model power spectra. We take $M_{\rm max} = 10^{16} h^{-1} M_{\sun}$ and $M_{\rm min} = 10^9 h^{-1} M_{\sun}$. In Figure \ref{fig:HaloModel} we present the numerical results for the density-density, pressure-pressure and density-pressure spectra along with the bias and spectral correlation coefficients. 

In analogy with the analysis presented above for the tSZ-WL cross-correlation, we can re-derive the same results within the halo model framework via a modified version of Eq.(\ref{eqn:Cl_phi-y})-Eq.(\ref{eqn:I_phi}). The power spectrum for the tSZ-WL cross-correlation in the halo model can therefore be written as (Figure \ref{fig:tSZ-WL-HaloModel})
\begin{align}
{\cal J}^{\phi}_\ell(k',k) &\equiv k' \int_0^{\infty} dr~r^2~j_{\ell}(k'\, r) \int_0^r dr'  F_{\rm K}(r,r')\, j_{\ell}(kr')
\sqrt{P^{\Phi\Phi}(k;r')}\;;  \\
{\cal J}^{y}_\ell(k) &\equiv {\bar T}_e \,  k^2 \, \sqrt{2 \over \pi} \int_0^{\infty} dr \, w_{\rm SZ}(r)\, j_{\ell}(k r)\, r_{\Pi}(k;r)b_{\Pi}(k; r)\, \sqrt{P^{\Phi\Phi}(k;r)}\;;  \\
\myC_\ell^{\phi y}(k) &= {4 \over \pi c^2} \int_0^{\infty} k'^2 {\cal J}^{\phi}_\ell(k, k'){\cal J}^{y}_\ell(k')dk'. \qquad
\label{Cl_phi_Halo}
\end{align}
\n
The bias $b_{\Pi}(k,r)$ in this formalism is not completely ad hoc but is instead an outcome of various inputs and assumptions that go into the halo model. 
Similar results can be obtained for the 3D cross-correlation of galaxy-surveys against the weak lensing surveys. We adopt the same configurations as for the tSZ-WL cross-correlation defined previously, namely $\ell = \lbrace 20, 50 \rbrace$ and $r = \lbrace 3600, 4600 \rbrace \, h^{-1} \rm Mpc$. The results are shown in Figure \ref{fig:tSZ-WL-HaloModel}. 

As we did before, we can also evaluate the cross-correlation spectra on small angular scales by invoking the Limber approximation. This results in a rather simplified expression for the various terms
\ben
&& \mathcal{J}_{\ell}^{\phi}(k,k^{\prime}) = {\pi \over 2}\; {\nu \over k k^{\prime}}\; 
F_{\rm K}\left ( {\nu \over k}, {\nu \over k'} \right ) \sqrt{P^{\Phi\Phi}\left (k^{\prime}; {\nu\over k^{\prime}} \right )}; \quad 
\mathcal{J}_{\ell}^y(k) = {k \sqrt \nu}\; {\bar T_e}\; w_{\rm SZ}\; \left ( {\nu \over k} \right ) r_{\Pi}\left (k; {\nu \over k} \right ) b_{\Pi}\left(k; {\nu \over k} \right); \\
&& \myC^{\phi y}_{\ell}(k) = {2 \over c^2}\; \sqrt{\nu}\; {\bar T_e}\; \int dk' F_K\left ({\nu \over k}, {\nu \over k'} \right ) w_{\rm SZ}({\nu \over k'}) 
r_{\Pi}\left(k', {\nu \over k'} \right) b_{\Pi}\left(k', {\nu \over k'} \right)  P^{\Phi\Phi}\left (k', {\nu \over k'} \right); \quad \nu = {\ell + {1\over 2}}.
\een
\subsubsection{Power and Cross Spectra: Mass Bins}
\begin{figure}
 \centering
 \textbf{Halo Model Power Spectra: Comparison of Mass Functions}\par\medskip
{\includegraphics[width=60mm]{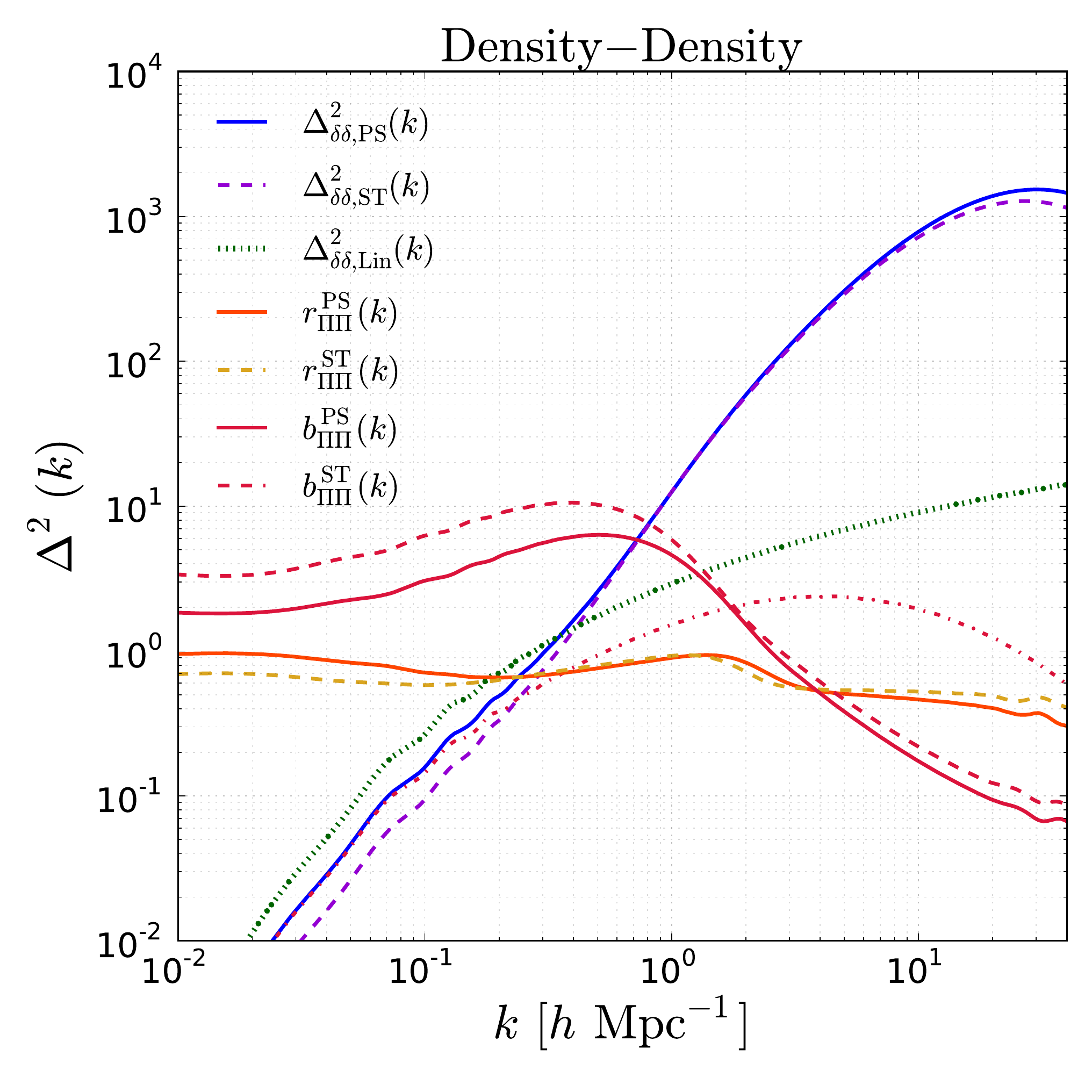}}
{\includegraphics[width=61mm]{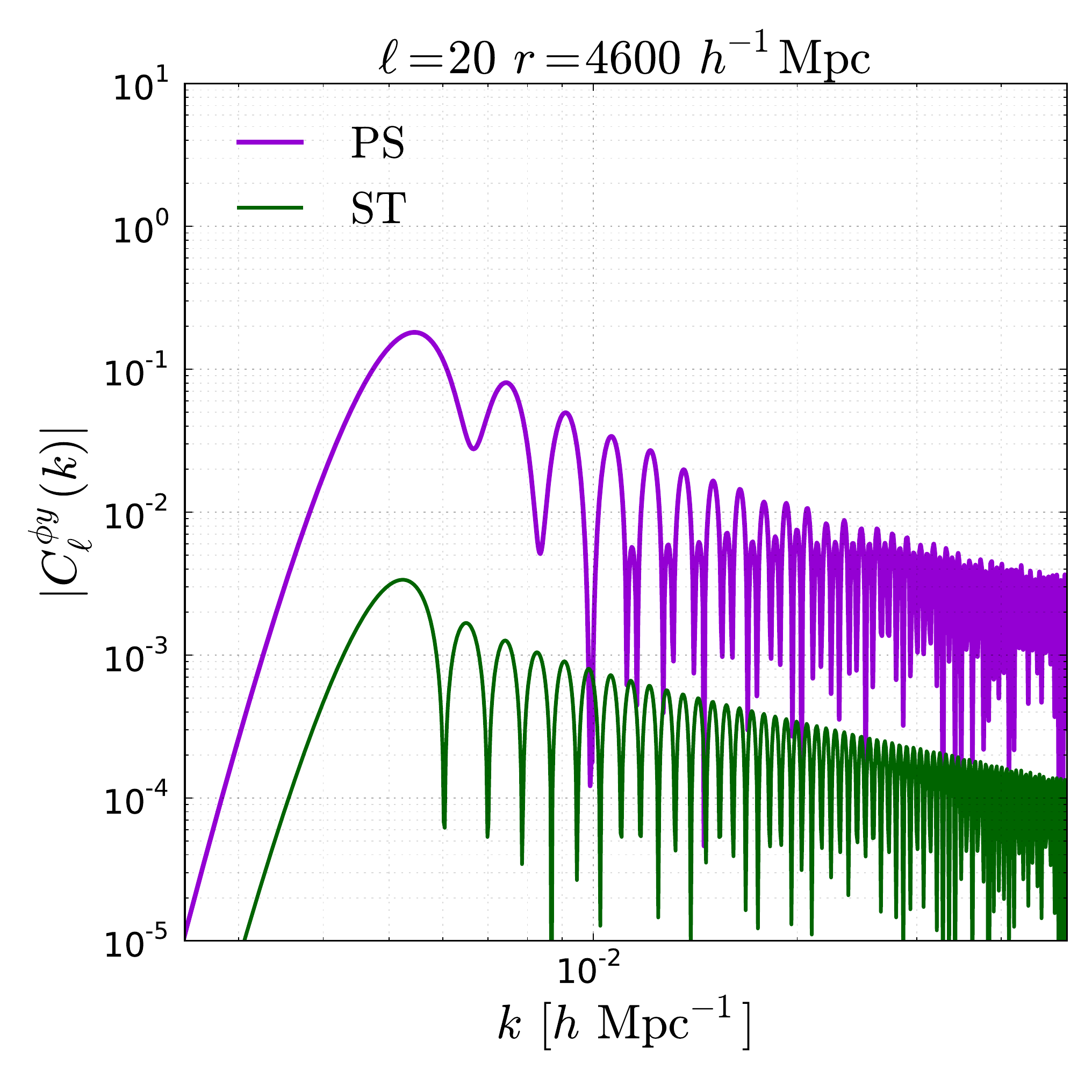}}
\caption{The results acquired by using the PS mass function are compared with that from ST mass function. The left panel shows the density-density power spectrum in both PS and ST along with the spectral and bias coefficients. Note that the ST power spectrum is suppressed at very small $k$ but has more power in the pressure-pressure and density-pressure spectra as seen in the cross-spectral coefficient $r_{\Pi \Pi}$. This results in the suppression of the tSZ-WL cross-correlation at the low $\ell$ modes considered in this paper seen in the right panel. Note that for $\ell \sim 20$ the peak of the spectra is on order $10^{-3} - 10^{-2} \, h \rm Mpc^{-1}$ and this corresponds to the suppressed regime for the ST mass function in the left panel. }
\label{fig:ST_tSZ-WL}
\end{figure}

In addition to the above, we can construct the cross-spectra as a function of mass bins. The halo model, as we saw, is fundamentally dependent on the underlying mass function, see Figure \ref{Fig:HaloPower_MassBin} for an example of three different mass bins. This means that at different masses we expect different physics to become more or less dominant. We expect the tSZ to be sensitive to the maximum mass scale and distribution of halos at high masses. This also means that we expect the tSZ to depend on the underlying mass function adopted in the study. Although we have used the Press-Schechter (PS) formalism, there are more modern alternatives that may be used. Examples include the extended Press-Schechter formalism \citep{Bond91}, the Sheth-Torman (ST) mass function \citep{ST99}, the Jenkins et al fit \citep{Jenkins01} and the Tinker et al fit \citep{Tinker08}. Each of these has their own pros and cons as well as the underlying assumptions that are fed into the models. For instance the Sheth-Torman mass 
function 
is thought to be more accurate at low masses and is a refinement of the Press-Schechter formalism which itself over-estimates the abundance of high mass halos and under-estimates the abundance of low mass halos. In Figure \ref{fig:ST_tSZ-WL} we plot the results obtained from using the ST mass function. These results are preliminary and a more in-depth study of the dependency of the tSZ-WL cross-correlation on the underlying mass function will be presented in a later paper. 
\begin{figure}
 \centering
 \textbf{Halo Model Power Spectra: Contributions from Mass Bins}\par\medskip
{\includegraphics[width=0.25\textwidth]{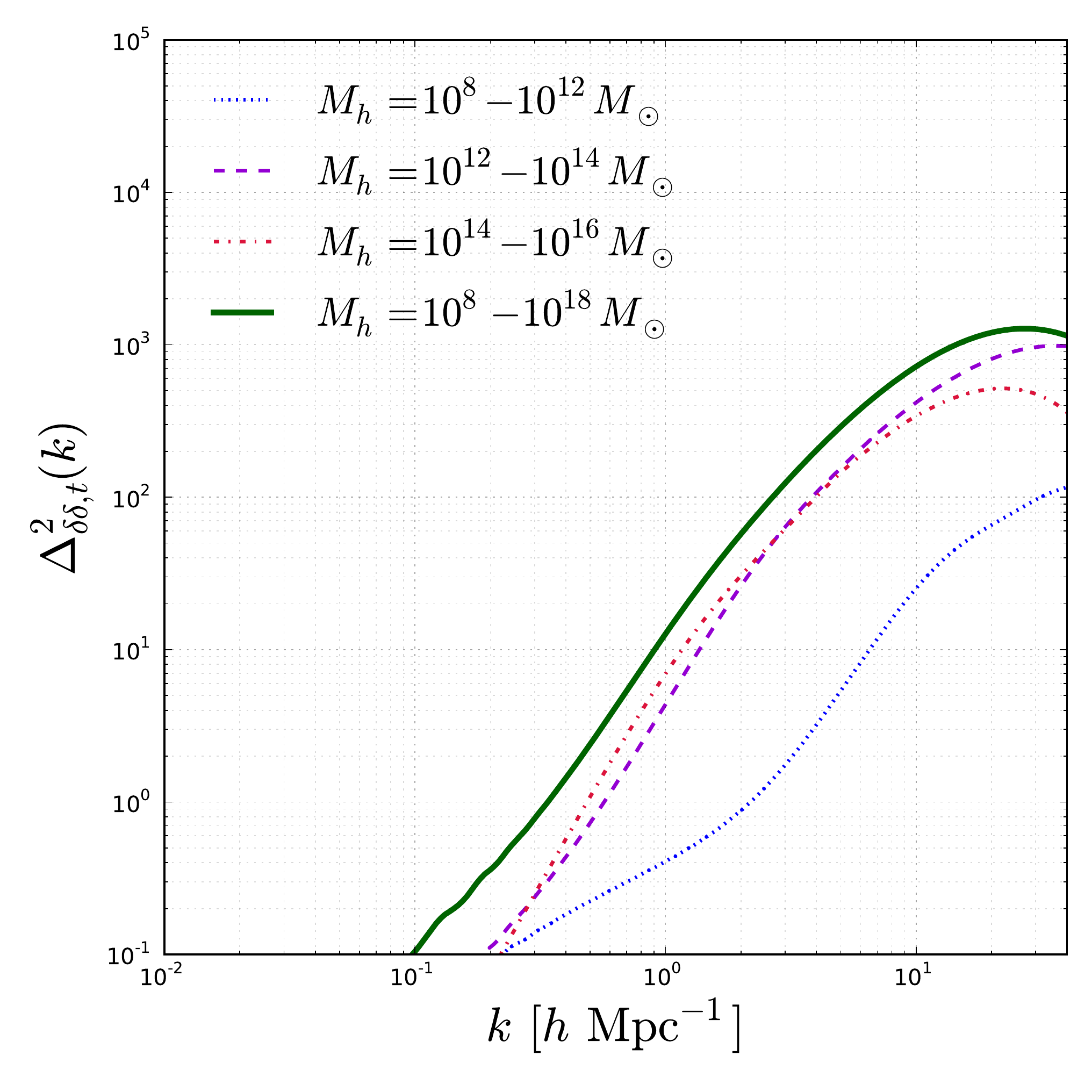}}
{\includegraphics[width=0.25\textwidth]{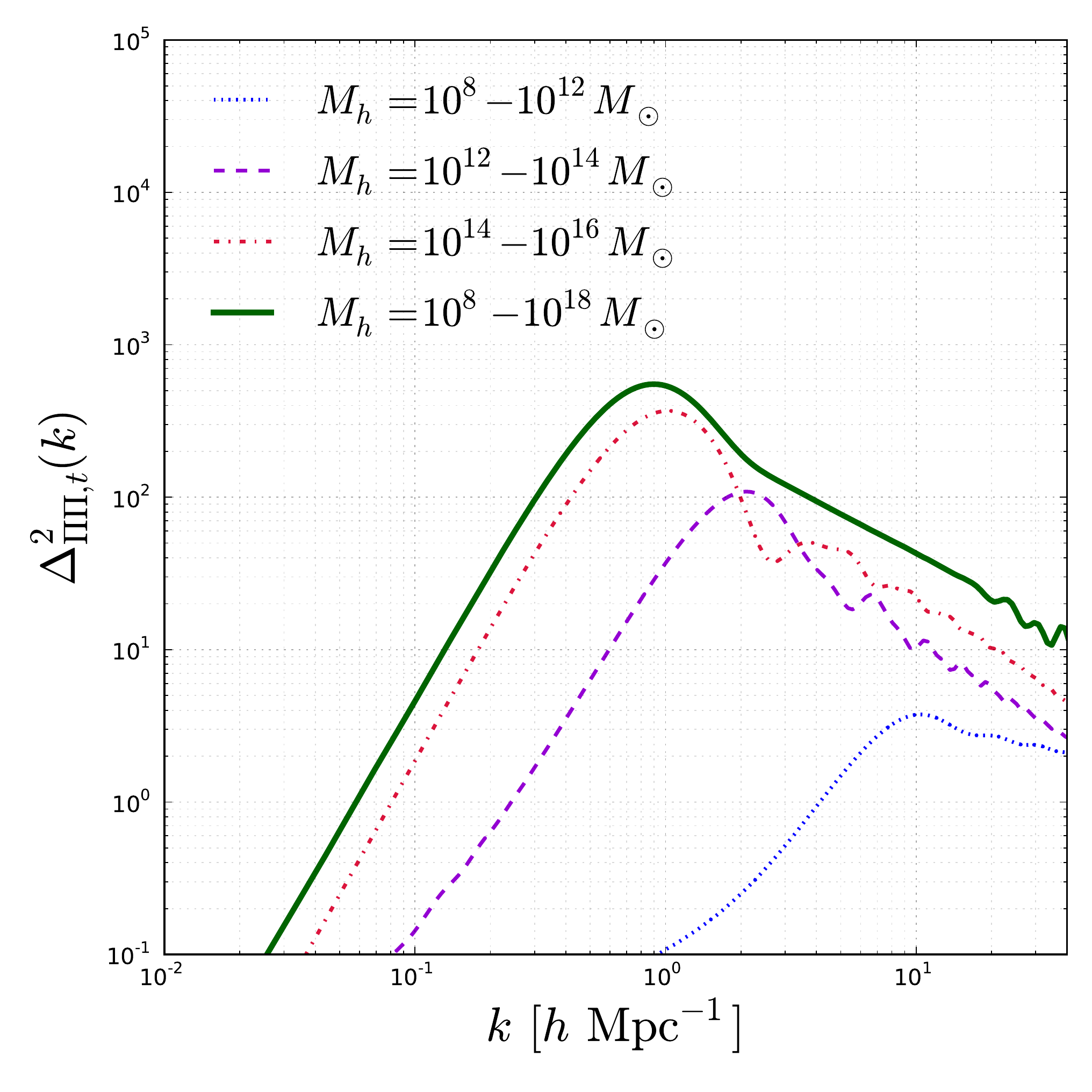}}
{\includegraphics[width=0.25\textwidth]{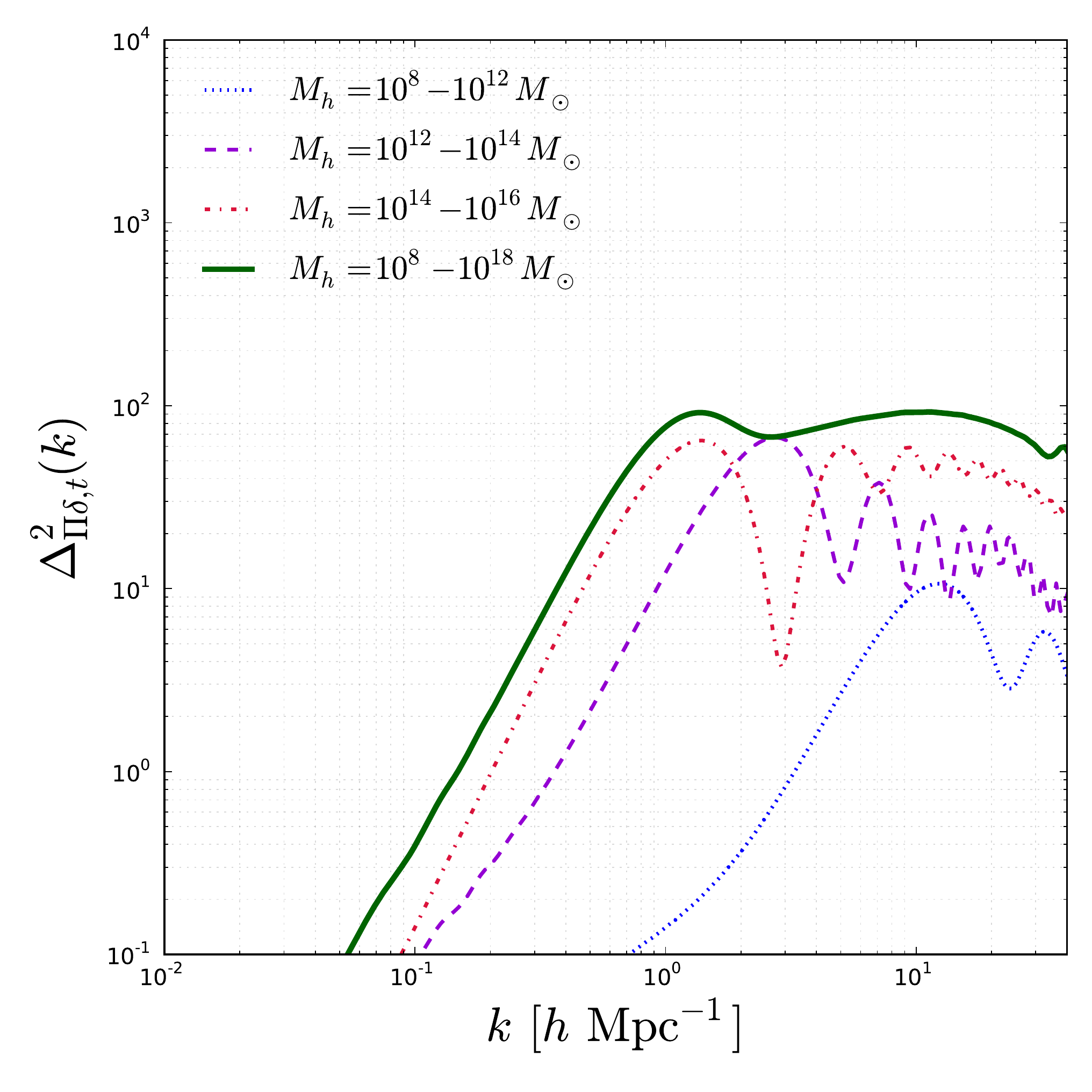}}
 \caption{The halo model allows us to consider the power spectrum as a function of mass. In this instance we take three different mass scales and look at the contributions to the power spectra in each case. It can clearly be seen that halos at low masses do not contribute significantly to the overall power. Additionally, the low masses are much more sub-dominant in the pressure-pressure spectra and pressure-density spectra than the density-density power alone. The tSZ effect is strongly dependent on the maximum mass. The three mass bins are: $M_h = 10^8 - 10^{12} M_{\sun}, M_h = 10^{12} - 10^{14} M_{\sun}$ and $M_h = 10^{14} - 10^{18} M_{\sun}$.}
 \label{Fig:HaloPower_MassBin}
\end{figure}
\begin{figure}
 \centering
 \textbf{tSZ-WL Cross-Correlation in the Halo Model: Differential Contributions from Mass Bins}\par\medskip
 \includegraphics[width=0.6\textwidth]{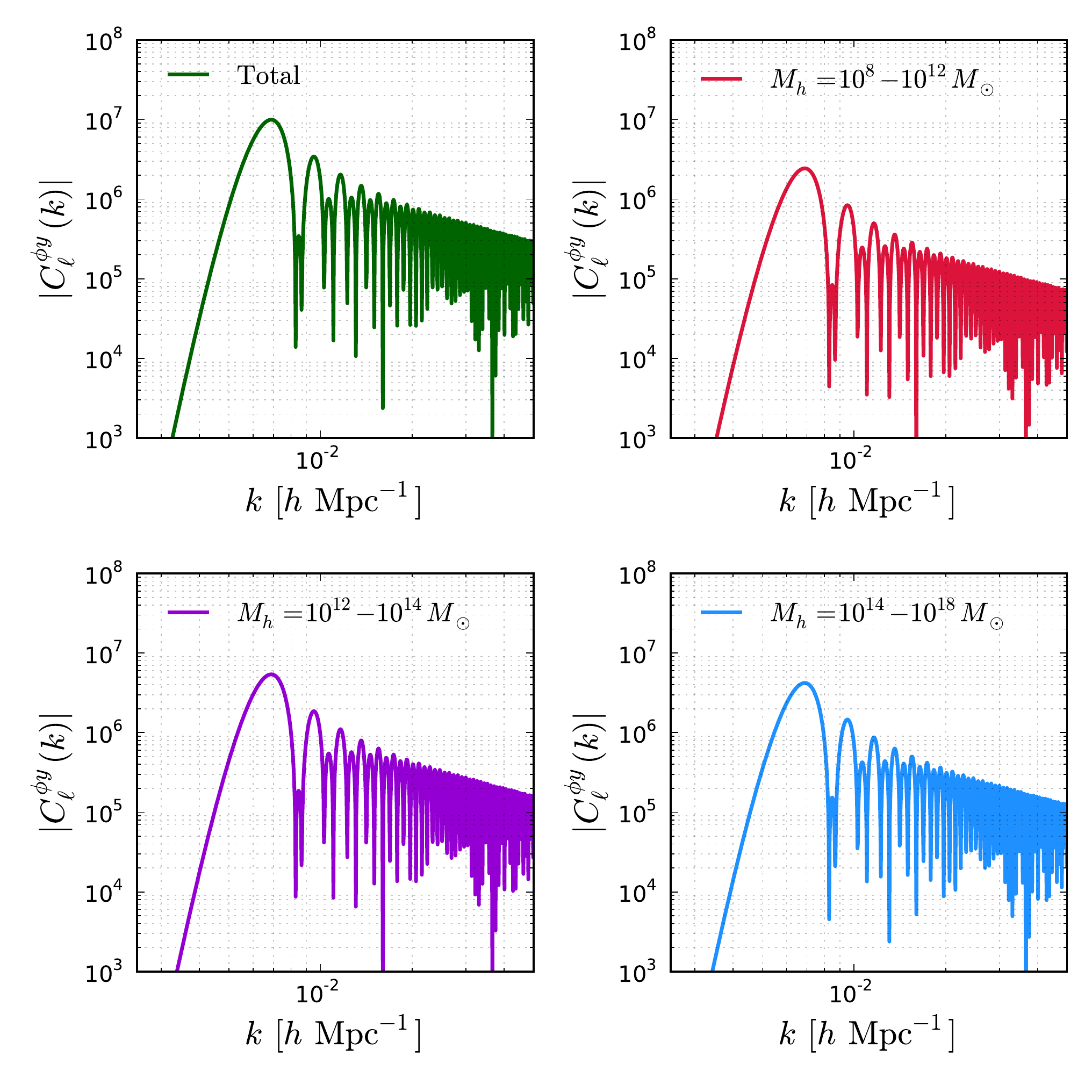}
 \caption{In this figure we plot the tSZ-WL cross correlation as a function of mass bin. The upper left panel is the total mass range. The upper right panel is the lowest mass bin $M_h = 10^8 M_{\sun}- 10^{12} M_{\sun}$. The lower left panel is an intermediate mass range $M_h = 10^{12} M_{\sun} - 10^{14} M_{\sun}$. The lower right panel is the largest mass bin at $M_h = 10^{14}M_{\sun} - 10^{18} M_{\sun}$. Note that at high masses the abundance of halos starts to drop off sharply and although the tSZ has strong contributions at high masses, the low abundance suppresses the overall contribution.}
 \label{Fig:HaloPower_MassBin}
\end{figure}

\subsection{Cross-correlating tSZ with Spectroscopic Redshift Surveys}
\begin{figure}
\centering
\textbf{tSZ-Spectroscopic Redshift Survey Cross-Correlation: Total Spectra}\par\medskip
{
  \includegraphics[width=55mm]{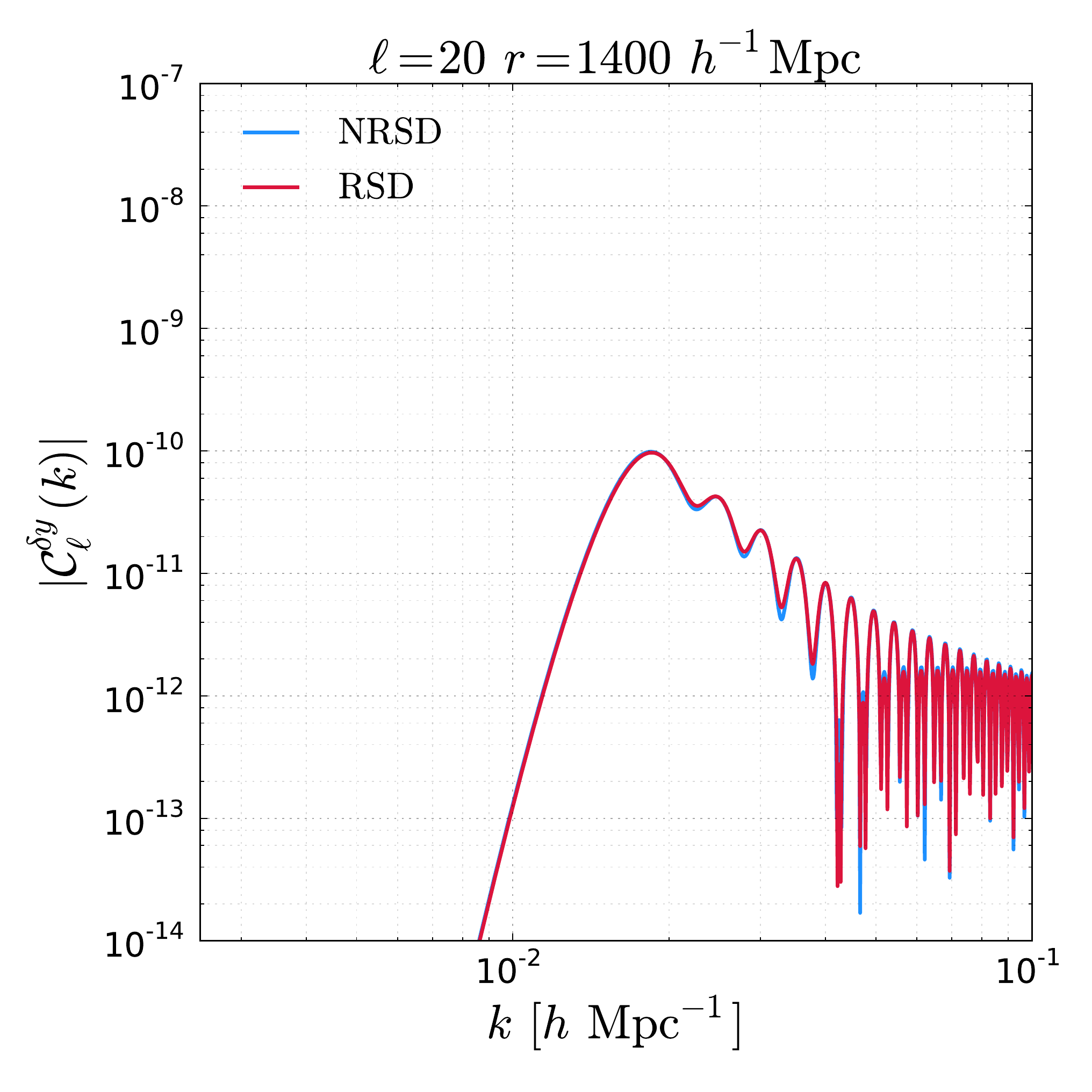}
}
{
  \includegraphics[width=55mm]{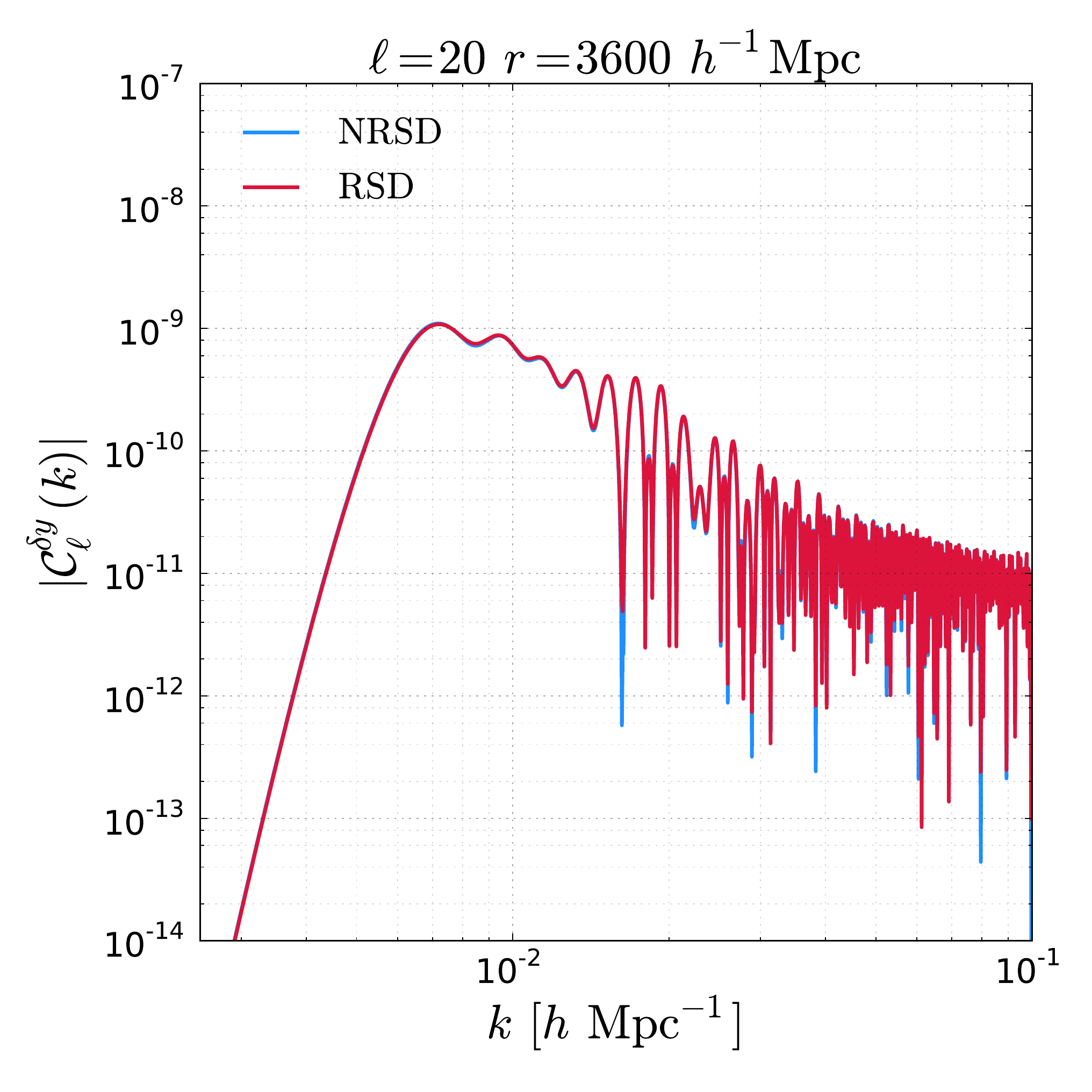}
}
{
  \includegraphics[width=55mm]{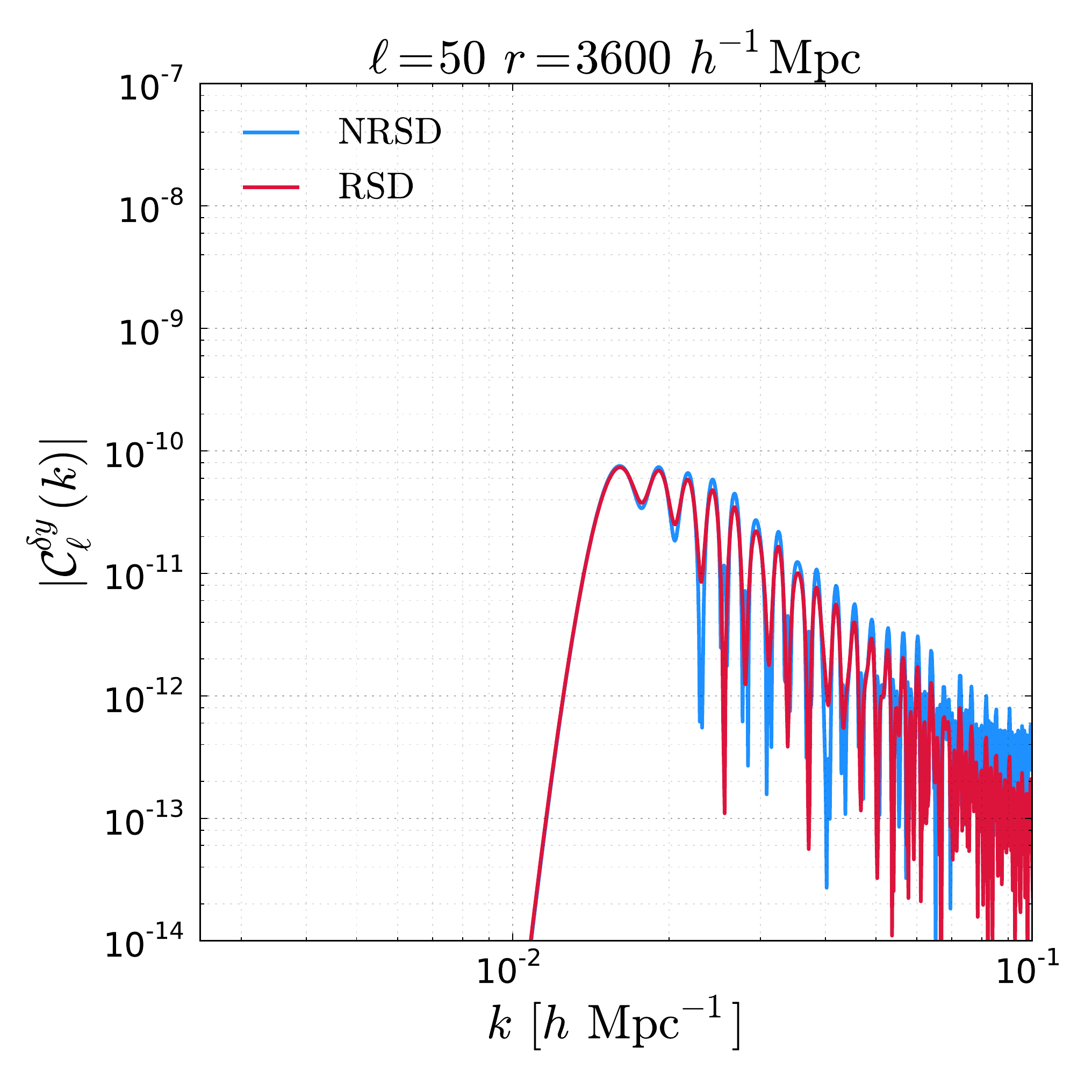}
}
\caption{Here we plot the effect of redshift space distortions (RSDs) on the galaxy-tSZ cross-correlation. As seen in previous studies \citep{PM13} the RSDs induce mode-mixing in the power spectra. This occurs due to the redshifted terms being related to derivatives of the spherical Bessel functions which do not form an orthogonal basis. This means that we have off-diagonal elements related to radial mode-mode coupling. As a result the power spectrum is damped. For the spectra shown here, the RSD contributions are rather negligible. For high $\ell$ we see more prominent contributions, as demonstrated in the right most panel. See Figure \ref{fig:tSZ-RSD-Sm} for the redshifted spectra divided by the unredshifted spectra isolating the modulations induced by the RSDs.}
\label{fig:tSZ-RSD}
\end{figure}

In order to consider the cross-correlation of the tSZ effect with spectroscopic redshift surveys, we need to take into account the partial observation effects that arise from finite survey volumes. In the case of Galaxy surveys, the observed field $\Psi^{\rm obs}({\bf r})$ is convolved with a radial selection function $\phi(r)$ that simply denotes the probability of including a galaxy within a given survey. The observed ({\em pseudo}) random field can be related to a 3D underlying random field via the survey dependent selection function $\phi(r)$ \citep{Rassat12,PM13}
\begin{align}
\Psi^{\rm obs}({\bf r}) = \phi(r)\Psi({\bf r}) .
\end{align}
\n
The observed power-spectrum and the underlying power-spectrum are linked through the following relation \citep{PM13}:
\begin{align}
\myC_{\ell}(k_1,k_2) &= \int_0^{\infty} dk' \, k'^2 \, {\cal I}^{(0)}_{\ell}(k_1,k') \, {\cal I}^{(0)}_{\ell}(k_2,k') \, P_{\delta \delta}(k') \\
{\cal I}^{(0)}_{\ell} &= \displaystyle\int_0^{\infty} dr \, r^2 \, \phi (r) \, k \,   j_{\ell}(k r)  j_{\ell}(k' r).
\label{eq:red}
\end{align}
\n
This power spectra will tend to decay rapidly as we move away from the diagonal $k = k^{\prime}$ and it is often most useful to focus on the diagonal contribution $\myC_{\ell} (k_1 , k_1)$. Following the procedure detailed in \citep{PM13} we can expand these results to include the effect of redshift-space distortions (RSDs). We briefly summarise the key steps but refer the reader to \citep{PM13} for further details and references. These distortions arise from the effects of a peculiar velocity, or departure from the Hubble flow, ${\bf{v}} ({\bf{r}})$ at ${\bf{r}}$ on the observed galaxy positions in redshift space ${\bf{s}}$
\begin{align}
 {\bf{s}} ({\bf{r}}) &= {\bf{r}} + {\bf{v}} ({\bf{r}}) \cdot \oh .
\end{align}
\n
We then construct the harmonics of the field $\Psi ({\bf{r}})$ convolved with the selection function $\phi (s)$:
\begin{align}
 \tilde \Psi_{\ell m}(k) &= \sqrt{2 \over \pi} \int_0^{\infty} s^2\,ds\,d\oh\, \phi(s) \Psi({\bf r})j_{\ell}(ks) Y^*_{\ell m}(\oh) .
\end{align}
\n
The Fourier transform of the velocity field is related to the Fourier transform of the density contrast via the linearised Euler equation:
\begin{align}
{\bf{v}} ({\bf{k}}) &= - i \beta {\bf{k}} \frac{\delta ({\bf{k}})}{k^2}.
\end{align}
\n
where $\beta = \Omega^{\gamma}_m / b$ and $b$ is a linear bias parameter. We take $b = 1$ and $\gamma \approx 0.55$ in our numerical calculations. This allows us to establish a series expansion in $\beta$ where the lowest order coefficient is obtained by neglecting RSD. The series expansion is schematically given by:
\begin{align}
 \tilde{\Psi}_{\ell m} (k) &= \tilde{\Psi}^{(0)}_{\ell m} (k) + \tilde{\Psi}^{(1)}_{\ell m} (k) + \dots \\
 \tilde{\Psi}^{(0)}_{\ell m} (k) &= \sqrt{\frac{2}{\pi}} \int^{\infty}_0 dk^{\prime} \, k^{\prime} \, \Psi_{\ell m} (k^{\prime}) \, {\cal I}^{(0)}_{\ell} (k^{\prime} , k) 
 \tilde{\Psi}^{(1)}_{\ell m} (k) = \sqrt{\frac{2}{\pi}} \int^{\infty}_0 dk^{\prime} \, k^{\prime} \, \Psi_{\ell m} (k^{\prime}) \, {\cal I}^{(1)}_{\ell} (k^{\prime} , k) .
\end{align}
\n
The kernels ${\cal I}^{(0)} (k^{\prime} , k)$ and ${\cal I}^{(1)} (k^{\prime} , k)$ define the convolution and do depend on the choice of selection function. These kernels are given by:
\begin{align}
{\cal I}_{\ell}^{(0)} (k,k') = \int_0^{\infty}\,dr\,r^2\,\phi(r)\,k\,j_{\ell}(kr)j_{\ell}(k'r); \quad
{\cal I}_{\ell}^{(1)} (k,k') = \frac{\beta}{k^{\prime}} \; \int_0^{\infty} dr \, r^2 k \, {d \over dr} [\phi(r) j_{\ell}(kr)] j^{\prime}_{\ell}(k^{\prime} r); \label{eq:phi2}
\end{align}
\n
The power spectra can be calculated from these harmonic coefficients as follows:
\begin{align}
 \left\langle \tilde{\Psi}_{\ell m}^{\alpha} (k) \tilde{\Psi}_{\ell^{\prime} m^{\prime}}^{\beta \ast} (k^{\prime})  \right\rangle &= \tilde{\myC}_{\ell}^{(\alpha \beta)} (k,k^{\prime}) \, \delta_{\ell \ell^{\prime}} \, \delta_{m m^{\prime}}.
\end{align}
\n
The total redshifted power spectrum is given as a sum over various contributions:
\begin{align}
\tilde\myC_{\ell}(k_1,k_2) &\equiv \sum_{\alpha,\beta}\tilde\myC^{(\alpha,\beta)}_{\ell}(k_1,k_2) = 
\tilde\myC^{(0,0)}_{\ell}(k_1,k_2)+2\tilde\myC^{(0,1)}_{\ell}(k_1,k_2)+ \tilde\myC^{(1,1)}_{\ell}(k_1,k_2)+\cdots;\\
\tilde\myC^{(\alpha,\beta)}_{\ell}(k_1,k_2) &\equiv \la \Psi_{\ell m}^{\alpha}\Psi_{\ell m}^{\beta*}\ra = \left ( {2 \over \pi} \right )^2 
\int_0^{\infty} k^2 dk \; {\cal I}^{(\alpha)}_{\ell}(k_1,k) \; {\cal I}^{(\beta)}_{\ell}(k_2,k) P_{\delta \delta}(k).
\end{align}
Now that we have the machinery in place to construct the spectroscopic redshift survey spectra, we can now construct the cross-correlation between the tSZ pressure fluctuations $y (\Omega)$ and the 3D density contrast $\delta$ (Figure \ref{fig:tSZ-RSD})
\begin{align}
\myC_\ell^{\delta y}(k) &= {4 \over \pi^2 c^{{4}}} \sum_{\alpha} \int_0^{\infty} dk' k'^2 {\cal I}^{\alpha}_\ell(k, k'){\cal I}^{y}_\ell(k') P_{\delta\delta}(k'); 
\label{eq:dy}
\end{align}
\n
In Figure \ref{fig:tSZ-RSD} we show features of the 3D tSZ-density power spectrum by taking slices through the full 3D space $(k_1 , k_2 , \ell)$. We consider diagonal contributions, $k_1 = k_2$, with a survey up to $r_{\textrm{max}} = 3600 h^{-1} \textrm{Mpc}$, a selection function with radial parameter $r_0 \in \lbrace 1400, 3600 \rbrace \, h^{-1} \, \textrm{Mpc}$ and multipoles $\ell = 5, 50$. 

Using Limbers' approximation, Eq.(\ref{eq:red}) can be simplified to:
\begin{align}
\myC_{\ell}(k,k') &= \delta_{\rm K}({k-k'})\left ( {\pi \over 2 k} \right )^2 \phi^2 \left ( {\nu \over k} \right ) P_{\delta\delta}(k); \quad\quad \nu={\ell+{1\over 2}}.
\end{align}
\n
and Eq.(\ref{eq:dy}) simplifies considerably as the RSD correction terms become negligible in the high $\ell$ limit. This can be seen from Eq. (\ref{eq:bess1}) and Eq.(\ref{eq:bess1}) when substituted into Eq.(\ref{eq:phi2}). Due to the choice of boundary conditions in the selection function $\phi (r)$ vanishes at $r = 0$ and $r = \infty$, this allows us to reverse the order of integration leading the the above simplifications. This leads to following expressions for the kernels $I^{(0)}_{\ell}(k,k')$ and $I^{(1)}_{\ell}(k,k')$
\ben
I^{(0)}_{\ell}(k,k') && \equiv {\pi \over 2\nu}{1 \over k} \phi\left ({\nu \over k}\right ) \delta_{1D}(k-k'); \\
I^{(1)}_{\ell}(k,k') && \equiv {\beta}{k \over k'}\int_0^{\infty} r^2 dr {d \over dr} \left [\phi(r) j_\ell(kr) \right ]
j^{\prime}_{\ell}(kr)\nn\\
&&  = \beta {k \over k'} \int \phi(r) j_{\ell}(kr) \left [ 2r j^{\prime}_{\ell}(k'r) +
r^2\,k\,j^{\prime\prime}_{\ell}(k^{\prime}r)\right ]
= {7 \over 8}\; {\pi \beta \over \nu^2}\; k \; \phi\left({\nu \over k}\right) \delta_{1D}(k-k').
\een
Where we have used the following approximate forms for $j^{\prime}_{\ell}(x)$ and $j^{\prime\prime}_{\ell}(x)$ defined in
Eq.(\ref{eq:bess1}) and Eq.(\ref{eq:bess2}):
\ben
&& j^{\prime}_{\ell}(x) \approx -{\pi^{1/2} \over (2\nu)^{3/2}} \delta_{1D}(\nu-x)
;\quad 
j_{\ell}^{\prime\prime}(x) \approx -{3 \pi^{1/2}\over (2\nu)^{5/2}  } \delta_{1D}(\nu-x).
\een
\section{Conclusion}
\label{sec:conclu}
In this paper we have extended in detail a study of 3D thermal Sunyaev-Zel'dovich cross correlations with cosmological weak lensing and spectroscopic redshift surveys. Most previous studies to date have focused on either projected studies or tomographic reconstruction. In projection studies, information is lost in the sense that by projecting onto the 2D sky we necessarily disregard information concerning distances to individual sources. An alternative approach is tomography, this is something of a hybrid method between 3D studies and 2D projection. In tomography the sources are divided into redshift slices on which a 2D projection is performed. This means that we foliate our sky with projections in a given redshift bin. This is a rather crude division and does not capture the full 3D information that will be possible in upcoming large scale structure surveys. The method proposed in this paper is based on a 3D spherical Fourier-Bessel expansion in which we aim to use distance information from the start. Note 
that certain parameters will less sensitive to this inclusion of distance information, such as the amplitude of the power spectra, but for others, notably those that depend on the line-of-sight of history of the Universe, 3D methods could be a very promising avenue of research. This paper encapsulates a few interesting results as well as summarising some of the key features present in the sFB formalism.
\begin{figure}
\centering
\textbf{tSZ-Spectroscopic Redshift Survey Cross-Correlation: Effect of RSD}\par\medskip
{
  \includegraphics[width=55mm]{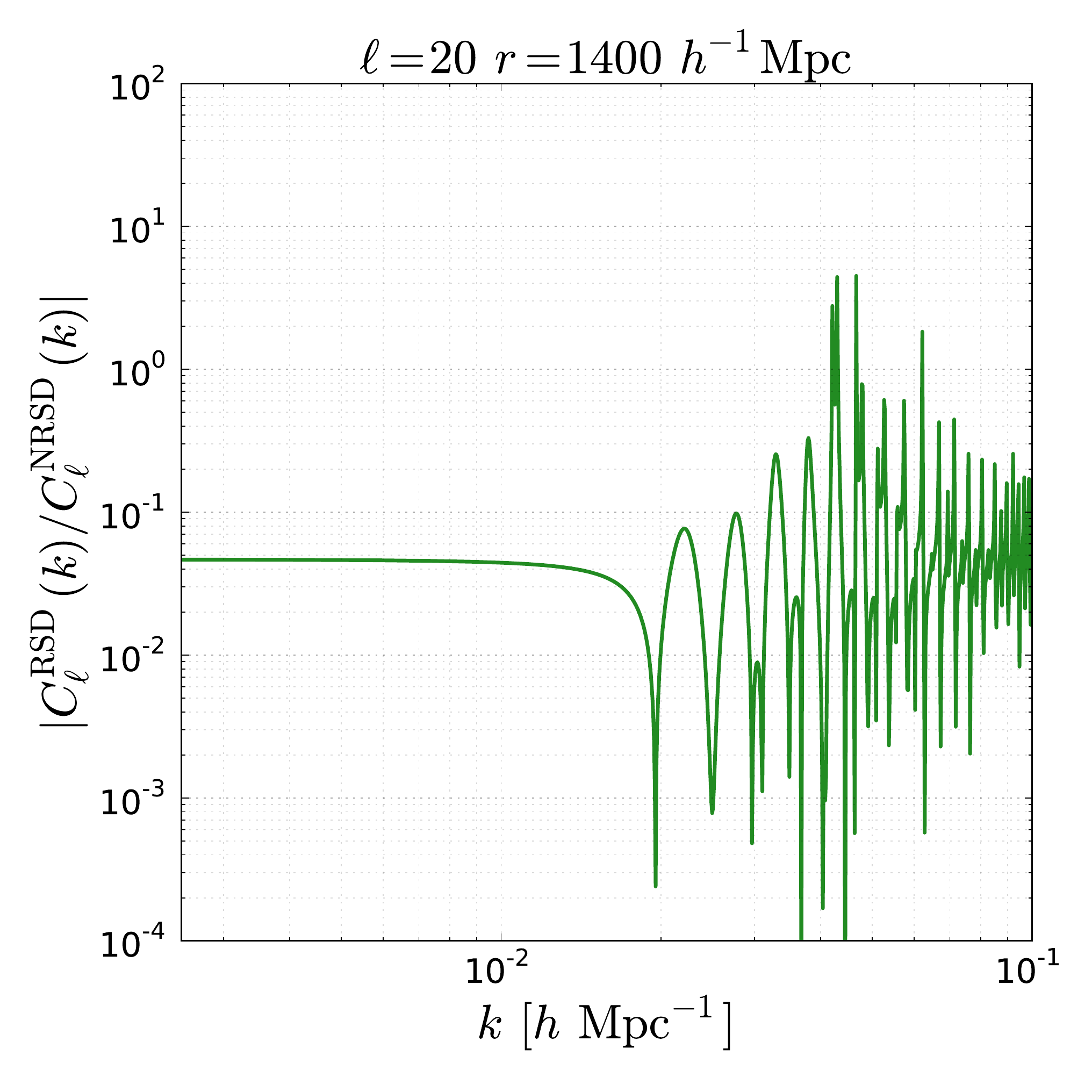}
}
{
  \includegraphics[width=55mm]{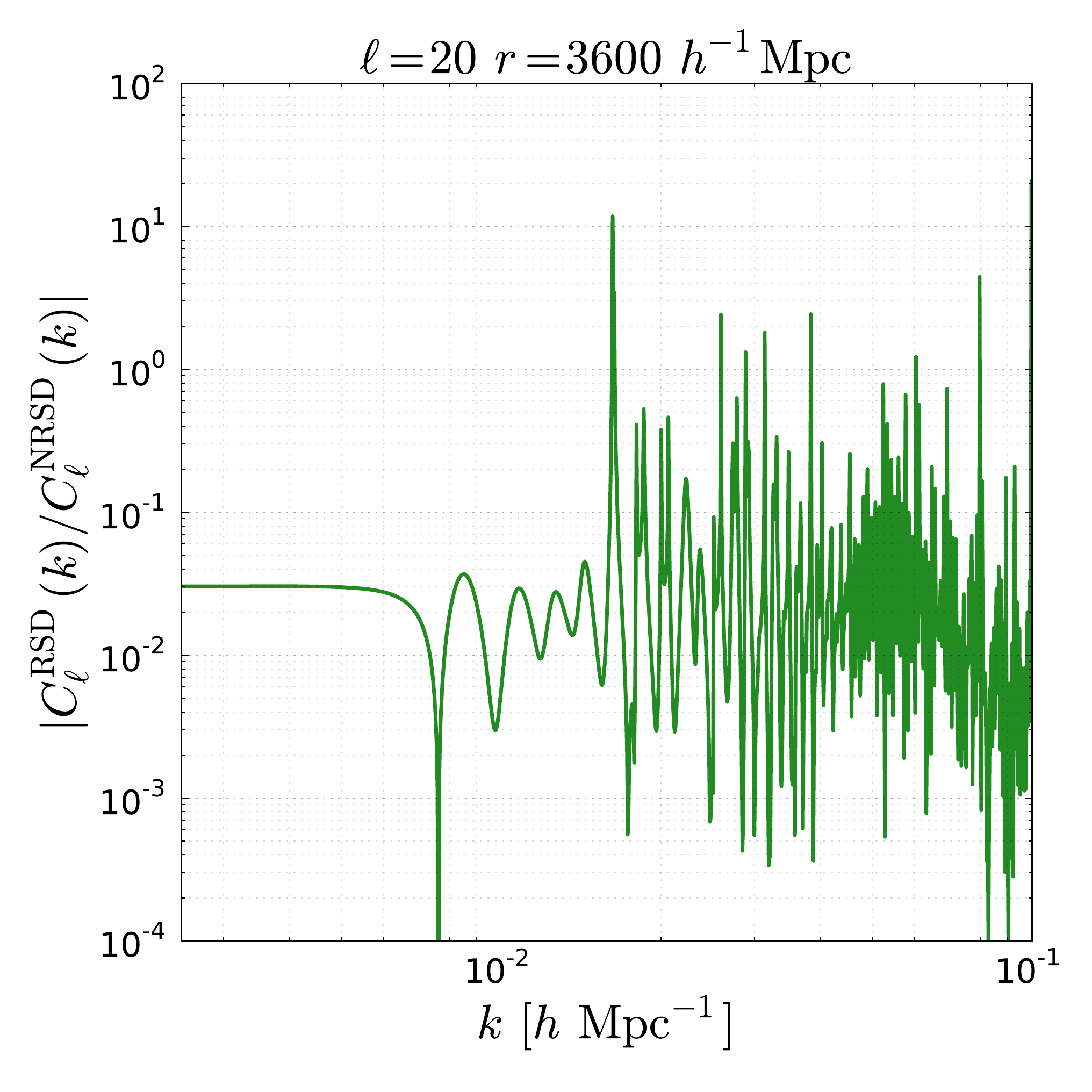}
}
{
  \includegraphics[width=55mm]{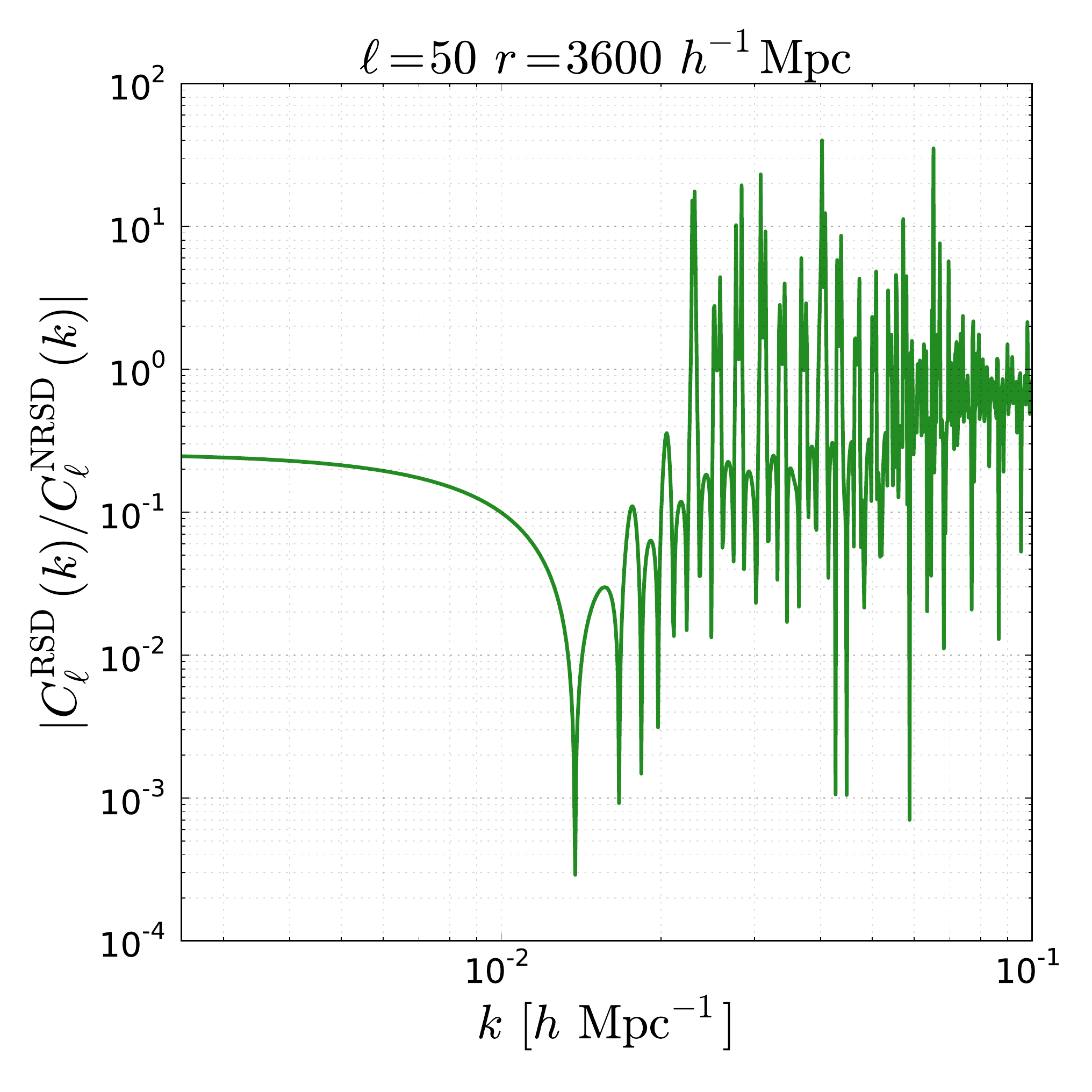}
}
\caption{Here we plot the effect of redshift space distortions (RSDs) smoothed by the unredshifted power spectra. Each of the panels corresponds to the panels shown in Figure \ref{fig:tSZ-RSD}. The RSD induce radial mode-mixing meaning that power is smoothed across the modes. This is seen at low $k$ where the spectra including RSD have less power than their unredshifted counterparts. At higher $k$ we hit oscillatory features that differ from those in the pure unredshifted contributions and beyond $k \sim 10^{-1}$ we are in a noise dominated regime where the oscillations of the Bessel functions are prominent and numerics becomes tedious.  }
\label{fig:tSZ-RSD-Sm}
\end{figure}
In order to study the tSZ-WL cross correlation, we adopted two different approaches. The first approach used the standard linear power spectrum in the analysis. The second approach used the halo model of large scale clustering to construct a non-linear power spectrum for the analysis. The halo model takes into account  a number of interesting physical inputs. These include (amongst other inputs): the dark matter density profile, the gas density profile, the electron temperature as a function of halo mass, the mass function of halos and the overdensity of collapse.  This allows us to connect the underlying physics to the predicted spectra in a more explicit manner than before. We know that the tSZ effect is sensitive to the higher mass halos and by combining the WL observations with the tSZ observations we can probe both the underlying baryonic and dark matter distributions as a function of halo mass. We expect the tSZ-WL cross-correlation to be sensitive to the halo mass function and density profiles. 

We introduce the conditional probability function of photometric redshifts to bridge survey-dependent observations to the cleaner theoretical predictions. True observations of galaxies have an intrinsic dispersion error on the measured redshifts. In the sFB formalism, redshift errors and errors in distance simply translate into radial errors. This results in a coupling of the modes and the observations become smoothed along our line of sight. We considered survey dependent parameters suitable for the DES and the LSST. 

Finally, we constructed the cross-correlation of the tSZ effect with spectroscopic redshift surveys in order to study the effects that redshift space distortions would have on a cross correlation of the tSZ effect with galaxy surveys. The procedure followed the procedure outlined in \citep{PM13}. 

In our modelling, we have used different redshift dependent 
linear biasing schemes at large angular scales for modelling of the diffused tSZ effect in association with
the halo-model for collapsed objects as a tool to investigate the tSZ-WL cross-correlations in 3D.
We use both the Press-Schechter (PS) as well as the Sheth-Tormen (ST) mass-functions in our calculations, finding that the results are
quite sensitive to detailed modelling as most of the contribution to the tSZ effect comes from the extended tail of the mass function (one-halo term).
We provide a detailed analysis of surveys with photometric redshifts. In the case of
cross-correlation with spectroscopic redshift surveys we provide detailed estimates of the contributions from redshift-space distortions.
The signal-to-noise (S/N) of the resulting cross-spectra $\myC_{\ell}(k)$ for individual 3D modes, defined by the radial and tangential wave numbers $(k,\ell)$, remains comparable to, but below, unity though optimal binning is expected to improve the situation.

In summary, the thermal Sunyaev-Zel'dovich effect acts as a probe of the thermal history of the Universe and the primary observable, the Compton y-parameter, appears to have no significant dependence on the redshift. The integrated nature of the tSZ effect means that redshift information can be lost diminishing our ability to probe the redshift evolution of the baryonic Universe. That is why we also study cosmological weak lensing as a complimentary tracer. Weak lensing is predominantly effected by the gravitational potential along the line of sight and is therefore an external tracer for the underlying dark matter field. By reconstructing the mass distribution of the Universe, we can hopefully recover redshift information and probe the baryonic and dark Universes in a complimentary way. Constraints on the dark sector, such as studies of decaying dark matter or dark matter-dark energy interactions, have recently attracted a lot of attention. Such effects could be probed by the tSZ or kSZ effects 
(e.g. \cite{Xu13}).
Similarly, the halo model for large scale clustering offers strong potential for testing different approaches to the various input ingredients: mass function, dark matter profile, gas density profile, etc. To this extent, we have seen that the tSZ is sensitive to high mass halos and a cross-correlation may be an interesting tool constraining and testing models for large scale clustering physical assumptions that enter the halo model, such as halo density profiles or the halo mass function. In our analysis we neglected general relativistic corrections which may be both important and interesting in their own right, especially in forthcoming surveys \citep{Umeh12,February13,Yoo13,Andrianomena14}. 

Finally, we would like to point out that it is known that the IGM is most likely have been preheated by non-gravitational sources. The feedback from SN or AGN can play an important role. The analytical modelling of such non-gravitational processes is rather difficult. 
Numerical simulations \citep{SWH01, Sel01, silva00, silva04, WHS02, Lin04}
have shown that the amplitude of the tSZ signal is sensitive to
the non-gravitational processes, e.g. the amount of radiative cooling and energy feedback. It is also not straightforward to disentangle contributions
from competing processes. The inputs from simulations are vital for any progress. Our analytical results should be treated as a first step in this
direction. We have focused mainly on large angular scales where we expect the gravitational process to dominate and such effects to be minimal. 
Thus the affect of additional baryonic physics can
be separated using the formalism developed here. To understand the effect of baryonic physics we can use the
techniques developed in \citep{MuJoCoSm11} for different
components and study them individually.

\section{Acknowledgements}
\label{acknow}
GP acknowledges support from an STFC Doctoral Training Grant. 
DM acknowledges support from the Science and Technology
Facilities Council (grant numbers ST/L000652/1).
\bibliography{paper.bbl}
\appendix
\section{Spherical Bessel Function}
\label{appendixA}
\subsection{Recurrence Relation}
The recurrence relations satisfied by the spherical Bessel functions is expressed through the
following relations:
\ben
&& {d  \over dx} \left [ x^{\ell+1} j_{\ell}(x) \right ] = 
x^{\ell +1} j_{\ell+1}(x) ; \quad\quad\quad  j_{\ell-1}(x)+j_{\ell+1}(x) = {2\ell +1 \over x} j_{\ell}(x); \\
&& {d  \over dx} \left [ x^{-\ell} j_{\ell}(x) \right ] = 
x^{-\ell} j_{\ell+1}(x) ; \quad\quad\quad  \ell j_{\ell-1}(x) - (\ell+1)j_{\ell+1}(x) = {(2\ell +1)} {dj_{\ell}(x) \over dx}.
\een
\subsection{Higher order Derivatives}
The first derivative of the spherical Bessel functions can be expressed 
using the following recursion relation:
\ben
j_\ell'(r)= {1 \over 2\ell+1}\Big [\ell j_{\ell-1}(r) - (\ell+1)j_{\ell+1}(r)\Big ].
\label{eq:bess1}
\een
The second- and higher-order derivatives are deduced by successive application
of the above expression: 
\ben
j_\ell''(r)= \Big [{(2\ell^2 +2\ell-1) \over (2\ell+3)(2\ell+1)}j_{\ell}(r) -
{\ell(\ell-1) \over (2\ell-1)(2\ell+1) }j_{\ell-2}(r) - {(\ell+1)(\ell+2)\over (2\ell+1)(2\ell+3)}j_{\ell+2}(r)   \Big ].
\label{eq:bess2}
\een
\subsection{Orthogonality} 
The orthogonality relationship for the spherical Bessel functions is given by the following expression:
\begin{align}
\int k^2 j_\ell(kr_a) j_\ell(kr_b) dk &= \left [ {\pi \over 2} \right ] \frac{\delta_{\rm 1D}(r_a-r_b)}{r^2_a}.
\label {eq:limber_approx1}
\end{align}
\n
\subsection{Limber Approximation and its Extension}
The extended Limber approximation is also implemented through the following approximate relation:
\begin{align}
\int k^2 \, F(k) \, j_\ell (k r_a) \, j_\ell(k r_b) dk &\approx  \left [ {\pi \over 2 \nu} \right ] F \left ( {\ell \over r_a} \right )
\frac{\delta_{\rm 1D}(r_a-r_b)}{r_a^2}; \qquad \nu = \ell + 1/2.
\label{eq:limber_approx2}
\end{align}
Thus for high $l$ the spherical Bessel functions can be replaced by a Dirac delta function $\delta_{1D}$:
\begin{align}
\displaystyle \lim_{\ell \to\infty} j_\ell(x) &= \sqrt {\pi \over 2\nu} \delta_{\rm 1D} \left ( \nu - x\right ).
\label{eq:limber_approx3}
\end{align}
\n
At small $\ell$ this approximation breaks down. We can move to higher order in the Limber approximation by incorporating the corrections detailed in \citep{LoVerde08} for which
\begin{align}
 \int dx F(x) J_{\nu} (x) &= \left[ F(x) - \frac{1}{2} \frac{x^2}{\nu^2} F^{\prime \prime} (x) - \frac{1}{6} \frac{x^3}{\nu^2} F^{\prime \prime \prime} (x) \right]_{x = \nu} + \mathcal{O}(\nu^{-4}) .
\end{align}
\n
The spherical Bessel functions $j_{\ell} (x)$ can be related the ordinary Bessel function $J_{\nu} (x)$ via
\begin{align}
j_{\ell} (x) &= \sqrt{\frac{\pi}{2x}} \, J_{\nu} (x).
\end{align}

\section{Spin-Weighted Spherical Harmonics}
The spin-weighted spherical harmonics are a generalisation of the normal spherical harmonics to higher spins. These harmonics are defined in terms of the Wigner D-matrices \citep{Varsh88}:
\begin{align}
 _{s}Y_{\ell m} (\oh) &= \sqrt{\frac{2 \ell + 1}{4 \pi}} \, D^{\ell}_{-s , m} \left( \theta , \phi , 0 \right) .
\end{align}
\n
The Wigner D-matrices were originally introduced in quantum mechanics as an eigenfunction of the Hamiltonian for spherical and symmetric rigid rotors. The matrix is intrinsically connected to the irreducible representation of the $SU(2)$ and $SO(3)$ groups. The matrices are defined by
\begin{align}
 D^j_{m m^{\prime}} \left( \alpha , \beta , \gamma \right) &= {\rm exp} ( - i m^{\prime} \alpha ) \, d^j_{m^{\prime} m} \, \rm{exp} (- i m \gamma) 
\end{align}
\n
where $\lbrace \alpha , \beta , \gamma \rbrace$ are Euler angles and Wigner's small d-matrix is defined by the following expression
\begin{align}
 d^j_{m^{\prime} m} &= \left[ \left( j + m \right) ! \left( j - m \right) ! \left( j + m \right) ! \left( j - m \right) ! \right]^{1/2} \, \displaystyle\sum_s \frac{(-1)^{m^{\prime} - m + s}}{\left( j + m - s \right) ! s! \left( m^{\prime} - m + s \right) ! \left( j - m^{\prime} - s \right) !} \, \\ &\qquad \qquad \qquad \times  \left[ \cos \left( \frac{\beta}{2} \right) \right]^{2j + m - m^{\prime} - 2s} \, \left[ \sin \left( \frac{\beta}{2} \right) \right]^{ m^{\prime} + 2s - m} . 
\end{align}
\n
Note that these formula make assumptions about the order and structure of rotations around the Euler angles. Having defined the spin weighted spherical harmonics, we generalise the orthogonality relationship of the spin-$0$ spherical harmonics to spherical harmonics of spin-$s$
\begin{align}
 \int d \oh \, \left[ _{s}Y_{\ell m} (\oh) \right] \, \left[ _{s^{\prime}}Y_{\ell^{\prime} m^{\prime} } (\oh) \right] \, \left[ _{s^{\prime \prime}}Y_{\ell^{\prime \prime} m^{\prime \prime}} ( \oh ) \right] &= \sqrt{ \frac{ (2 \ell + 1) (2 \ell^{\prime} + 1) (2 \ell^{\prime \prime} + 1) }{4 \pi} } 
 \left ( \begin{array}{ c c c }
     \ell & \ell^{\prime} & \ell^{\prime \prime} \\
     m & m^{\prime} & m^{\prime \prime}
  \end{array} \right)
 \left ( \begin{array}{ c c c }
     \ell & \ell^{\prime} & \ell^{\prime \prime} \\
     -s & -s^{\prime} & -s^{\prime \prime}
  \end{array} \right) .
\end{align}
\n
Likewise we can express the completeness relations as follows:
\begin{align}
 \displaystyle\sum_{\ell m} \left[ _{s}Y_{\ell m} (\oh) \right] \, \left[ _{s}Y_{\ell m} (\oh^{\prime}) \right] &= \delta_{2D} \left( \oh - \oh^{\prime} \right) .
\end{align}
The orthogonality relation takes the following form:
\begin{align}
\int d\oh\; _{s}Y_{\ell m} (\oh) \; _{s^{\prime}}Y_{\ell^{\prime} m^{\prime}} (\oh) = \delta^{\rm K}_{ss^{\prime}}\delta^{\rm K}_{\ell\ell^{\prime}}\delta^{\rm K}_{mm^{\prime}}.
\end{align}
\n
Finally, the complex conjugate of a spin-$s$ harmonics is given by
\begin{align}
 _{s}Y_{\ell m}^{\ast} &= \left( - 1 \right)^m \, _{-s}Y_{\ell -m} (\oh) .
\end{align}
\subsection{$\edth$ and $\edthbar$}
Given the 2D Riemannian manifold $S^2$, we can define a null tetrad $\lbrace m , \bar{m} \rbrace$ constructed from the orthonormal basis vectors  $\lbrace \hat{e}_1 , \hat{e}_2 \rbrace$ spanning $S^2$
\begin{align}
 m^a = \frac{1}{\sqrt{2}} \left( \hat{e}^a_1 + i \hat{e}^a_2 \right); \quad
 \bar{m}^a = \frac{1}{\sqrt{2}} \left( \hat{e}^a_1 - i \hat{e}^a_2 \right)
\end{align}
\n
where $m^a \bar{m}_a = +1$ and $m^a m_a = \bar{m}^a \bar{m}_a = 0$. Using these results we can formally define the $\edth$ and $\edthbar$ operators as totally projected convective covariant derivatives with respect to the null tetrad. First, let $\eta_{a \dots b y \dots z}$ be a tensor on $S^2$, meaning that it has been projected into $S^2$ on every index such that there are $p$ indices in the first index and $q$ indices in the second. This means that we can define spin $s$ quantities as follows:
\begin{align}
 _{s}\eta &= \eta_{a_1 \dots a_s} \, m^{a_1} \dots m^{a_s} \\
 _{-s}\eta &= \eta_{a_1 \dots a_s} \, \bar{m}^{a_1} \dots \bar{m}^{a_s} \\
 \eta_{a \dots b y \dots z} &= \left[ _{s}\eta \right] m^{a} \dots m^b + \left[ _{-s}\eta \right] \bar{m}^y \dots \bar{m}^z .
\end{align}
\n
These objects transform under rotations as $_{s}\eta \rightarrow _{s}\eta \, e^{i s \psi}$, where $s = p-q$ denotes the spin weight of $\eta$. Now we are in a position to define \textit{edth} $\edth$ and \textit{edth bar} $\edthbar$ by
\begin{align}
 \edth \left[ _{s}\eta \right] &= m^a \dots m^b \bar{m}^y \dots \bar{m}^z \, m^c \nabla_c \, \eta_{a \dots b y \dots z} \label{eqn:edth} \\
 \edthbar \left[ _{s}\eta \right] &= m^a \dots m^b \bar{m}^y \dots \bar{m}^z \, \bar{m}^c \nabla_c \, \eta_{a \dots b y \dots z} \label{eqn:edthbar}  ,
\end{align}
\n
where $\nabla_c$ is the covariant derivative defined with respect to $S^2$. If $\eta$ has a spin weight of $+1$ then $\edth \eta$ has a spin weight of $s+1$ and $\edthbar \eta$ a spin weight of $s-1$. This is why $\edth$ is known as a spin raising operator and $\edthbar$ a spin lowering operator.

It is often most convenient to jump straight into a coordinate dependent definition for a general spin-$s$ field $_{s}\eta$. We assume that we are dealing with the standard unit sphere in spherical polar coordinates with a metric given by:
\begin{align}
 g_{ab} &= \rm diag \left( 1 , \sin^2 \theta \right) 
\end{align}
\n
for which we have the following non-zero Christoffel symbols:
\begin{align}
 \Gamma_{11}^0 = - \sin \theta \, \cos \theta; \quad \Gamma_{01}^1 = \Gamma^1_{10} = \cot \theta .
\end{align}
\n
In these coordinates, the natural null tetrad vectors are given by:
\begin{align}
 m^a = \frac{1}{\sqrt{2}} \, \left( 1 , i \csc \theta \right); \quad
 m_a = \frac{1}{\sqrt{2}} \, \left( 1 , i \sin \theta \right) 
\end{align}
\n
where the conjugate vectors are trivially defined. Using these definitions, we simply substitute these into Eqn. (\ref{eqn:edth}) and Eqn. (\ref{eqn:edthbar}). This results in the following general formula for $\edth$ and $\edthbar$ on the 2D sky:
\begin{align}
 \edth _{s}\eta &= - \sin^s \theta \, \left( \partial_{\theta} + i \csc \theta \, \partial_{\phi} \right) \, \left( \sin^{-s} \theta \right) \, _{s}\eta; \\
 \edthbar _{s}\eta &= - \sin^{-s} \theta \, \left( \partial_{\theta} - i \csc \theta \,  \partial_{\phi} \right) \, \left( \sin^{s} \theta \right) \, _{s}\eta.
\end{align}
\end{document}